%% file: paper.tex
\newif\ifarxiv
\arxivtrue 
\ifarxiv
\documentclass{article}
\usepackage{arxiv}
\else
\documentclass[review,onefignum,onetabnum]{siamonline220329}
\fi

\pdfoutput=1
\ifarxiv
\else
\ifpdf
\hypersetup{
  pdftitle={Let them have CAKES: A Cutting-Edge Algorithm for Scalable, Efficient, and Exact Search on Big Data},
  pdfauthor={M. E. Prior, T. J. Howard III, O. McLaughlin, T. Ferguson, N. Ishaq, N. M. Daniels}
}
\fi
\fi

\input{cakes_shared}
\begin{document}

\maketitle

\begin{abstract}
  The Big Data explosion has created a demand for efficient and scalable algorithms for similarity search.
  While much recent work has focused on \textit{approximate} $k$-NN search, \textit{exact} $k$-NN search has not kept up.
  We present CAKES, a set of three novel algorithms for exact $k$-NN search.
  CAKES's algorithms are generic over \textit{any} distance function, and do not scale with the cardinality or embedding dimension of the dataset. Instead, they scale with geometric properties of the dataset--namely, metric entropy and fractal dimension--thus providing immense speed improvements over existing exact $k$-NN search algorithms when the dataset conforms to the manifold hypothesis.
  We demonstrate these claims by contrasting performance on a randomly-generated dataset against that on some datasets from the ANN-Benchmarks suite under commonly-used distance functions, a genomic dataset under Levenshtein distance, and a radio-frequency dataset under Dynamic Time Warping distance. CAKES exhibits near-constant running time on data conforming to the manifold hypothesis as cardinality grows, and has perfect recall on data in metric spaces. CAKES also has significantly higher recall than state-of-the-art $k$-NN search algorithms even when the distance function is not a metric.
  We conclude that CAKES is a highly efficient and scalable algorithm for exact $k$-NN search on Big Data.
  We provide a Rust implementation of CAKES under an MIT license at https://github.com/URI-ABD/clam.

\end{abstract}
%

\maketitle

    \input{sections/1-introduction.tex}
    \input{sections/2-methods.tex}
    \input{sections/3-datasets.tex}
    \input{sections/4-results.tex}
    \input{sections/5-discussion.tex}


    \FloatBarrier
    \bibliographystyle{siamplain}
    \bibliography{references}
\end{document}



\maketitle

\section{Supplementary Methods}

\subsection{\texorpdfstring{$\rho$}{p}-Nearest Neighbors Search}

We conduct $\rho$-NN search as described in~\cite{ishaq2019clustered}, but with the following improvement:
when a cluster overlaps with the query ball, instead of always proceeding to search both of its children, we proceed only with those children that might contain points in the query ball.

\begin{figure}
    \centering
    \includegraphics[scale=0.75]{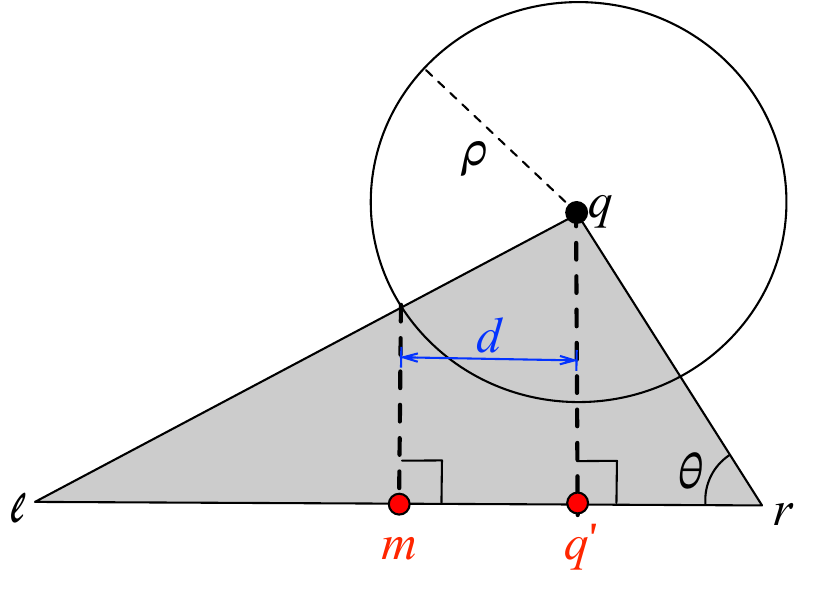}
    \caption{The geometry of a query ball overlapping with a cluster and either one or both of its children. Here, $l$ is the left pole, $r$ is the right pole, and $q$ is the query. Other points and distances are described in the text.}
    \label{fig:supplement:overlapping-children}
    \caption{CAKES uses geometric properties of clusters.}
\end{figure}

To determine whether both children can contain points in the query ball, we consider Figure~\ref{fig:supplement:overlapping-children}.
Here, we overload the notation for $\overline{x y}$ to refer both to the line segment joining points $x$ and $y$ as well as to the length of that line segment.

Let $q$ denote the query, $\rho$ denote the search radius, and $l$ and $r$ denote the cluster's left and right poles respectively.
Without loss of generality, we assume that $\overline{q r \vphantom{l}} \leq \overline{q l}$.
Now let $q'$ be the projection of $q$ onto $\overline{l r}$, $m$ be the midpoint of $\overline{l r}$, and $d$ be the distance from $q'$ to $m$.
As a consequence of how we assign a point in the parent cluster to the left child in the Partition algorithm, if $\rho < d$, then the left child cannot contain points inside the query ball.
In such a case we proceed to search only the right child.
Otherwise, we proceed with both children.

To check whether $d \leq \rho$, we note that $d = \overline{m q' \vphantom{l}} = \overline{m r \vphantom{l}} - \overline{q' r \vphantom{l}} = \frac{\overline{l r}}{2} - \overline{q' r \vphantom{l}}$.
Let $\theta$ denote $\angle l r q$, as shown in Figure~\ref{fig:supplement:overlapping-children}.
By the Law of Cosines on $\triangle l r q$, we have that $\text{cos}(\theta) = \tfrac{\overline{l r}^2 + \ \overline{q r \vphantom{l}}^2 - \ \overline{q l}^2}{2 \cdot \overline{l r} \cdot \overline{q r \vphantom{l}}}$.
Since $\triangle r q q'$ is a right triangle, we also have that $\text{cos}(\theta) = \tfrac{\overline{q' r \vphantom{l}}}{\overline{q r \vphantom{l}}}$.
Combining the previous two equations and solving for $\overline{q' r \vphantom{l}}$, we have that $\overline{q' r \vphantom{l}} = \tfrac{\overline{q r \vphantom{l}}^2 + \ \overline{l r}^2 - \ \overline{q l}^2}{2 \cdot \overline{l r}}$.
Substituting for $\overline{q' r \vphantom{l}}$ in the equation for $d$, we have that $d = \tfrac{\overline{l r}}{2} - \tfrac{\overline{q r \vphantom{l}}^2 + \ \overline{l r}^2 - \ \overline{q l}^2}{2 \cdot \overline{l r}} = \tfrac{\overline{q l}^2 - \overline{q r \vphantom{l}}^2}{2 \cdot \overline{l r}}$.

Thus, $d \leq \rho \iff (\overline{q l} + \overline{q r \vphantom{l}})(\overline{q l} - \overline{q r \vphantom{l}}) \leq 2 \cdot \overline{l r} \cdot \rho$. Note, in particular, that this only requires distances between actual points from the dataset, and so it can be used with any distance function, even when $q'$ and $m$ are not actual points or cannot be imputed from the data.

To perform $\rho$-NN search, we first perform a coarse \textit{tree-search}, to find the leaf clusters that overlap with the query ball or any clusters which lie entirely within the query ball.
Then, for all such clusters, we perform a finer-grained \textit{leaf-search}, to find all points that are no more than a distance $\rho$ from the query.
The asymptotic complexity of $\rho$-NN is the same as in~\cite{ishaq2019clustered} and shown in Equation~\ref{eq:supplement:rnn-search-complexity}.

\begin{gather}
    \mathcal{O}
    \Bigg(
        \underbrace{
            \log~\overbrace{\mathcal{N}_{\hat{r}}(X)}^{\textrm{metric entropy}}
        }_{\textrm{tree-search}}
        \ + \
        \underbrace{
            \overbrace{ \big| B_X(q, \rho) \big|}^{\textrm{output size}}
            \overbrace{ \left( \frac{\rho + 2 \cdot \hat{r}}{ \rho} \right) ^ d}^{\textrm{scaling factor}}
        }_{\textrm{leaf-search}}
    \Bigg)
    \label{eq:supplement:rnn-search-complexity}
\end{gather}
where $\hat{r}$ is the \textit{mean} radius of leaf clusters, $\mathcal{N}_{\hat{r}}(X)$ is the metric entropy at that radius, $B_X(q, \rho)$ is a ball of radius $\rho$ around the query $q$, and $d$ is the LFD around the query at the length scale of $\rho$ and $\rho + 2 \cdot \hat{r}$.

\section{Supplementary Results}

\subsection{Indexing and Tuning}

\begin{figure}
    \captionsetup[subfigure]{aboveskip=-15pt,belowskip=-3pt}
    \begin{subfigure}[b]{0.47\textwidth}
        \includegraphics[width=0.9\textwidth]{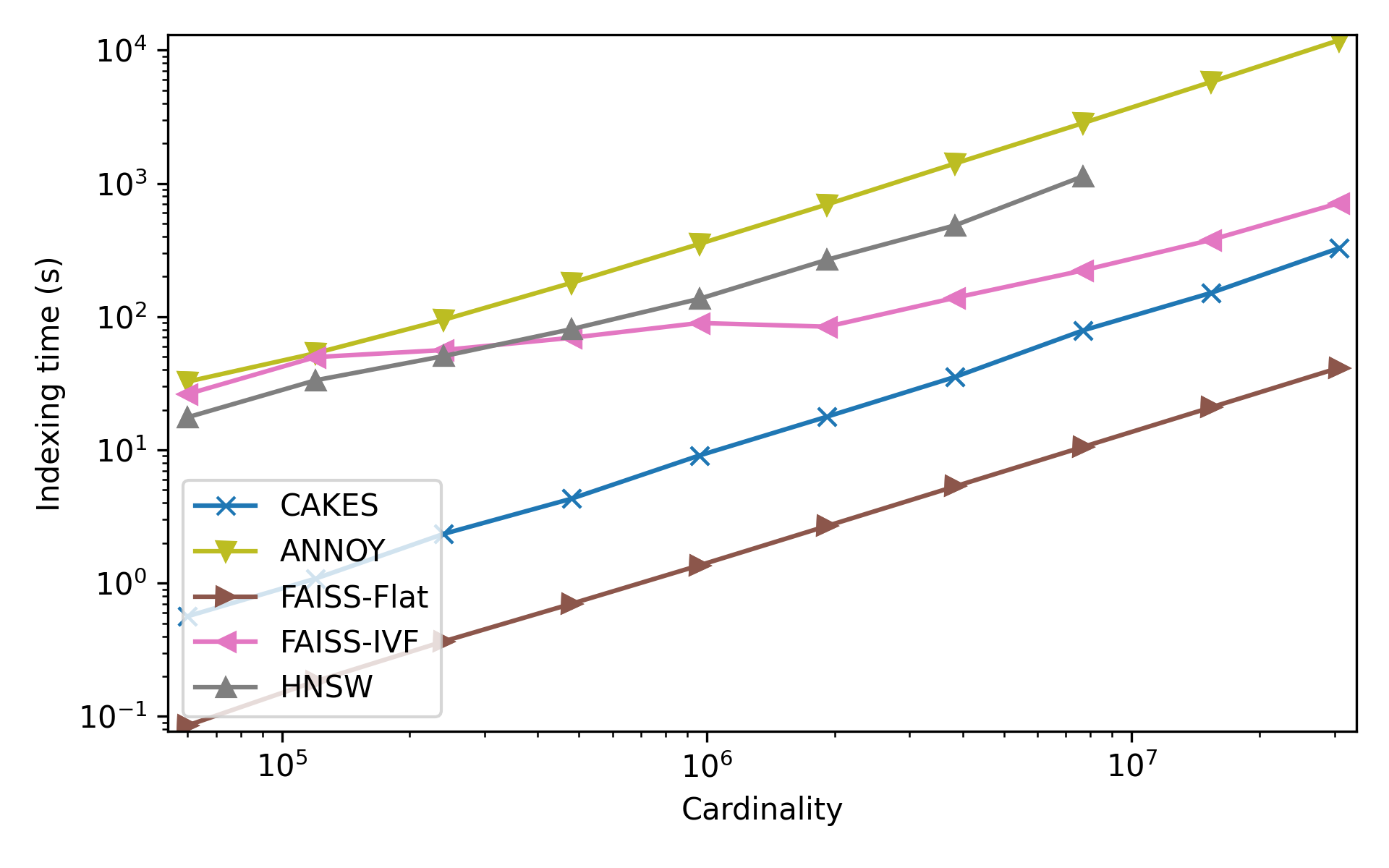}\\
        \subcaption{Fashion-mnist}
        \label{fig:supplement:fashion-mnist-indexing}
    \end{subfigure}%
    \begin{subfigure}[b]{0.47\textwidth}
        \includegraphics[width=0.9\textwidth]{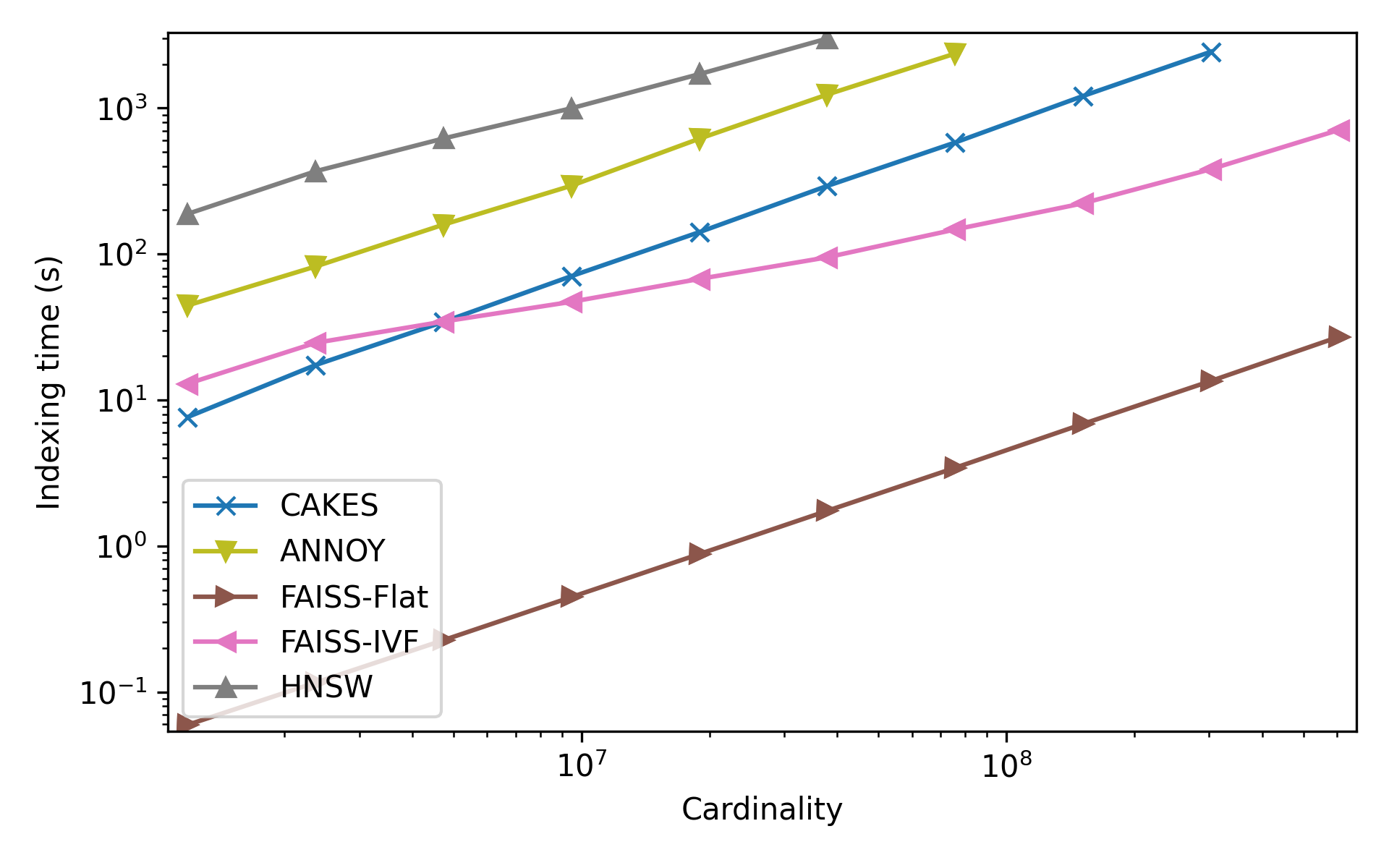}\\
        \subcaption{Glove-25}
        \label{fig:supplement:glove-25-indexing}
    \end{subfigure}
    \\
    \begin{subfigure}[b]{0.47\textwidth}
        \includegraphics[width=0.9\textwidth]{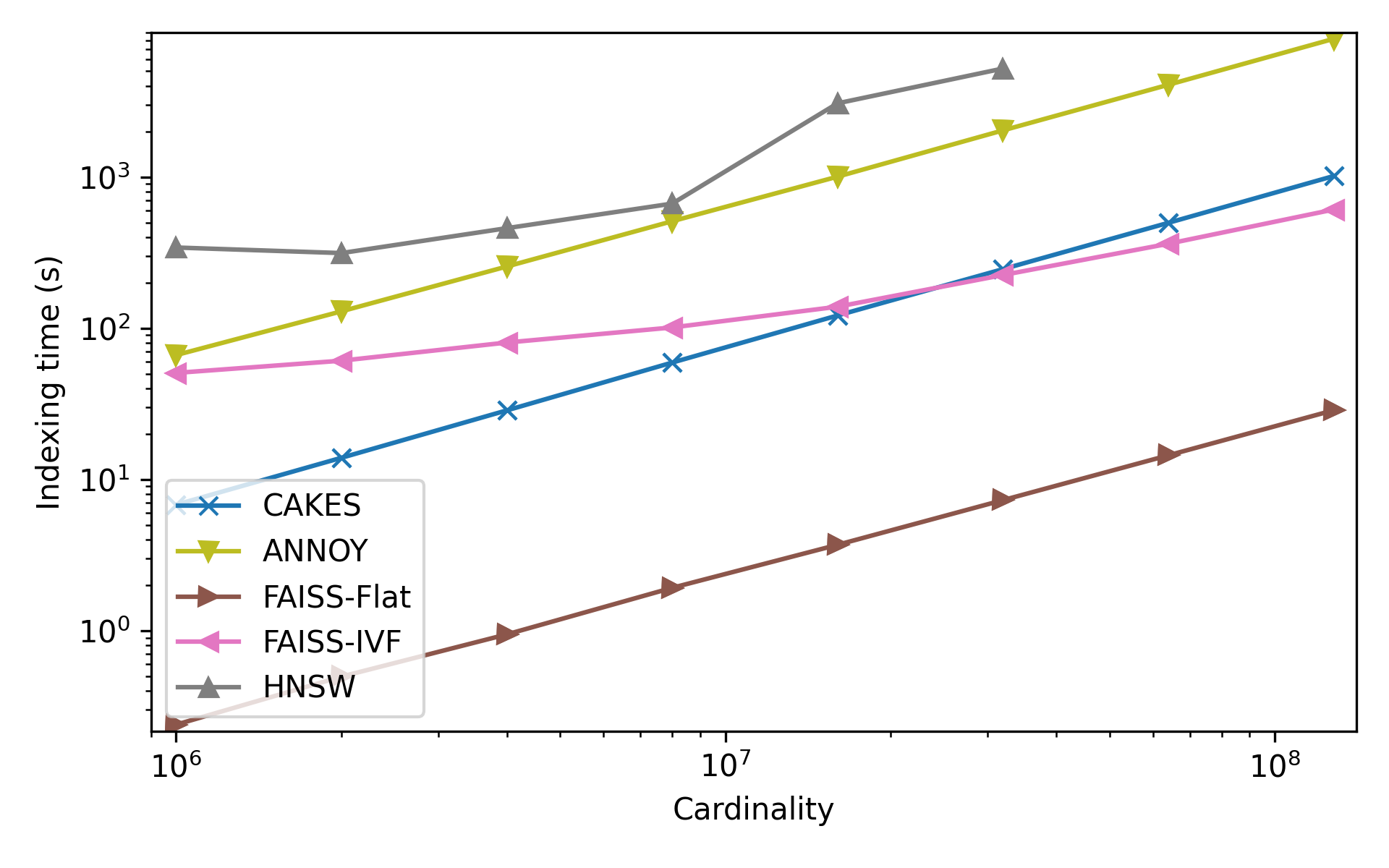}\\
        \subcaption{Sift}
        \label{fig:supplement:sift-indexing}
    \end{subfigure}%
    \begin{subfigure}[b]{0.47\textwidth}
        \includegraphics[width=0.9\textwidth]{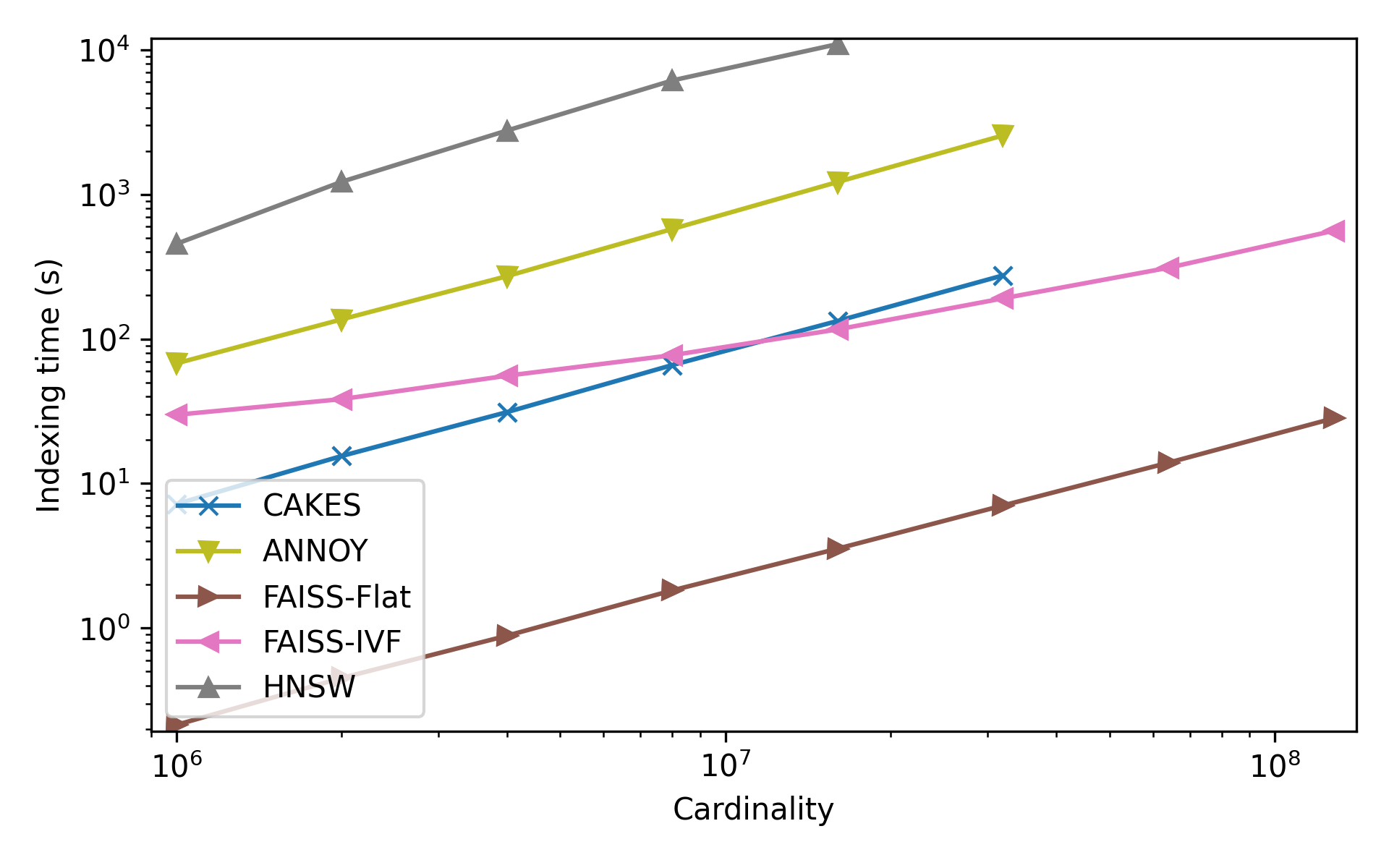}\\
        \subcaption{Random}
        \label{fig:supplement:random-indexing}
    \end{subfigure}%
    \\
    \caption{Indexing and tuning time for each algorithm with each of the ANN benchmark datasets and the Random dataset.}
    \label{fig:supplement:indexing}
\end{figure}

The plots in Figure~\ref{fig:supplement:indexing} show the results of these benchmarks.
The horizontal axis in each subplot shows the cardinality of the dataset augmented with synthetic points.
The left-most point on each line is at the cardinality of the original dataset without any synthetic augmentation.
The vertical axis denotes the sum of indexing and tuning time in seconds.
Both axes are on a logarithmic scale.
Hereafter, when we refer to the ``indexing time'' of an algorithm, we are implicitly referring to the sum of indexing and tuning time for said algorithm.

On all datasets, we observe that the indexing time for CAKES increases roughly linearly as cardinality increases.
HNSW and ANNOY have the slowest indexing times across all the algorithms we benchmarked for each of the four datasets, at each cardinality.
On some datasets, HNSW and ANNOY exhibit indexing times which are orders of magnitude slower than that of CAKES.
FAISS-Flat exhibits the fastest indexing time on each dataset.
This is not surprising, however, given that FAISS-Flat is a na\"{\i}ve linear search algorithm and is not building an index.

We also highlight some differences in indexing time between different datasets.
With Fashion-Mnist, as shown in Figure~\ref{fig:supplement:fashion-mnist-indexing}, we observe that the indexing time for CAKES is faster than that of FAISS-IVF for all cardinalities.
With Glove-25 (see Figure~\ref{fig:supplement:glove-25-indexing}), however, at cardinalities greater than $10^7$, FAISS-IVF has faster indexing time than CAKES.
With Sift, CAKES's indexing time is faster than that of FAISS-IVF until a cardinality of nearly $10^8$, and with the Random dataset, we observe that CAKES has faster indexing time than FAISS-IVF until a cardinality of nearly $10^7$.

\begin{figure}
    \begin{subfigure}[b]{0.47\textwidth}
        \includegraphics[width=1.0\textwidth]{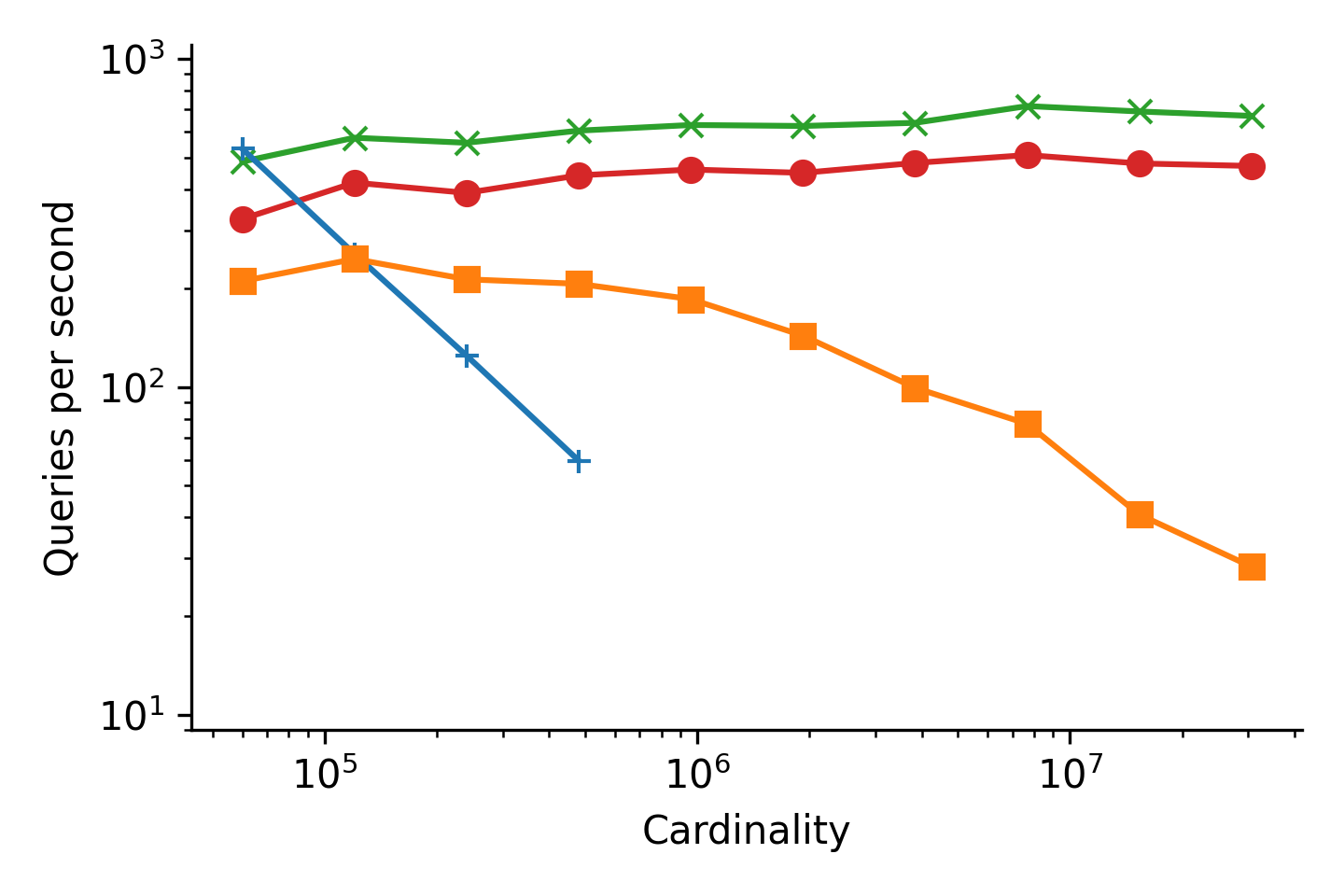}
        \subcaption{Fashion-Mnist for $k=100$.}
        \label{fig:supplement:fashion-mnist-scaling}
    \end{subfigure}%
    \begin{subfigure}[b]{0.47\textwidth}
        \includegraphics[width=1.0\textwidth]{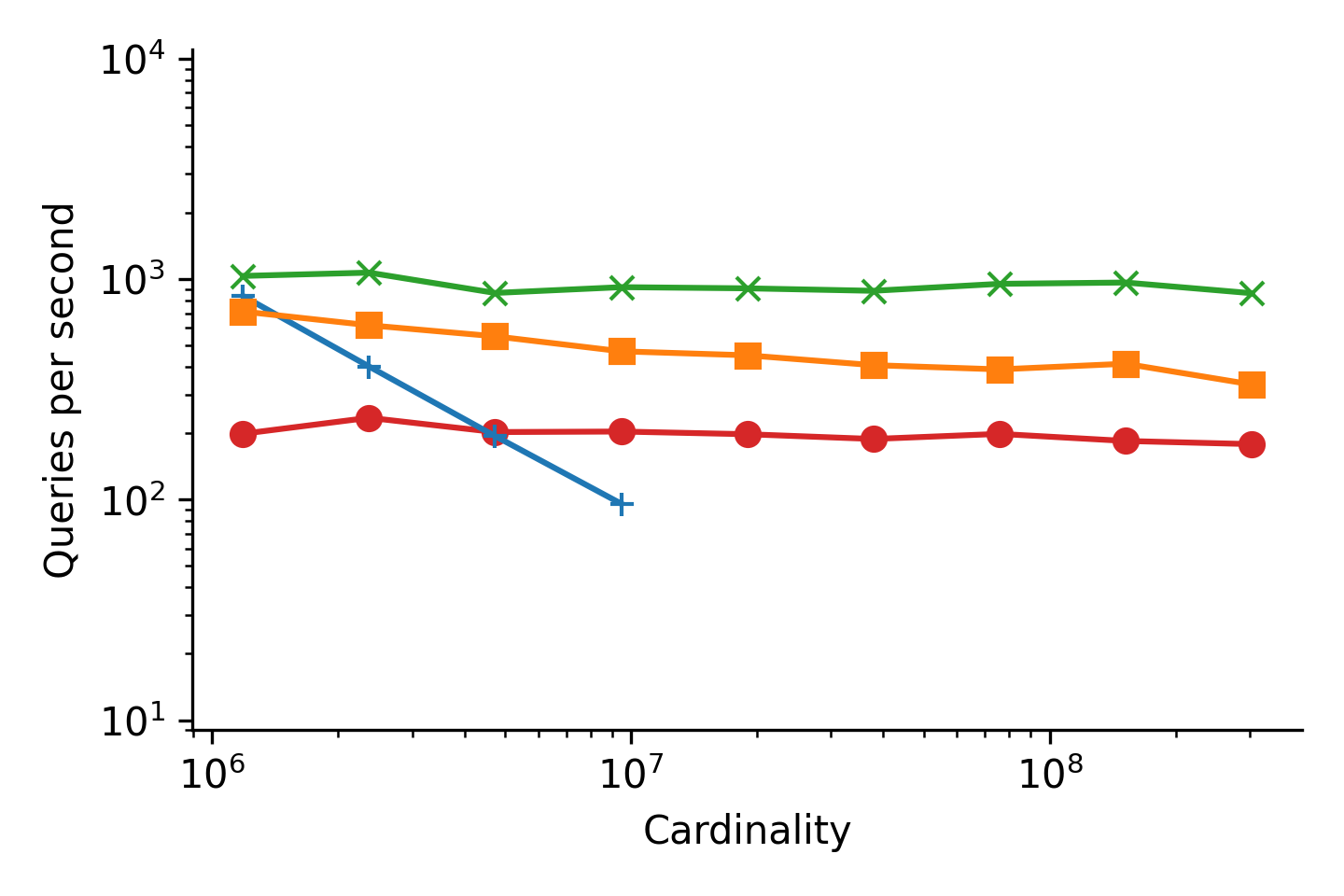}
        \subcaption{Glove-25 for $k=100$.}
        \label{fig:supplement:glove-25-scaling}
    \end{subfigure}%
    \\
    \begin{subfigure}[b]{0.47\textwidth}
        \includegraphics[width=1.0\textwidth]{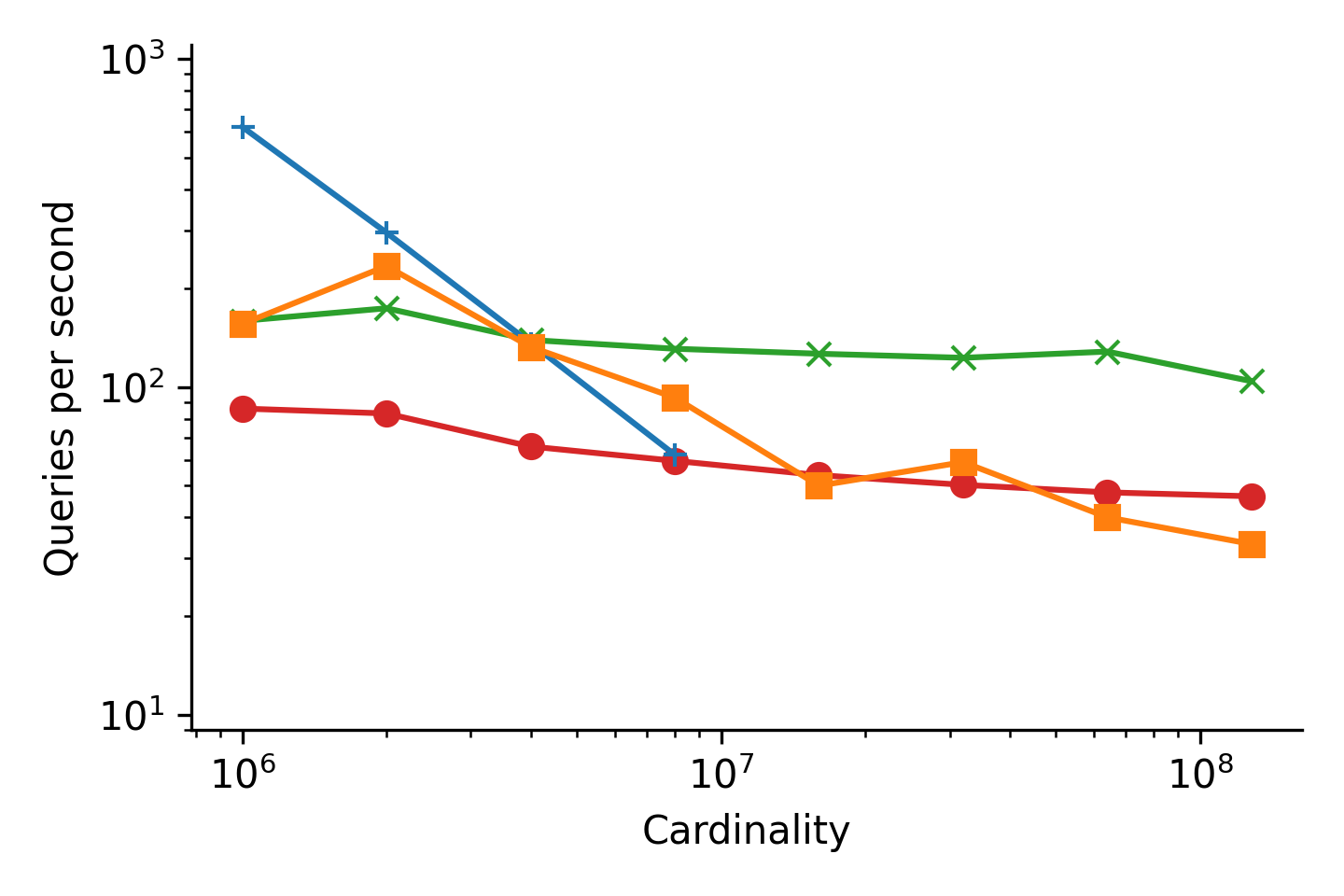}
        \subcaption{Sift for $k=100$.}
        \label{fig:supplement:sift-scaling}
    \end{subfigure}%
    \begin{subfigure}[b]{0.47\textwidth}
        \includegraphics[width=1.0\textwidth]{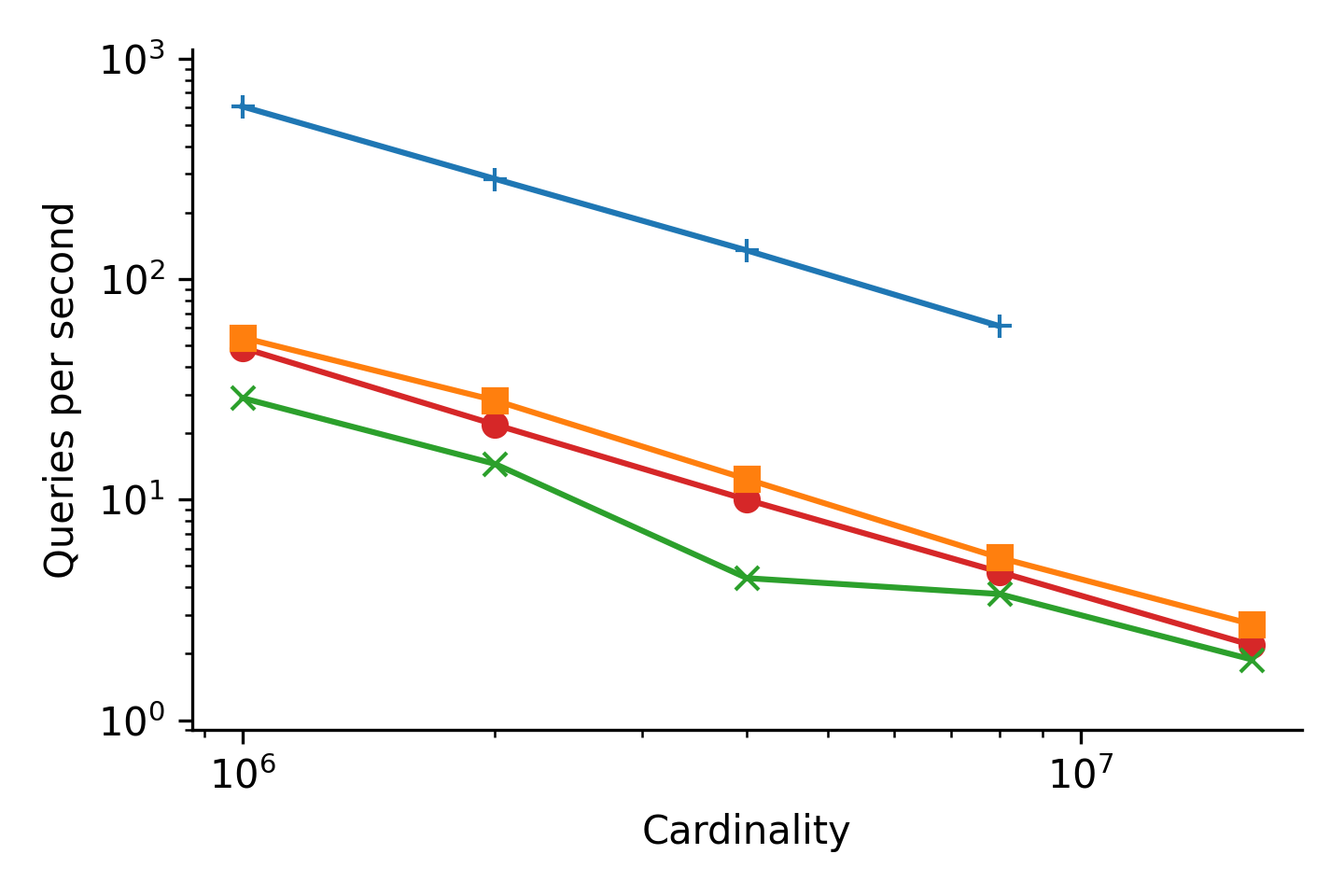}
        \subcaption{A random dataset for $k=100$.}
        \label{fig:supplement:random-scaling}
    \end{subfigure}%
    \\
    \begin{subfigure}[b]{0.47\textwidth}
        \includegraphics[width=1.0\textwidth]{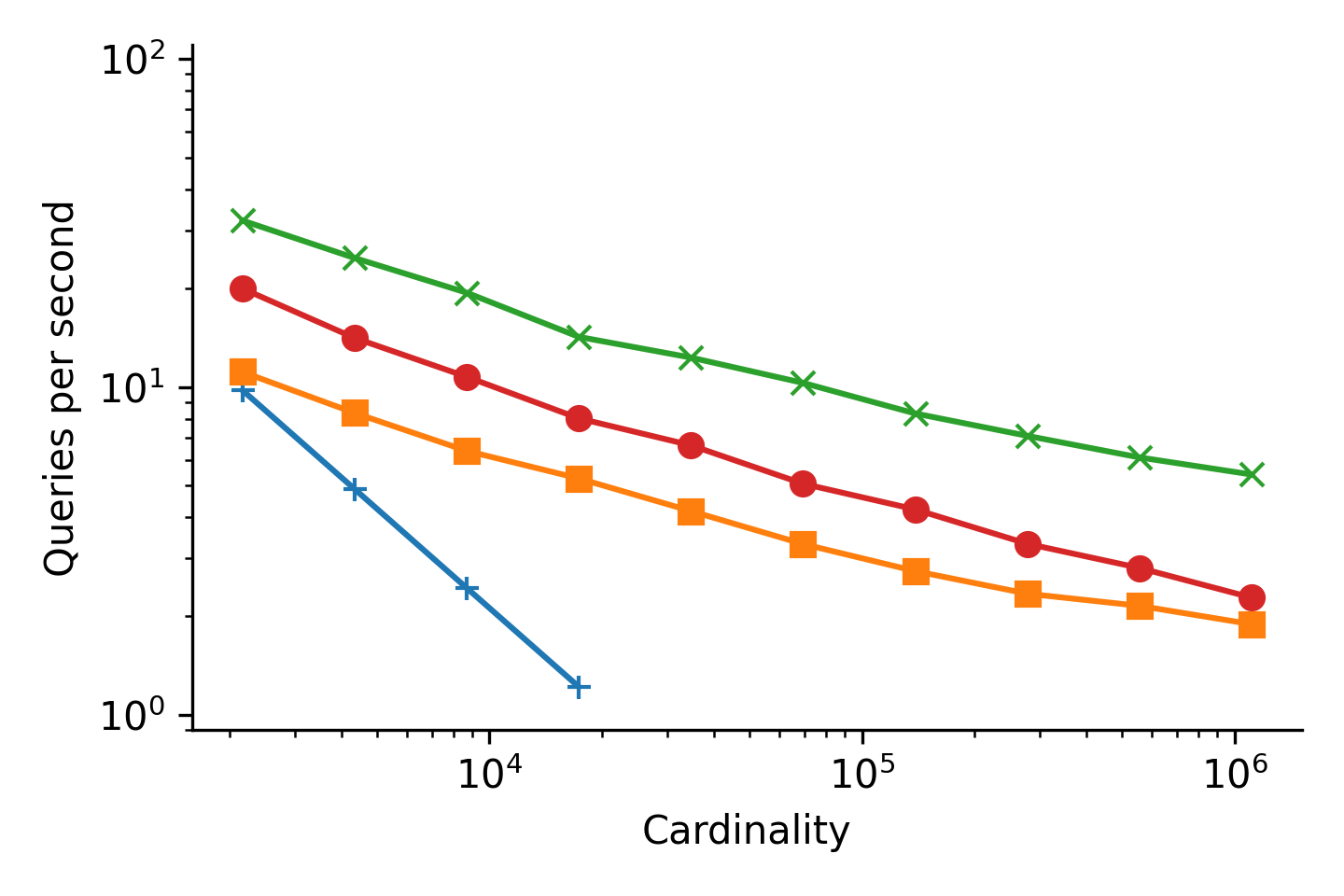}
        \subcaption{Silva for $k=100$.}
        \label{fig:supplement:silva-scaling}
    \end{subfigure}%
    \begin{subfigure}[b]{0.47\textwidth}
        \includegraphics[width=1.0\textwidth]{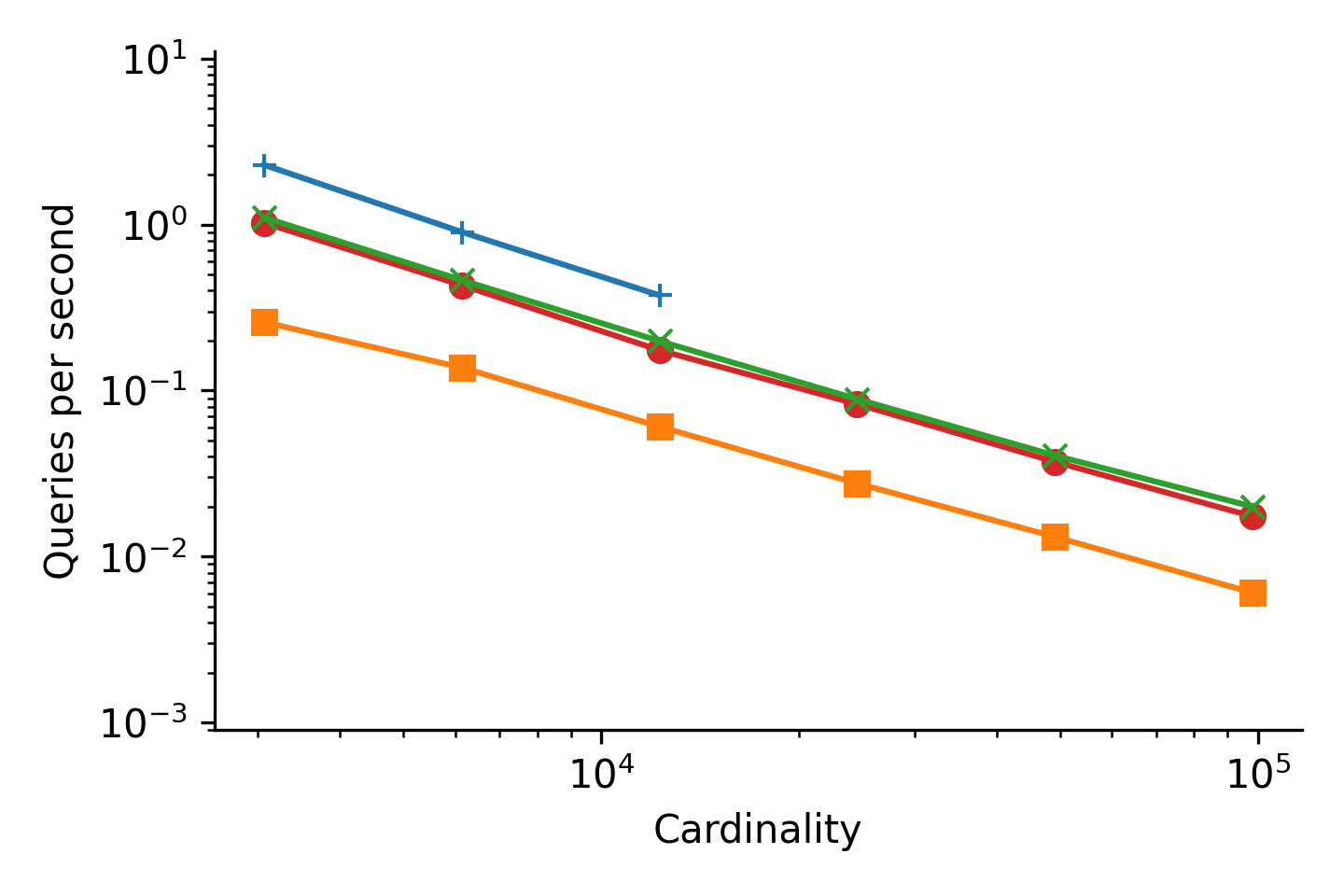}
        \subcaption{RadioML for $k=100$ at SnR = 10dB.}
        \label{fig:supplement:radioml-scaling}
    \end{subfigure}%
    \\
    \begin{subfigure}[b]{0.94\textwidth}
        \centering
        \includegraphics[width=0.7\textwidth]{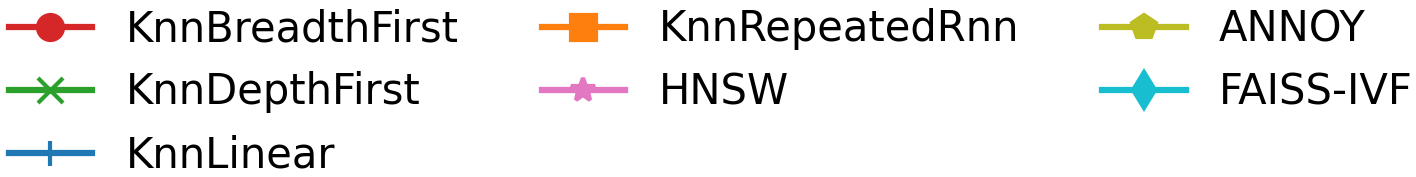}
        \label{fig:supplement:scaling-legend}
    \end{subfigure}%
    \caption{Throughput (with $k$=100) across six datasets, including a randomly-generated dataset.
    In each plot, the horizontal axis represents increasing cardinality of the dataset, while the vertical axis represents the throughput in queries per second (higher is better).
    For linear search with CAKES, we only report the throughput for a few of the initial multipliers because the trend is clear.}
    \label{fig:supplement:scaling-plots}
\end{figure}

\subsection{Clustering}

\begin{figure}
    \captionsetup[subfigure]{aboveskip=-15pt,belowskip=-3pt}
    \begin{subfigure}[b]{0.47\textwidth}
        \includegraphics[width=0.9\textwidth]{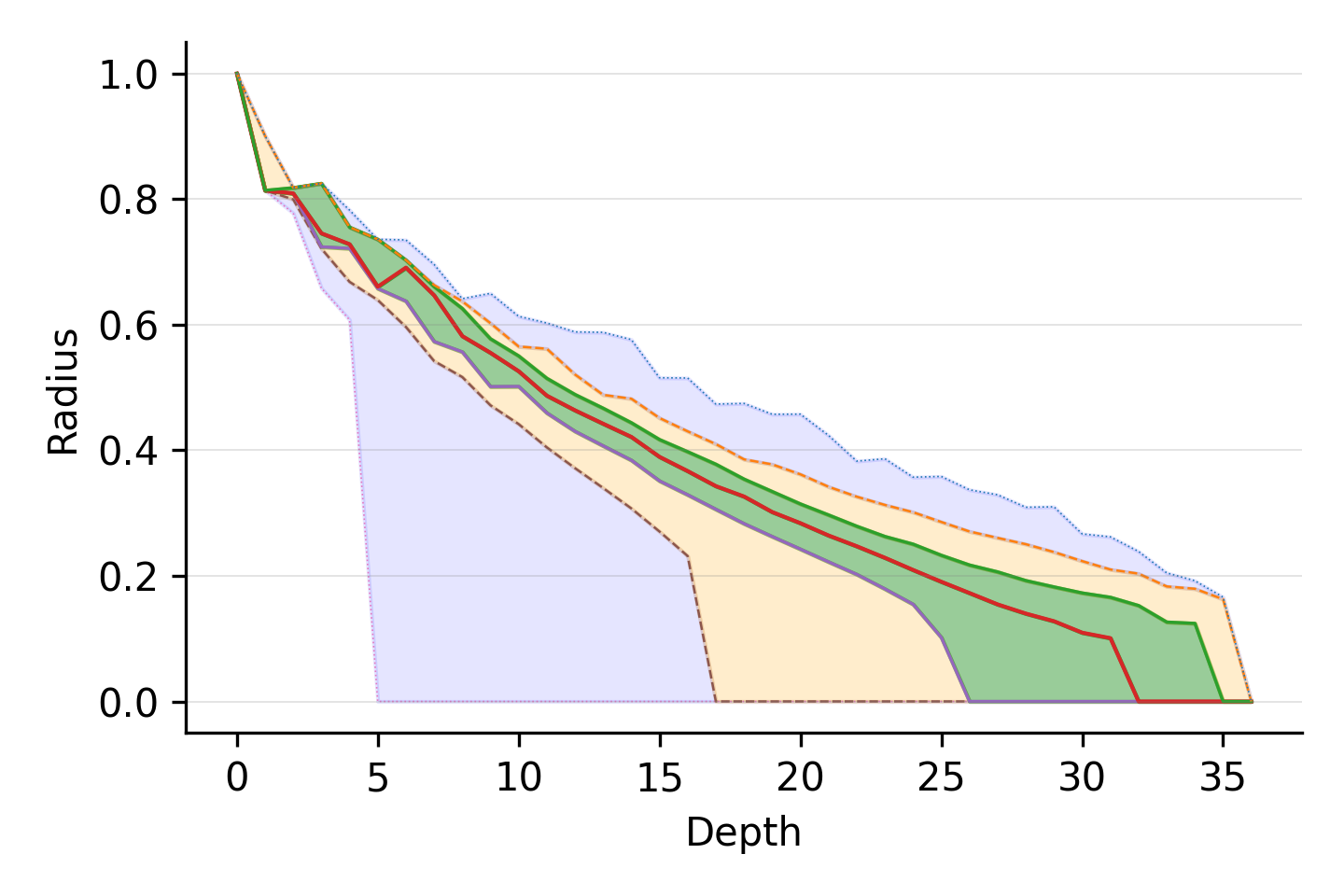}\\
        \subcaption{Fashion-mnist}
        \label{fig:supplement:fashion-mnist-radius}
    \end{subfigure}%
    \begin{subfigure}[b]{0.47\textwidth}
        \includegraphics[width=0.9\textwidth]{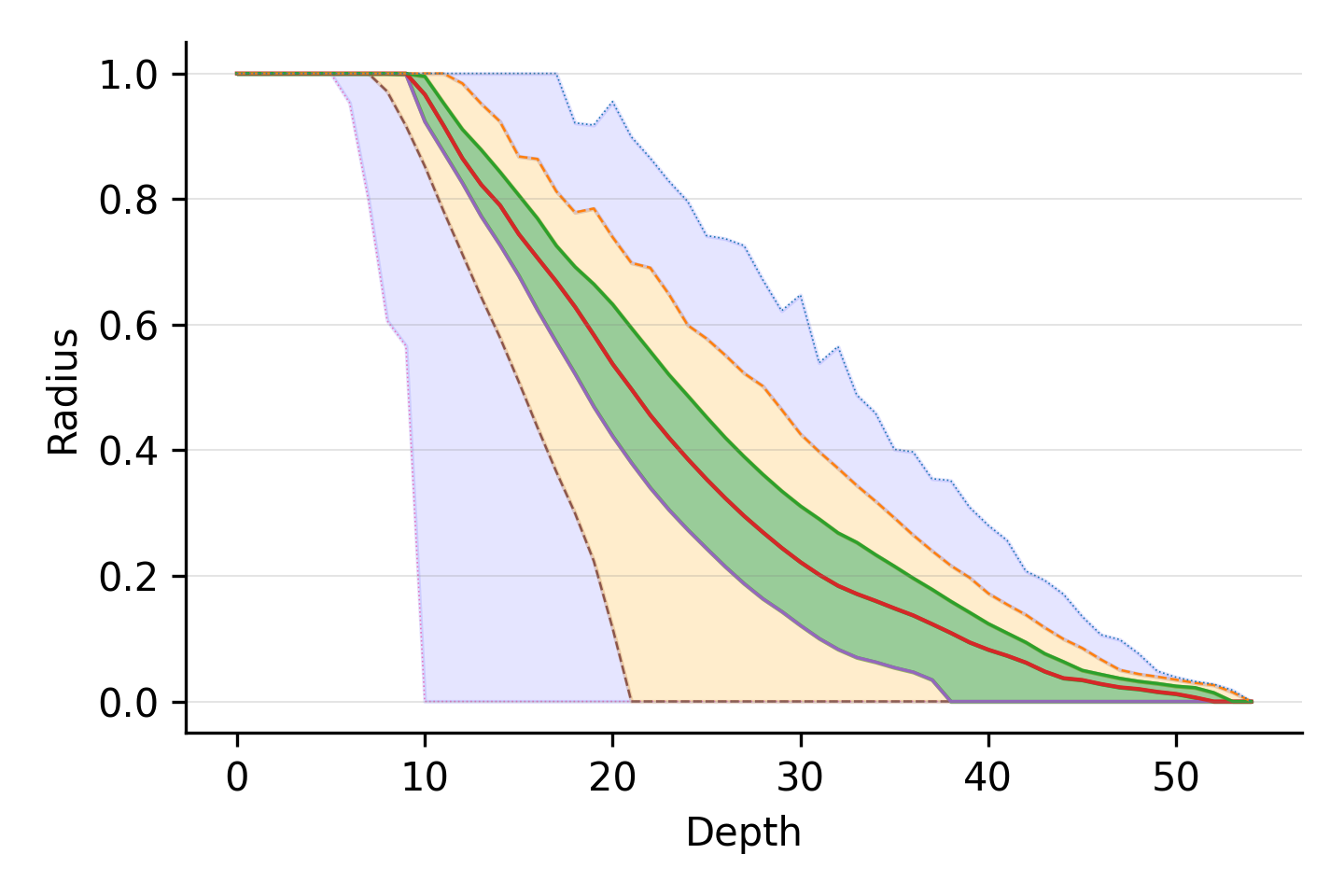}\\
        \subcaption{Glove-25}
        \label{fig:supplement:glove-25-radius}
    \end{subfigure}
    \\
    \begin{subfigure}[b]{0.47\textwidth}
        \includegraphics[width=0.9\textwidth]{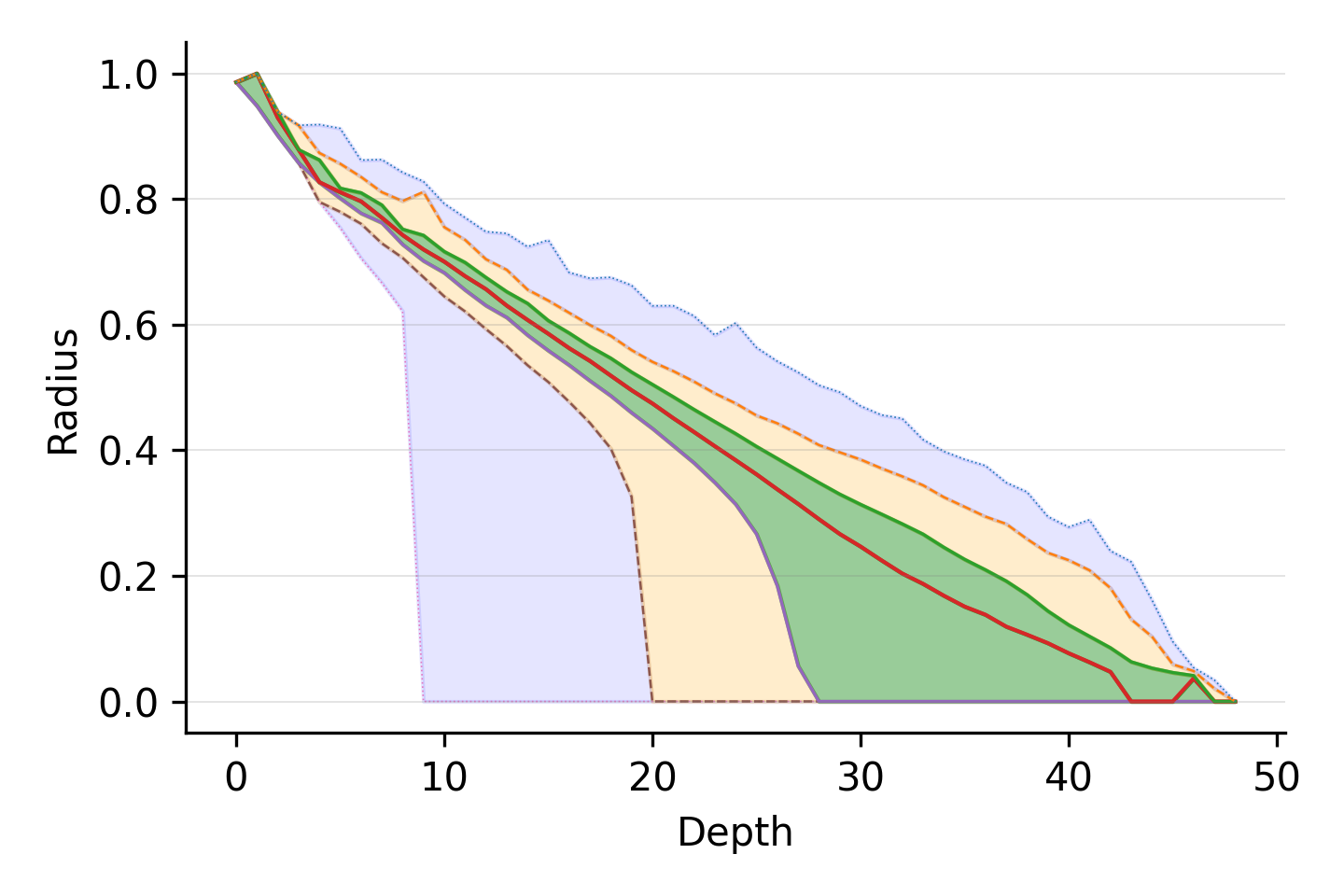}\\
        \subcaption{Sift}
        \label{fig:supplement:sift-radius}
    \end{subfigure}%
    \begin{subfigure}[b]{0.47\textwidth}
        \includegraphics[width=0.9\textwidth]{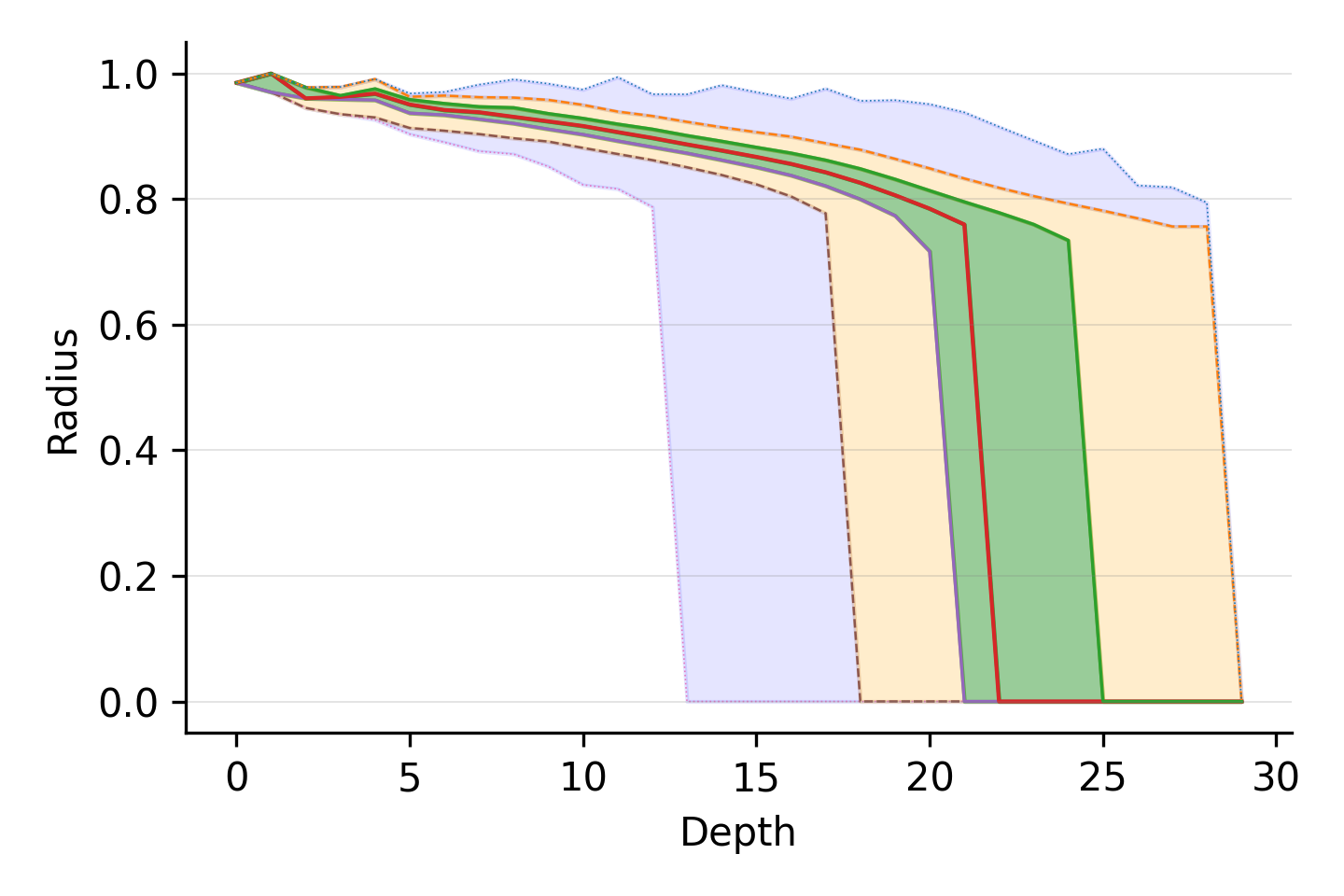}\\
        \subcaption{A random dataset}
        \label{fig:supplement:random-radius}
    \end{subfigure}
    \\
    \begin{subfigure}[b]{0.47\textwidth}
        \includegraphics[width=0.9\textwidth]{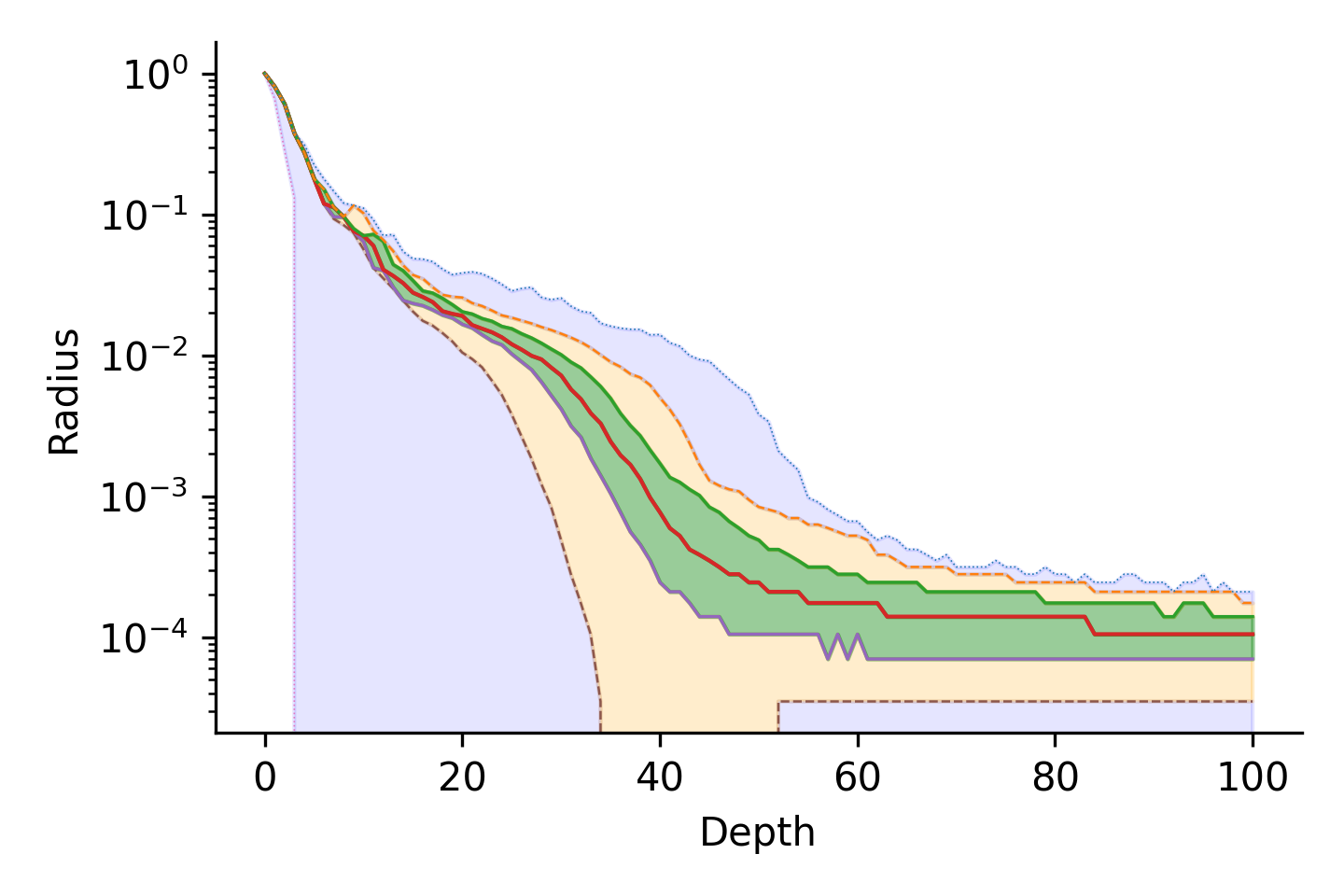}\\
        \subcaption{Silva 18S}
        \label{fig:supplement:silva-radius}
    \end{subfigure}%
    \begin{subfigure}[b]{0.47\textwidth}
        \includegraphics[width=0.9\textwidth]{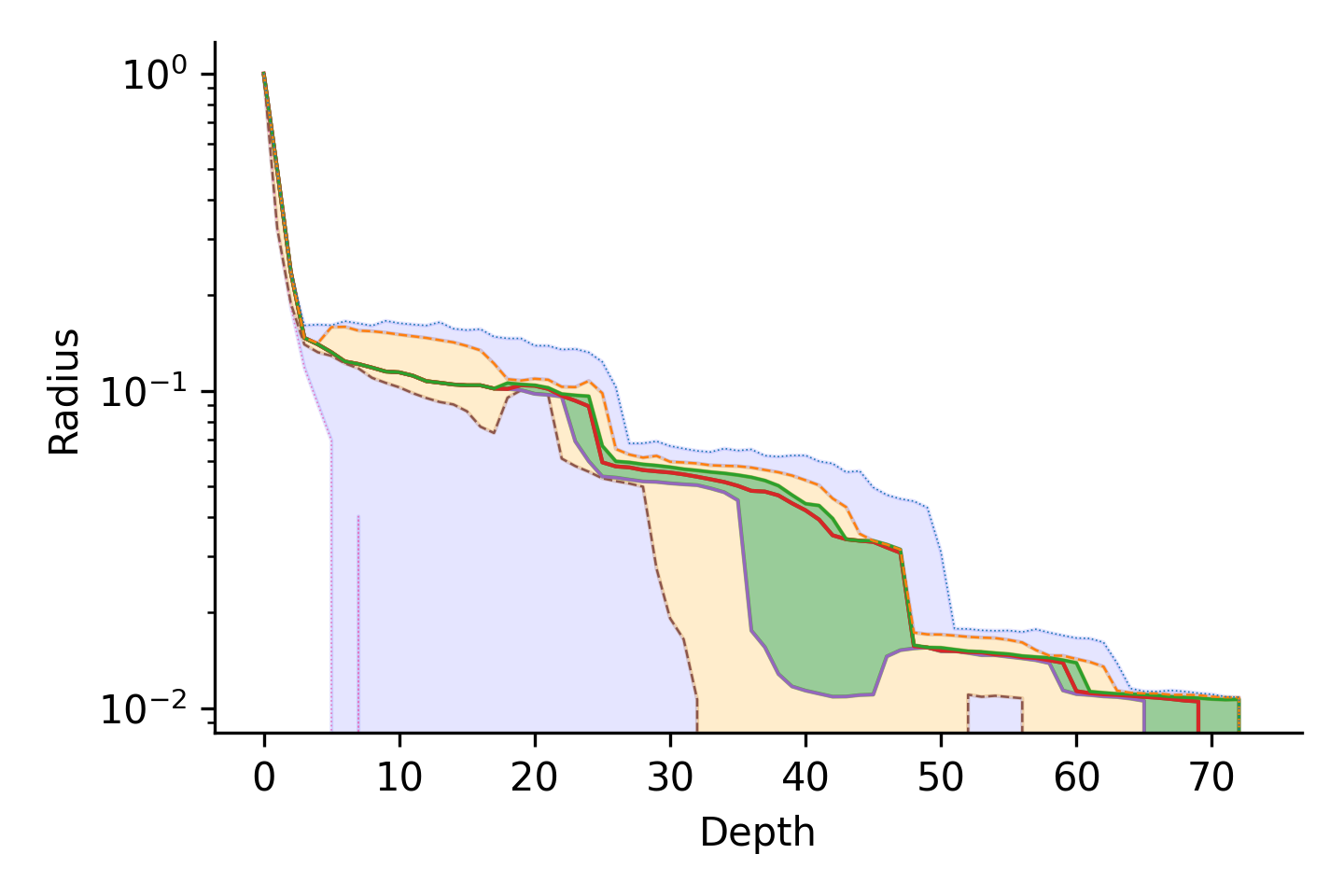}\\
        \subcaption{RadioML}
        \label{fig:supplement:radioml-radius}
    \end{subfigure}%
    \\
    \vskip 0.005in
    \begin{subfigure}[b]{0.94\textwidth}
        \centering
        \includegraphics[width=0.7\textwidth]{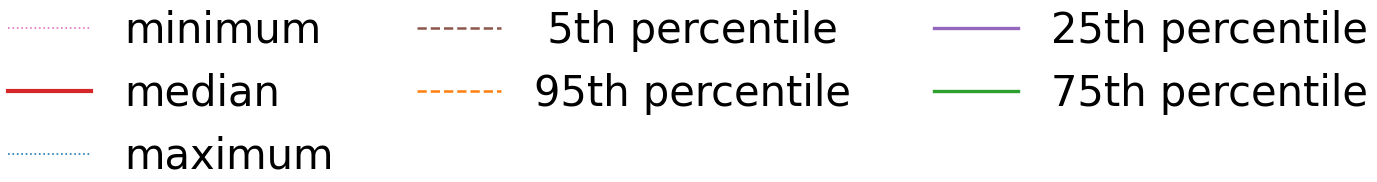}
        \label{fig:supplement:radius-legend}
    \end{subfigure}%
    \caption{Radius vs. cluster depth across six datasets, grouped by percentile of radius and weighted by the cardinalities of the clusters.
    In order to use the same y-axis for all plots, we divided the radii of all clusters by the maximum radius of any cluster in the dataset.
    Note that for the Silva 18S and the RadioML datasets, we use a logarithmic scale for the y-axis.}
    \label{fig:supplement:radius-plots}
\end{figure}

\begin{figure}
    \captionsetup[subfigure]{aboveskip=-15pt,belowskip=-3pt}
    \begin{subfigure}[b]{0.47\textwidth}
        \includegraphics[width=0.9\textwidth]{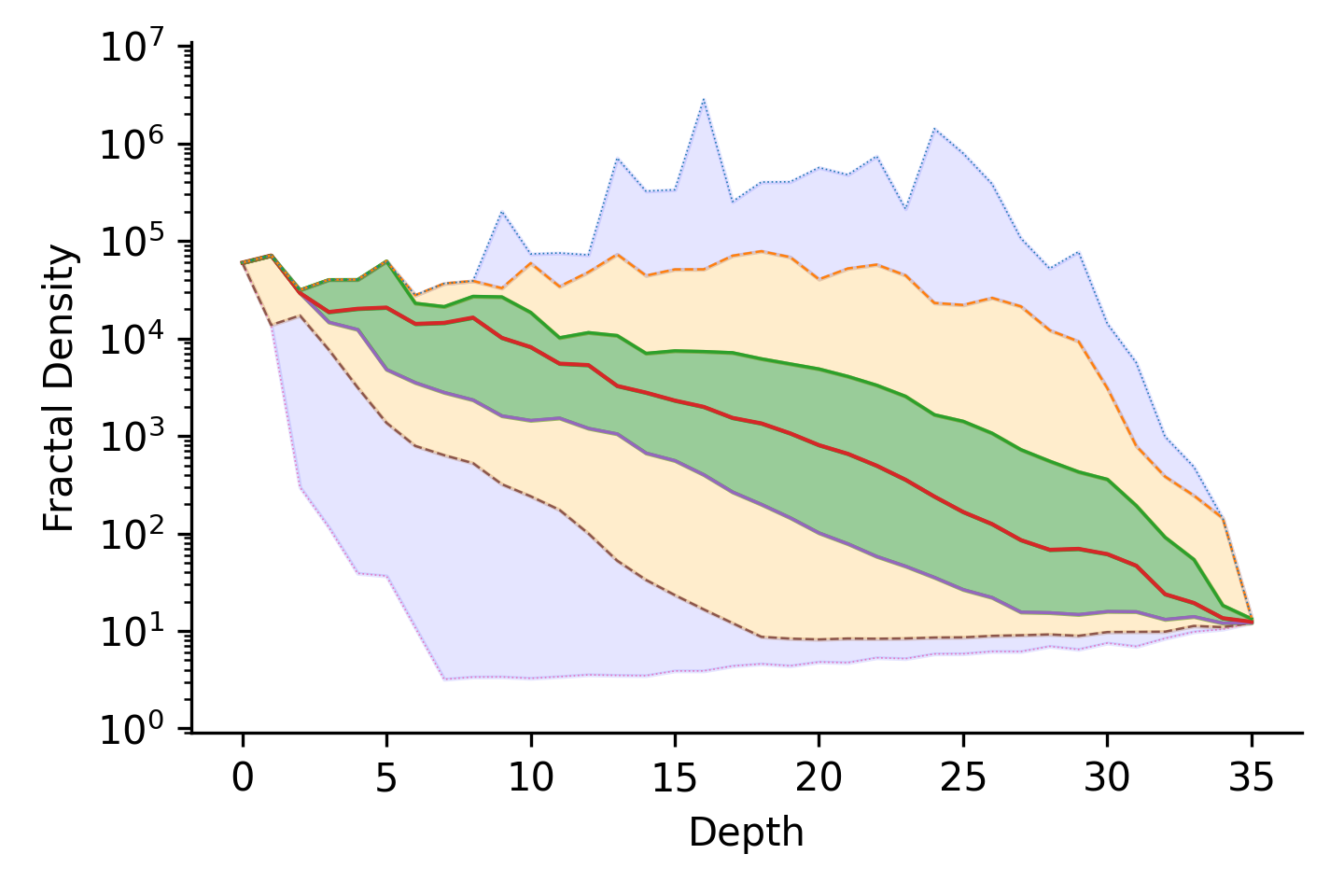}\\
        \subcaption{Fashion-mnist}
        \label{fig:supplement:fashion-mnist-fractal_density}
    \end{subfigure}%
    \begin{subfigure}[b]{0.47\textwidth}
            \includegraphics[width=0.9\textwidth]{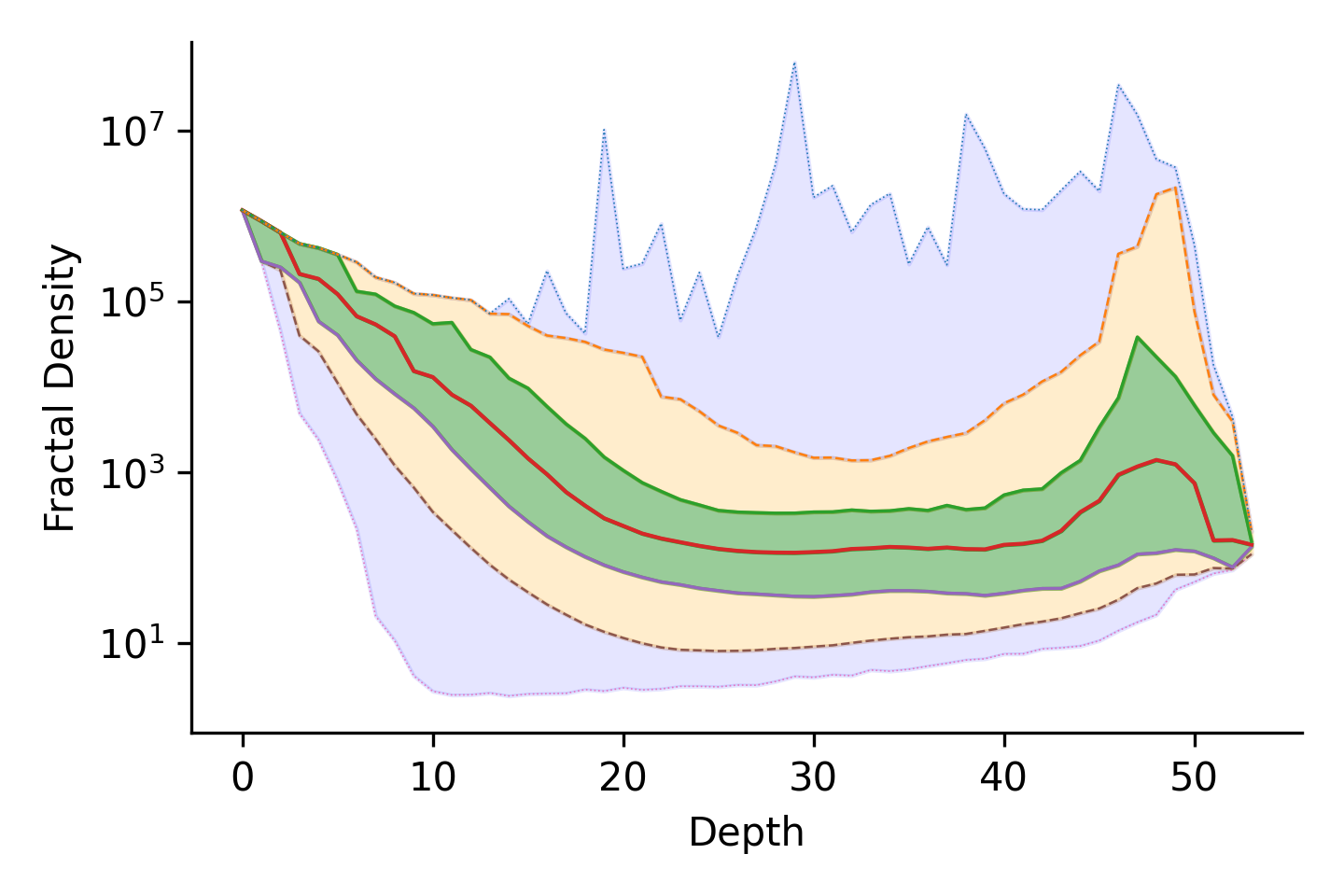}\\
            \subcaption{Glove-25}
            \label{fig:supplement:glove-25-fractal_density}
    \end{subfigure}
    \\
    \begin{subfigure}[b]{0.47\textwidth}
        \includegraphics[width=0.9\textwidth]{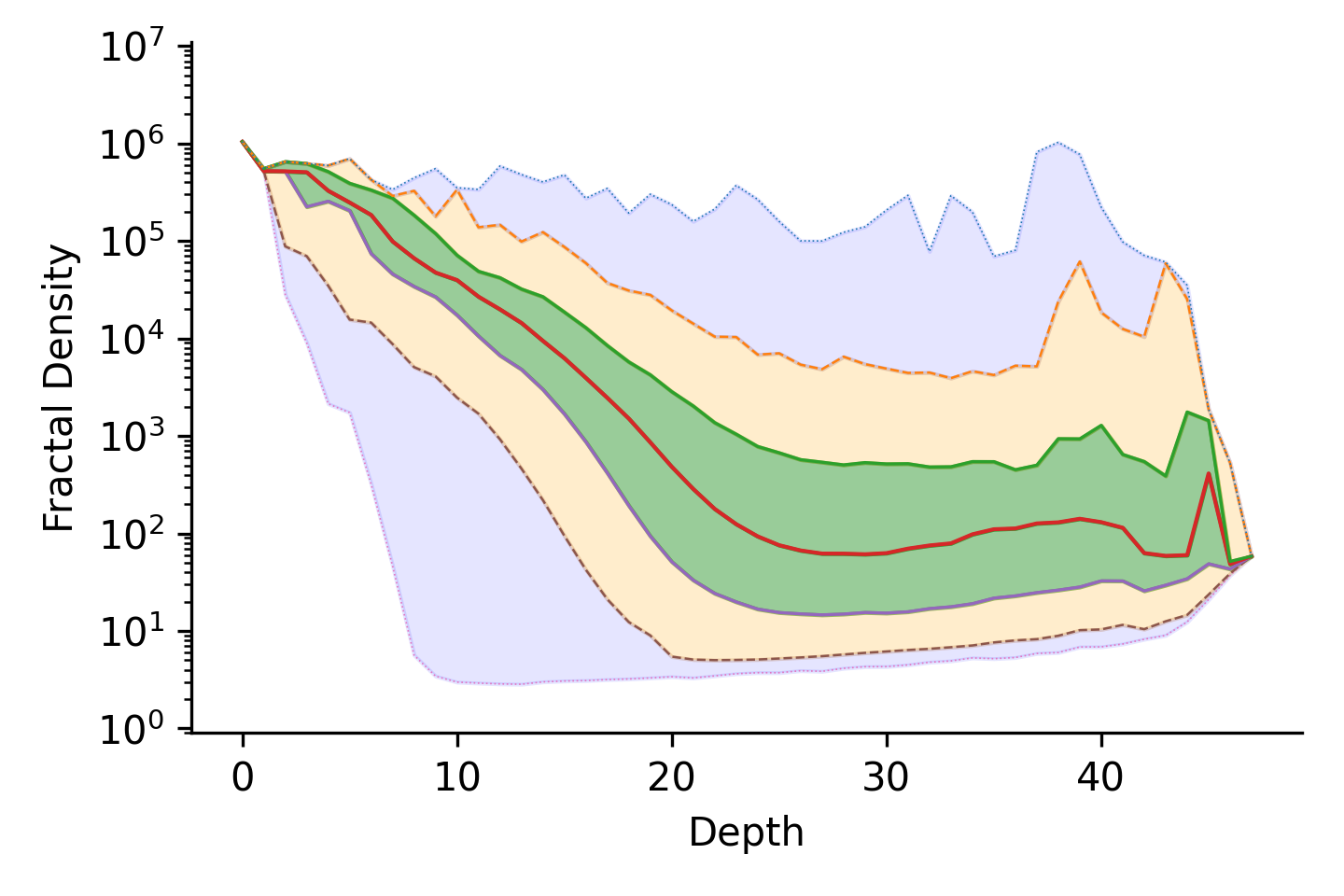}\\
        \subcaption{Sift}
        \label{fig:supplement:sift-fractal_density}
    \end{subfigure}%
    \begin{subfigure}[b]{0.47\textwidth}
        \includegraphics[width=0.9\textwidth]{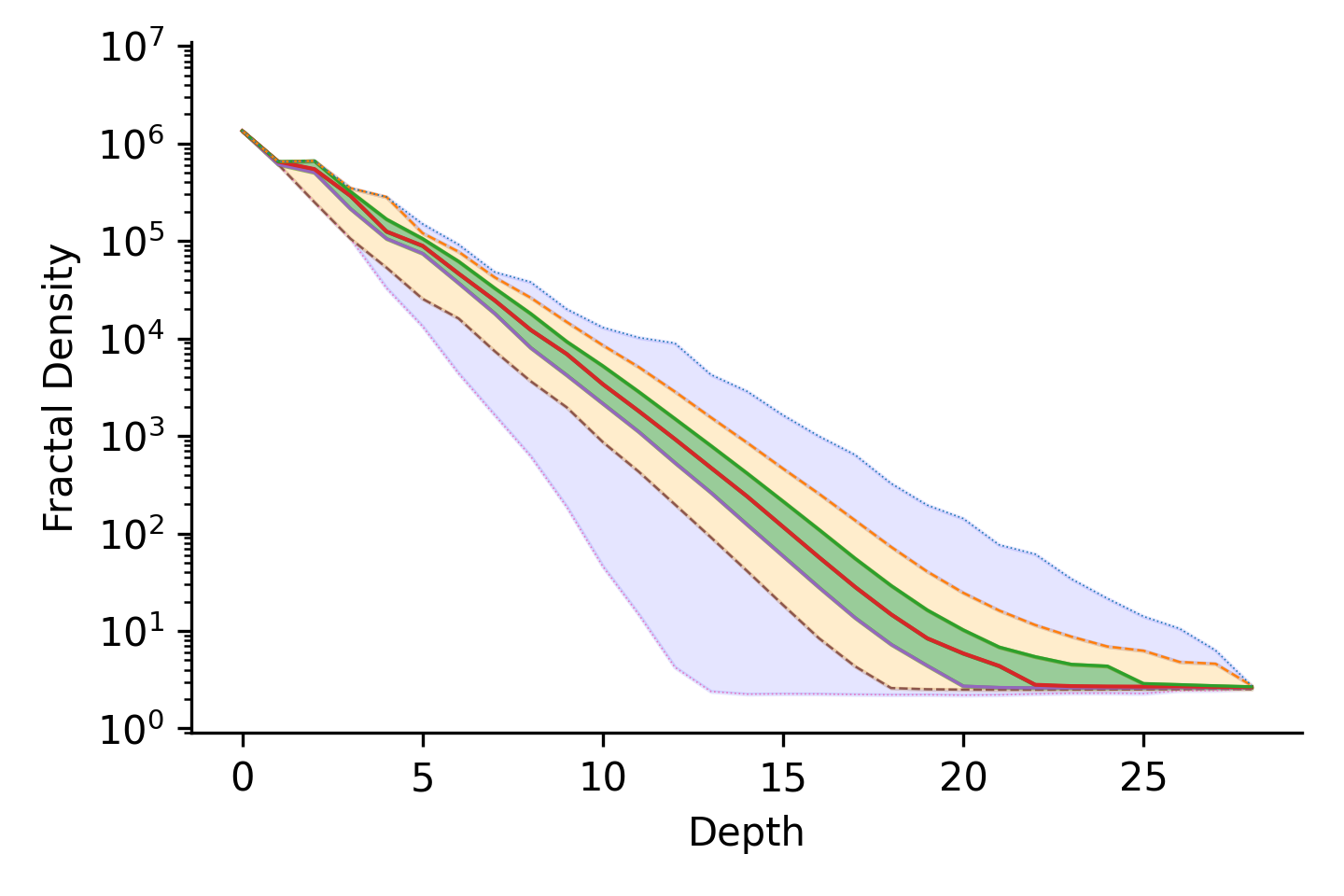}\\
        \subcaption{A random dataset}
        \label{fig:supplement:random-fractal_density}
    \end{subfigure}
    \\
    \begin{subfigure}[b]{0.47\textwidth}
        \includegraphics[width=0.9\textwidth]{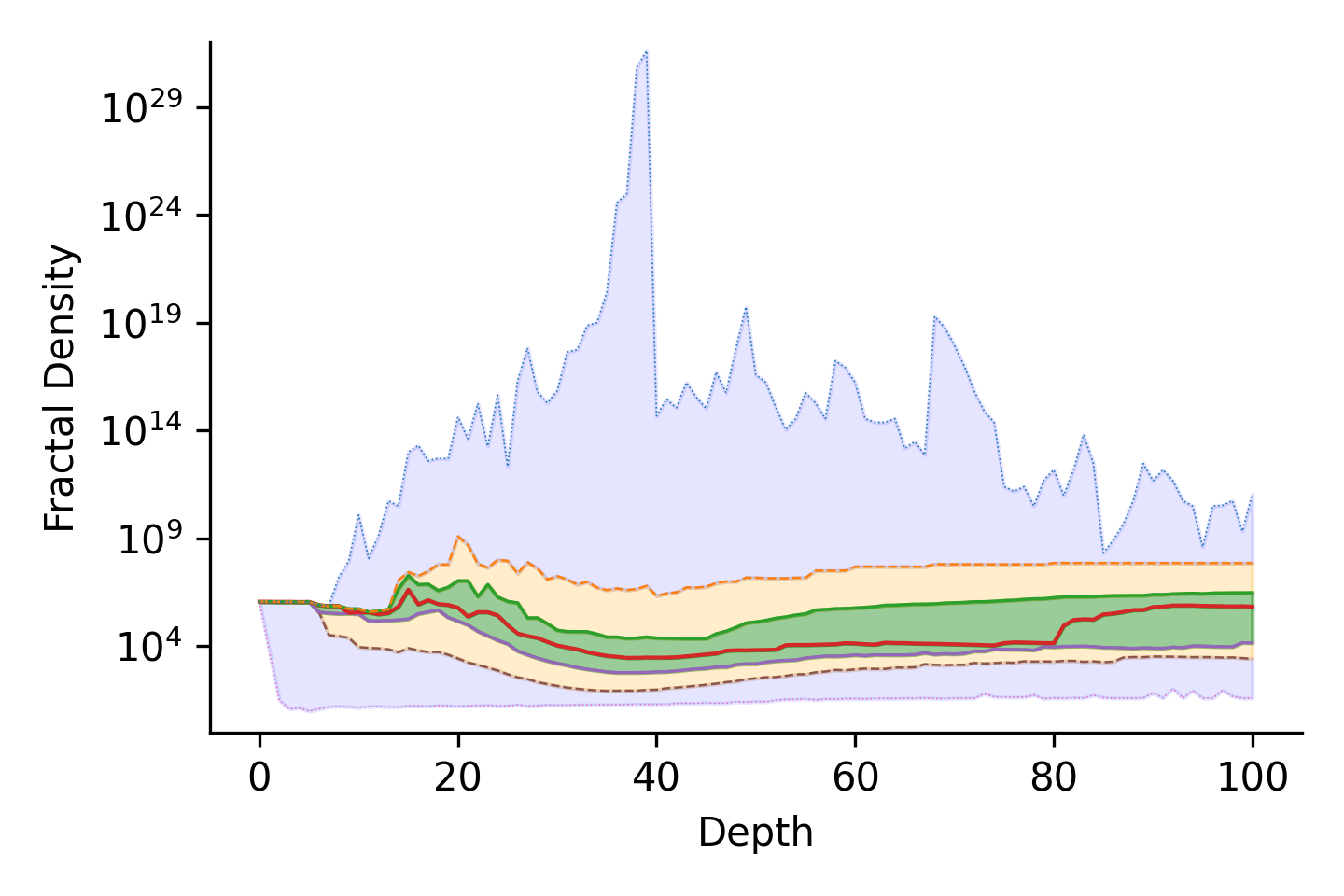}\\
        \subcaption{Silva 18S}
        \label{fig:supplement:silva-fractal_density}
    \end{subfigure}%
    \begin{subfigure}[b]{0.47\textwidth}
        \includegraphics[width=0.9\textwidth]{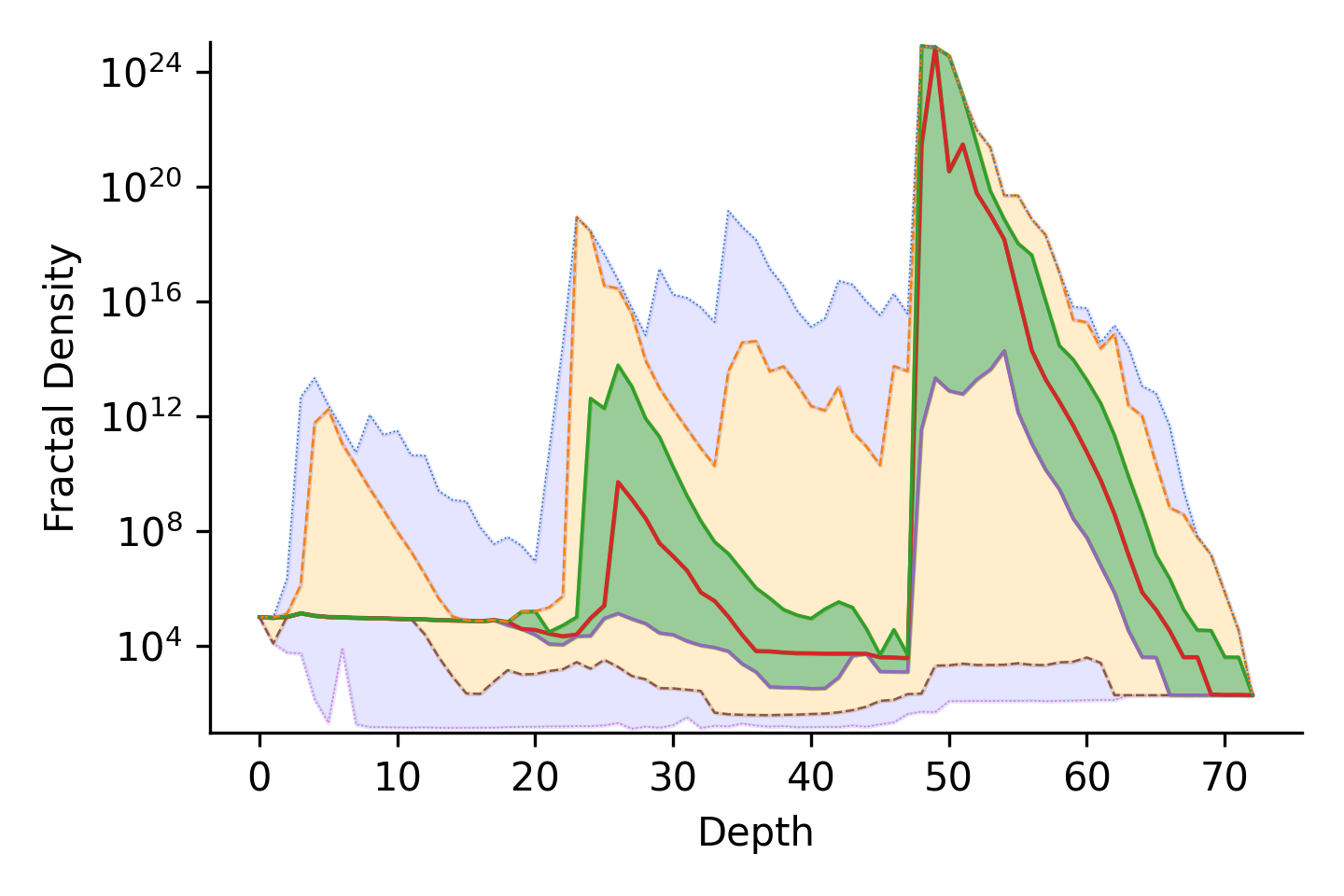}\\
        \subcaption{RadioML}
        \label{fig:supplement:radioml-fractal_density}
    \end{subfigure}%
    \\
    \vskip 0.005in
    \begin{subfigure}[b]{0.94\textwidth}
        \centering
        \includegraphics[width=0.7\textwidth]{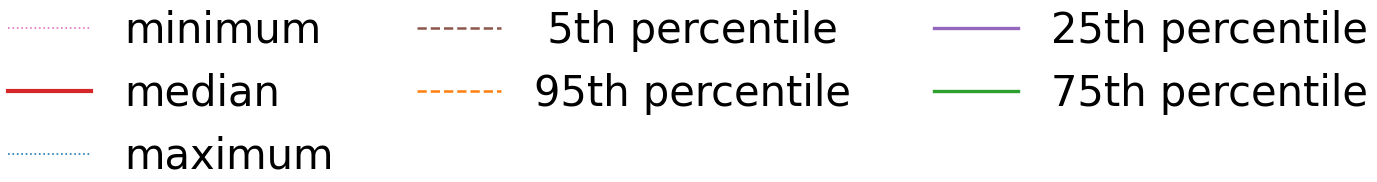}
        \label{fig:supplement:fractal_density-legend}
    \end{subfigure}%
    \caption{Fractal Density vs. cluster depth across six datasets, grouped by percentile of fractal density and weighted by the cardinalities of the clusters.
    Fractal Density is defined as $\frac{cardinality}{radius^{LFD}}$.
    As with the radius plots, we normalized the cluster radii before calculating the Fractal Density.
    We also excluded all clusters with normalized radii smaller than $10^{-6}$, as they distort the scale of the plots.}
    \label{fig:supplement:fractal_density-plots}
\end{figure}

\bibliographystyle{siamplain}
\bibliography{references}

%% file: cakes_shared.tex
\usepackage[utf8]{inputenc} 
\usepackage[T1]{fontenc}    
\usepackage{url}            
\usepackage{booktabs}       
\usepackage{amsfonts}       
\usepackage{nicefrac}       
\usepackage{microtype}      
\usepackage{lipsum}
\usepackage{graphicx}
\let\subcaption\relax
\usepackage{caption}
\usepackage{subcaption}
\usepackage{amsmath}
\usepackage{amssymb}
\usepackage{mathtools}
\usepackage{siunitx}
\ifarxiv
\usepackage{algorithm}
\usepackage{amsthm}
\newtheorem{theorem}{Theorem}
\fi

\usepackage{hyperref}
\usepackage{xr}
\usepackage{xcolor}
\usepackage{multirow}

\DeclareMathOperator*{\argmax}{arg\,max}

\usepackage{placeins}
\usepackage{xr}
\usepackage{hyperref}
\usepackage{lipsum}
\usepackage{amsfonts}
\usepackage{graphicx}
\usepackage{epstopdf}
\usepackage{algpseudocode}
\usepackage{float}
\ifpdf
  \DeclareGraphicsExtensions{.eps,.pdf,.png,.jpg}
\else
  \DeclareGraphicsExtensions{.eps}
\fi

\usepackage{enumitem}
\setlist[enumerate]{leftmargin=.5in}
\setlist[itemize]{leftmargin=.5in}

\title{Let them have CAKES: A Cutting-Edge Algorithm for Scalable, Efficient, and Exact Search on Big Data\thanks{Submitted to the editors DATE.
}}

\ifarxiv

\author{
    Morgan E. Prior \\
    Department of Computer Science \\
    Tufts University \\
    Medford, MA \\
    \texttt{morgan.prior@tufts.edu } \\
    \And
    Thomas J. Howard III \\
    Department of Computer Science and Statistics\\
    University of Rhode Island\\
    Kingston, RI\\
    \texttt{thoward27@uri.edu} \\
    \And
    Oliver McLaughlin \\
    Department of Computer Science and Statistics\\
    University of Rhode Island\\
    Kingston, RI\\
    \texttt{olwmc@gmail.com} \\
    \And
    Terry Ferguson \\
    Department of Computer Science and Statistics\\
    University of Rhode Island\\
    Kingston, RI\\
    \texttt{fergusontr@gmail.com} \\
    \And
    Najib Ishaq \\
    Department of Computer Science and Statistics\\
    University of Rhode Island\\
    Kingston, RI\\
    \texttt{najib\_ishaq@zoho.com} \\
    \And
    Noah M. Daniels \\
    Department of Computer Science and Statistics\\
    University of Rhode Island\\
    Kingston, RI\\
    \texttt{noah\_daniels@uri.edu} \\
}

\fi

\usepackage{amsopn}


%% file: sections/1-introduction.tex
\section{Introduction}
\label{sec:introduction}

Researchers are collecting data at an unprecedented rate.
In many fields, datasets are growing exponentially, and this increase in the rate of data collection outpaces improvements in computing performance as predicted by Moore's Law~\cite{kahn2011future}.
This indicates that the performance of computing systems will not keep pace with the growth of data.
Often dubbed ``the Big Data explosion,'' this phenomenon has created a need for better algorithms to analyze large datasets.

Examples of large datasets include genomic databases, time-series data such as radio frequency signals, and neural network embeddings.
Large language models such as GPT~\cite{2020arXiv200514165B, OpenAI2023GPT4TR} and LLAMA-2~\cite{Touvron2023Llama2O}, and image embedding models~\cite{radford2021learning, dosovitskiy2020image} are a common source of neural network embeddings.
Among biological datasets,  SILVA 18S~\cite{10.1093/nar/gks1219} contains ribosomal DNA sequences of approximately 2.25 million genomes; the longest of these individual sequences is 3,712 amino acids, but in a multiple sequence alignment, the length becomes 50,000 letters.
Among time-series datasets, the RadioML dataset~\cite{oshea2018radioml} contains approximately 2.55 million samples of synthetically-generated signals of different modulation modes.

Many researchers are interested in similarity search on these datasets.
Similarity search enables a variety of applications, including recommendation~\cite{annoy} and classification systems~\cite{suyanto2022knnclassifier}.
As the cardinalities and dimensionalities of datasets have grown, however, efficient and accurate similarity search has become challenging;
even state-of-the-art algorithms exhibit a steep tradeoff between recall and throughput~\cite{malkov2016hnsw, johnson2019billion, annoy, aumuller2020ann}.

Given some measure of similarity between data points, there are two common definitions of similarity search: $k$-nearest neighbor search ($k$-NN) and $\rho$-nearest neighbor search ($\rho$-NN).
$k$-NN search aims to find the $k$ most similar points to a query, while $\rho$-NN search aims to find all points within a similarity threshold $\rho$ of a query.
Previous works have used the term \textit{approximate} search to refer to $\rho$-NN search, but in this paper, we reserve the term \textit{approximate} for algorithms that do not exhibit perfect recall.
In contrast, an \textit{exact} search algorithm exhibits perfect recall.

$k$-NN search is one of the most ubiquitous classification and recommendation methods in use~\cite{fix1952discriminatory, cover1967nearest}.
Na\"{i}ve implementations of $k$-NN search, whose time complexity is linear in the dataset's cardinality, prove prohibitively slow for large datasets.
While fast algorithms for $k$-NN search on large datasets exist, they are often approximate~\cite{gao2023high}, and while approximate search may be sufficient for some applications, the need for efficient and \textit{exact} search remains~\cite{ukey2023survey}.
For example, for a majority voting classifier, approximate $k$-NN search may agree with exact $k$-NN search for large values of $k$, but may be sensitive to local perturbations for smaller values of $k$.
This is especially true when classes are not well-separated~\cite{zhang2022imbalanced}.
Further, there is evidence that distance functions that do not obey the triangle inequality, such as cosine distance, perform poorly for $k$-NN search in biomedical settings~\cite{hu2016distance};
this suggests that approximate $k$-NN search could perform poorly in such contexts.

This paper introduces CAKES (CLAM-Accelerated $K$-NN Entropy-Scaling Search), a set of three novel algorithms for exact $k$-NN search.
We also compare CAKES against several current algorithms; namely, FAISS~\cite{johnson2019billion}, HNSW~\cite{malkov2016hnsw}, and ANNOY~\cite{annoy}, on datasets from the ANN-benchmarks suite~\cite{aumuller2020ann}.
We further benchmark CAKES on a large genomic dataset, the SILVA 18S dataset~\cite{10.1093/nar/gks1219}, using Levenshtein~\cite{levenshtein1966binary} distance on unaligned genomic sequences, and a radio frequency dataset, RadioML~\cite{oshea2018radioml}, using Dynamic Time Warping (DTW)~\cite{gold2018dynamic} distance on complex-valued time-series.
Finally, we compare real-world datasets to an artificially-generated dataset obeying simple statistical properties.

\subsection{Related Works}
\label{sec:intoduction:related-works}

Recent $k$-nearest neighbor search algorithms designed to scale with the exponential growth of data include Hierarchical Navigable Small World networks (HNSW)~\cite{malkov2016hnsw}, InVerted File indexing (FAISS-IVF)~\cite{faissivf}, random projection and tree building (ANNOY)~\cite{annoy}, and entropy-scaling search~\cite{yu2015entropy, ishaq2019clustered}. However, some of these algorithms do not provide exact search (as defined in Section \ref{sec:introduction} above).

Hierarchical Navigable Small World networks~\cite{malkov2016hnsw} rely on navigable small world (NSW) networks~\cite{kleinberg2000navigation, boguna2009navigability} and skip lists.
HNSW builds a multi-layered graph of the dataset.
A query point and each data point are inserted into the graph and joined by an edge to the $M$ nearest nodes in the in the graph, where $M$ is a tunable parameter.
The highest layer in which an element can be placed is determined randomly with an exponentially-decaying probability distribution.
Search starts at the highest layer and descends to the lowest layer, greedily following a path of edges to the nearest node, until reaching the query point.
To improve accuracy, the $efSearch$ hyperparameter can be changed to specify the number of closest nearest neighbors to the query vector to be found at each layer.

InVerted File indexing (IVF)~\cite{faissivf, sacks1987multikey, kent1990signature} clusters data into high-dimensional Voronoi cells, and whichever cell a query point falls into is then searched exhaustively, similarly to an early precursor to our work~\cite{yu2015entropy}.
The number of cells used is governed by the $n_{list}$ parameter.
Increasing this parameter decreases the number of points being exhaustively searched, so it improves speed at the cost of accuracy.
To mitigate accuracy issues caused by a query point falling near a cell boundary, the algorithm has a tunable parameter $n_{probe}$ that specifies the number of additional adjacent or nearby cells to search.

ANNOY~\cite{annoy} relies on random projection and tree building for approximate $k$-NN search.
At each intermediate node of the tree, two points are randomly sampled from the space, and the hyperplane equidistant from them is chosen to divide the space into two subspaces.
This process iterates to create a forest of trees, and the number of iterations is a tunable parameter.
At search time, one can increase the number of trees to be searched to improve recall at the cost of speed.

\subsection{Entropy-Scaling Search}
\label{sec:intoduction:entropy-scaling-search}

The entropy-scaling search paradigm exploits the geometric and topological structure inherent in large datasets.
Importantly, as suggested by their name, entropy-scaling search algorithms exhibit asymptotic complexity that scales with topological properties of the dataset, instead of its cardinality, when the dataset has a manifold structure.
In 2019, we introduced CHESS (Clustered Hierarchical Entropy-Scaling Search)~\cite{ishaq2019clustered}, which extended entropy-scaling $\rho$-NN search from a flat clustering approach to a hierarchical clustering approach.
CLAM (Clustering, Learning and Approximation with Manifolds), originally developed to allow ``manifold mapping'' for anomaly detection~\cite{ishaq2021clustered}, is a refinement of the clustering algorithm from CHESS.
In this paper, we introduce CAKES, a set of three entropy-scaling algorithms for $k$-NN search, implemented in the Rust programming language.

We also provide a theoretical analysis of the time complexity of CAKES's algorithms in Sections~\ref{sec:methods:knn-search:repeated-rnn-complexity} and~\ref{sec:methods:knn-search:complexity-of-sieve-methods}.
These analyses are not worst-case analyses in the traditional sense, as they do not assume the worst possible dataset, namely, a uniform distribution.
Given that CAKES's algorithms are intended to be used on datasets with a manifold structure, complexity analysis assuming a uniform distribution of data would be uninformative.
As a result, we assume in our analyses that the dataset has a manifold structure, as quantified by low fractal dimension and metric entropy.
For our purposes, it makes more sense to analyze the expected performance of CAKES's algorithms under these assumptions.

%% file: sections/2-methods.tex
\section{Methods}
\label{sec:methods}

In this manuscript, we are primarily concerned with $k$-NN search in a finite-dimensional space.
Given a dataset $\textbf{X} = \{x_1 \dots x_n\}$ of cardinality $|\textbf{X}| = n$, we define a \textit{point} or \textit{datum} $x_i \in \textbf{X}$ as a singular observation. Examples include the neural-network embedding of an image, the genome of an organism, and a measurement of a radio frequency signal.

We define a \textit{distance function} $f : \textbf{X} \times \textbf{X} \mapsto \mathbb{R}^+ \ \cup \ \{0\}$ which, given two points, deterministically returns a finite and non-negative real number.
A distance value of zero defines an identity among points (i.e.,\,$f(x, y) = 0 \Leftrightarrow x = y$) and larger values indicate greater dissimilarity among points.
We also require that the distance function be symmetric (i.e.,\,$f(x, y) = f(y, x) \ \forall \ x, y \in \textbf{X}$).
In addition to these constraints, if the distance function obeys the triangle inequality (i.e.,\,$f(x, y) \leq f(x, z) + f(z, y) \ \forall \ x, y, z \in \textbf{X}$), then it is also a \textit{distance metric}.
Similar to~\cite{yu2015entropy}, when used with distance metrics, all search algorithms in CAKES are exact.
For example, Euclidean, Levenshtein~\cite{levenshtein1966binary} and Dynamic Time Warping (DTW)~\cite{muller2007dynamic} distances are all distance metrics, while cosine distance is not a metric because it violates the triangle inequality (e.g.,\,consider the points $x = (1, 0)$, $y = (0, 1)$ and $z = (1, 1)$ on the Cartesian plane).

The choice of an appropriate distance function varies by dataset and domain.
For example, with neural-network embeddings, one could use Euclidean or Cosine distance.
With genomic or proteomic sequence data, Levenshtein and Hamming distances are useful.
With time-series data, one could use Dynamic Time Warping (DTW) or Wasserstein distance.

CAKES derives its advantages from the manifold hypothesis~\cite{fefferman2016testing}, the notion that high-dimensional data collected from constrained generating phenomena typically only occupy a low-dimensional manifold within their embedding space.
We say that such data are \textit{manifold-constrained} and have low \textit{local fractal dimension} (LFD).
In other words, we assume that the dataset is embedded in a $D$-dimensional space, but that the data only occupy a $d$-dimensional manifold, where $d \ll D$.
While we sometimes use Euclidean notions to describe the geometric and topological properties of the clusters and manifold, CLAM and CAKES do not rely on such notions;
they serve merely as convenient and intuitive vocabulary to discuss the underlying mathematics.
CAKES exploits the low LFD of such datasets to accelerate search.
We define LFD at some length scale around a point in the dataset as:
\begin{equation}
    \text{LFD}(q, r_1, r_2) = \frac{\text{log} \left( \frac{|B(q, r_1)|}{|B(q, r_2)|} \right) }{\text{log} \left( \frac{r_1}{r_2} \right) }
    \label{eq:methods:lfd-original}
\end{equation}
where $B(q, r)$ is the set of points contained in the metric ball of radius $r$ centered at a point $q$ in the dataset $\textbf{X}$.

We use a simplified version of Equation~\ref{eq:methods:lfd-original} by using a length scale where $r_1 = 2 \cdot r_2$.
\begin{equation}
    \text{LFD}(q, r) = \text{log}_2 \left( \frac{|B(q, r)|}{|B(q, \frac{r}{2})|} \right)
    \label{eq:methods:lfd-half}
\end{equation}

Intuitively, LFD measures the rate of change in the number of points in a ball of radius $r$ around a point $q$ as $r$ increases. When the vast majority of points in the dataset have low ($\ll D$) LFD, we can simply say that the dataset has low LFD.
We stress that this concept differs from the \textit{embedding dimension} of a dataset.
To illustrate the difference, consider the SILVA 18S rRNA dataset that contains genomic sequences with unaligned lengths of up to 3,712 base pairs and aligned length of 50,000 base pairs.
Hence, the \textit{embedding dimension} of this dataset is at least 3,712 and at most 50,000; as used in this paper it is 3,712.
However, physical constraints (namely, biological evolution and biochemistry) constrain the data to a lower-dimensional manifold within this space.
LFD is an approximation of the dimensionality of that lower-dimensional manifold in the ``vicinity'' of a given point.
Section~\ref{sec:results:lfd-of-datasets} discusses this concept on a variety of datasets, showing how real datasets uphold the manifold hypothesis.
For real-world datasets, we expect the LFD to be locally uniform, i.e.,\,when $r$ is small, but potentially highly variable at larger length scales, i.e.,\,when $r$ is large.

\subsection{Clustering}
\label{sec:methods:clustering}

We define a \textit{cluster} as a set of points with a \textit{center} and a \textit{radius}.
The center is the geometric median of the points in the cluster, i.e.,\,it is the point that minimizes the sum of distances to all other points in the cluster.
In cases where the cardinality of the cluster is large, we take a random subsample of $\sqrt{|C|}$ points and compute the geometric median of that subsample~\cite{ishaq2019clustered}.
The center, therefore, is one of the points in the cluster and is used as a representative of the cluster.
The radius is the maximum distance from the center to any point in the cluster.
Each non-leaf cluster has two child clusters in much the same way that a node in a binary tree has two child nodes.
Note that clusters can have overlapping volumes and, in such cases, points in the overlapping volume are assigned to exactly one of the overlapping clusters.
As a consequence, a cluster can be a proper subset of the metric ball at the same center and radius, i.e.,\,$C(c, r) \subset B(c, r)$. We denote the cluster tree by $\mathcal{T}$ and the root by $\mathcal{R}$ (see Section~\ref{sec:methods:clustering:building-the-tree}).

Hereafter, when we refer to the LFD of a cluster, it is estimated at the length scale of the cluster radius and half that radius, i.e.,\,using Equation~\ref{eq:methods:lfd-half}.
We also only use points that are in $C(c, r)$ instead of all those in $B(c, r)$.

The \textit{metric entropy} $\mathcal{N}_{r}(X)$ for some radius $r$ is the minimum number of clusters of a uniform radius $r$ needed to cover the data~\cite{yu2015entropy} for a flat clustering with clusters of radius $r$.
In this paper, where the clustering is hierarchical rather than flat, we define the metric entropy $\mathcal{N}_{\hat{r}}(X)$ as the number of leaf clusters in the tree where $\hat{r}$ is the mean radius of all leaf clusters.

\subsubsection{Building the Tree}
\label{sec:methods:clustering:building-the-tree}

We start by performing a divisive hierarchical clustering on the dataset using CLAM to obtain a cluster tree $\mathcal{T}$.
The procedure is almost identical to that outlined in CHESS~\cite{ishaq2019clustered}, but with better selection of poles for partitioning (see Algorithm~\ref{alg:methods:partition}).
After building $\mathcal{T}$ with CLAM, we also perform a depth-first reordering of the dataset (see Section~\ref{sec:methods:clustering:depth-first-reordering}).

\begin{algorithm} 
    \caption{Partition($C$, $criteria$)} 
    \label{alg:methods:partition} 
    \begin{algorithmic} 
        \Require $f: X \times X \mapsto \mathbb{R}^+ \cup \{0\}$, a distance function
        \Require $C$, a cluster
        \Require $criteria$, user-specified continuation criteria

        \State $seeds \Leftarrow$ random sample of $\left\lceil \sqrt{|C|} \right\rceil$ points from $C$
        \State $c \Leftarrow$ geometric median of $seeds$
        \State $l \Leftarrow \argmax f(c, x) \ \forall \ x \in C$
        \State $r \Leftarrow \argmax f(l, x) \ \forall \ x \in C$
        \State $L \Leftarrow \{x \ | \ x \in C \land f(l, x) \le f(r, x)\}$
        \State $R \Leftarrow \{x \ | \ x \in C \land f(r, x) < f(l, x)\}$

        \If{$|L| > 1$ \textbf{and} $L$ satisfies $criteria$}
            \State Partition($L$, $criteria$)
        \EndIf

        \If{$|R| > 1$ \textbf{and} $R$ satisfies $criteria$}
            \State Partition($R$, $criteria$)
        \EndIf
    \end{algorithmic}
\end{algorithm}

Given a cluster $C$ with $|C|$ points, we define its two children by the following process. We take a random subsample $S$ of $\sqrt{|C|}$ of $C$'s points, and compute pairwise distances between all points in $S$.
Using these distances, we compute the \textit{geometric median} of $S$; in other words, we find the point that minimizes the sum of distances to all other points in $S$.
We define the \textit{center} of $C$ to be this geometric median.

The \textit{radius} of $C$ is the maximum distance from the center to any other point in $C$.
The point that is responsible for that radius (i.e.,\,the furthest point from the center) is designated the \textit{left pole} and the point that is furthest from the left pole is designated the \textit{right pole}.
We then partition the cluster into a \textit{left child} and a \textit{right child}, where the left child contains all points in the cluster that are closer to the left pole than to the right pole, and the right child contains all points in the cluster that are closer to the right pole than to the left pole.
Without loss of generality, we assign to the left child those points that are equidistant from the two poles.
Starting from a root cluster $\mathcal{R}$ containing the entire dataset, we repeat this procedure until each leaf contains only one datum, or we meet some other user-specified stopping criteria (e.g., minimum cluster radius, minimum cluster cardinality, maximum tree depth, etc).
This process is described in Algorithm \ref{alg:methods:partition}.
During the partitioning process, we also compute (and cache) the LFD of each cluster using Equation~\ref{eq:methods:lfd-half}.

\subsubsection{Depth-First Reordering}
\label{sec:methods:clustering:depth-first-reordering}

In CHESS~\cite{ishaq2019clustered}, each cluster stored a list of indices into the dataset.
This list was used to retrieve the clusters' points during search.
Although this approach allowed us to retrieve the points in constant time, its memory cost was prohibitively high.
With a dataset of cardinality $n$ and each cluster storing a list of indices for its points, we stored a total of $n$ indices at each depth in the tree $\mathcal{T}$.
Assuming $\mathcal{T}$ is balanced, and thus $\mathcal{O}(\log n)$ depth, this approach had a memory overhead of $\mathcal{O}(n \log n)$.
In this work, we introduce a new approach wherein, after building $\mathcal{T}$, we reorder the dataset so that points are stored in a depth-first order.
Then, within each cluster, we need only store its \textit{cardinality} and an \textit{offset} to access its points from the dataset.
The root cluster $\mathcal{R}$ has an offset of zero and a cardinality equal to the number of points in the dataset.
A left child has the same offset its parent, and the corresponding right child has an offset equal to the left child's offset plus the left child's cardinality.
With no additional memory cost nor time cost for retrieving points during search, depth-first reordering offers the same time complexity as CHESS but with $\mathcal{O}(n)$ memory overhead.

\subsubsection{Time Complexity}
\label{sec:methods:clustering:time-complexity}

The asymptotic complexity of Partition is the same as described in~\cite{ishaq2019clustered}.
By using an approximate partitioning with a $\sqrt{n}$ sample, we achieve $\mathcal{O}(n)$ cost of partitioning and $\mathcal{O}(n \log n)$ cost of building $\mathcal{T}$.
This is a significant improvement over exact partitioning and tree-building, which cost $\mathcal{O}(n^2)$ and $\mathcal{O}(n^2 \log n)$ respectively.

This complexity analysis ignores the cost of checking the continuation criteria.
While a user could specify criteria of any complexity (our implementation allows for this), for this paper we continue clustering until each cluster is a singleton, i.e.,\,it contains only one point or duplicates of the same point and has a radius of zero.
This criterion costs $\mathcal{O}(1)$ to check per cluster, and so does it not affect the overall complexity of the algorithm as used in this paper.

We also note here that this $\mathcal{O}(n \log n)$ complexity assumes that $\mathcal{T}$ is balanced.
In practice, for real datasets, we expect Algorithm~\ref{alg:methods:partition} to produce anything but a balanced tree; the varying sampling density in different regions of the manifold and the low dimensional ``shape'' of the manifold itself will cause it to be unbalanced.
The only case in which we would expect a balanced tree is if the dataset were uniformly distributed, e.g.,\,in a $d$-dimensional hyper-cube.

When building a tree for a real dataset, the depth would be larger than $\log n$, which would seem to increase the cost of tree-building.
However, because of the large numbers of leaf clusters at shallow depths, each subsequent level of $\mathcal{T}$ would also have fewer points, and so the cost of building each subsequent level would be lower.
Any analysis of unbalanced trees would be highly dependent on the specific dataset, and so we do not provide one here.
We do, however, explore empirical results based on forcing a balanced tree in Section~\ref{sec:results:clustering-strategies-and-number-of-distance-computations}.
In any case, the tree is built only once, so this cost is amortized over all search queries and other applications.

\begin{figure}[H]
    \vskip -0.2in
    \centering
    \includegraphics[scale=0.65]{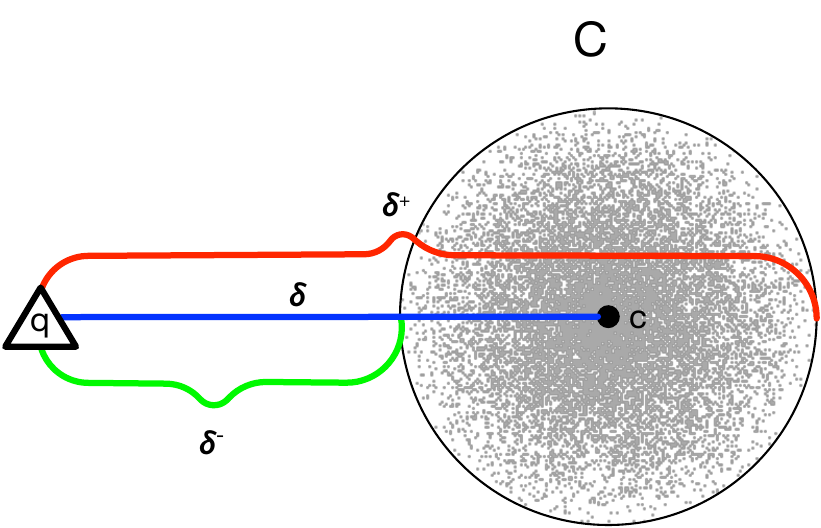}
    \caption{
        {\color{blue}$\delta$}, {\color{red}$\delta^{+}$}, and {\color{green}$\delta^{-}$} for a cluster $C$ and a query $q$.
        ${\color{blue}\delta} = f(q, c)$ is the distance from the query to the cluster center $c$.
        ${\color{red}\delta^{+}} = \delta + r$ is the distance from the query to the theoretically farthest point in $C$.
        ${\color{green}\delta^{-}} = \text{max}(0, \delta - r)$ is the distance from the query to the theoretically closest point in $C$.
    }
    \label{fig:methods:deltas}
    \vskip -0.5in
\end{figure}

\begin{minipage}{.425\textwidth}
    \begin{algorithm}[H]
        \caption{tree-search($C$, $q$, $r$)}
        \label{alg:methods:rnn-search:tree-search}
        \begin{algorithmic}
            \Require $f$, a distance function
            \Require $C$, a cluster
            \Require $q$, a query
            \Require $r$, a search radius
            \If{$\delta^+_C \leq r$}
                \State \textbf{return} $\{C\}$
            \Else
                \State $[L, R]$ $\Leftarrow$ \textit{children} of $C$
                \State \textbf{return} tree-search($L, q, r$) \\
                \ \ \ \ \ \ \ \ \ \ \ \ \ \ \ $\cup$ tree-search($R, q, r$)
            \EndIf
        \end{algorithmic}
    \end{algorithm}
\end{minipage}
\hfill
\begin{minipage}{.475\textwidth}
    \begin{algorithm}[H]
        \caption{leaf-search($Q$, $q$, $r$)}
        \label{alg:methods:rnn-search:leaf-search}
        \begin{algorithmic}
            \Require $f$, a distance function
            \Require $Q$, a set of clusters
            \Require $q$, a query
            \Require $r$, a search radius
            \State $H \Leftarrow \emptyset$
            \For{$C \in Q$}
                \If{$\delta^+_C \leq r$}
                    \State $H$ $\Leftarrow$ $H \cup C$
                \Else
                    \For{$p \in C$}
                        \If{$f(p, q) \leq r$}
                            \State $H$ $\Leftarrow$ $H \cup \{p\}$
                        \EndIf
                    \EndFor
                \EndIf
            \EndFor
            \State \textbf{return} $H$
        \end{algorithmic}
    \end{algorithm}
\end{minipage}

\subsection{\texorpdfstring{$k$}{k}-Nearest Neighbors Search}
\label{sec:methods:knn-search}

Given a dataset $\textbf{X}$, a distance function $f$ and the root cluster $\mathcal{R}$ of the tree $\mathcal{T}$ (constructed by Algorithm~\ref{alg:methods:partition}), we can now formally pose the $k$-NN search problem.
Given a user specified integer $k$ and query $q$, $k$-NN search aims to find the $k$ closest points to $q$ in $\textbf{X}$.
In other words, $k$-NN search aims to find the set $H$ such that $|H| = k$ and $H = B_X(q, \rho_k)$ where $\rho_k = \max \left\{ f(q, p) \ \forall \ p \in H \right\}$ is the distance from $q$ to the $k^{th}$ nearest neighbor in $\textbf{X}$.

In this section, we present three novel algorithms for exact $k$-NN search:
Repeated $\rho$-NN, Breadth-First Sieve, and Depth-First Sieve.
In these algorithms, we use $H$, for \textit{hits}, to refer to the data structure that stores the closest points to the query found so far, and $Q$ to refer to the data structure that stores the clusters and points that are still in contention for being one of the $k$ nearest neighbors.
These algorithms also use some terminology defined and illustrated in Figure~\ref{fig:methods:deltas}.

\begin{algorithm} 
    \caption{Repeated $\rho$-NN($\mathcal{R}$, $q$, $k$)} 
    \label{alg:methods:repeated-rnn} 
    \begin{algorithmic} 
        \Require $\mathcal{R}$, the root cluster
        \Require $q$, a query
        \Require $k$, the number of neighbors to find
        \State{$H \Leftarrow [\ ]$, a max-heap by $\delta$ of size $k$}
        \State $r \Leftarrow radius$ of $\mathcal{R}$
        \State $r \Leftarrow$ $\frac{r}{|\mathcal{R}|}$
        \State $Q \Leftarrow$ tree-search($\mathcal{R}$, $q$, $r$)
        \While{$\sum_{C \in Q} |C| < k$}
            \If{$Q = \emptyset$}
                \State $r \Leftarrow 2 \cdot r$
            \ElsIf{$\sum_{C \in Q} |C| >= k$}
                \State $\mu \Leftarrow \frac{1}{|Q|} \cdot \sum_{C \in Q} \big( LFD(C)^{-1} \big)$
                \State $r \Leftarrow r \cdot \min \bigg( 2, \left( {\frac{k}{\sum_{C \in Q} |C|}} \right)^{\mu} \bigg)$
            \EndIf
            \State $Q \Leftarrow$ tree-search($\mathcal{R}$, $q$, $r$)
        \EndWhile
        \State $H \Leftarrow \text{leaf-search}(Q, q, r)$
        \State \textbf{return} $H$ as a list
    \end{algorithmic}
\end{algorithm}

\subsubsection{Repeated \texorpdfstring{$\rho$}{p}-NN}
\label{sec:methods:knn-search:repeated-rnn}

This algorithm relies on the \textit{tree-search} (Algorithm~\ref{alg:methods:rnn-search:tree-search}) and \textit{leaf-search} (Algorithm~\ref{alg:methods:rnn-search:leaf-search}) as described in~\cite{ishaq2019clustered} and reproduced here for completeness.

For Repeated $\rho$-NN search (Algorithm~\ref{alg:methods:repeated-rnn}), we begin by performing \textit{tree-search} with a search radius $r$ equal to the radius of the root $\mathcal{R}$ divided by the cardinality of the dataset.
If no clusters are found, then we double $r$ and perform tree-search again, repeating until we find at least one cluster.

Now, so long as $\sum_{C \in Q} |C| < k$, we continue to perform tree-search, but instead of doubling $r$ on each iteration, we multiply it by a factor determined by the LFD in the vicinity of the query ball.
In particular, we increase the radius by a factor of
\begin{equation}
    \min \left(2, \left( {\frac{k}{\sum_{C \in Q} |C|}} \right)^{\mu} \right)
    \label{eq:methods:repeated-rnn-factor}
\end{equation}
where $\mu$ is the multiplicative inverse of the harmonic mean of the LFD of the clusters in $Q$, i.e.,\,$\mu = \frac{1}{|Q|} \cdot \sum_{C \in Q} \big( LFD(C)^{-1} \big)$.
We use the harmonic mean to ensure that $\mu$ is not dominated by outlier clusters with very high LFD.
We cap the radial increase at 2 to ensure that we do not increase the radius too quickly in any single iteration.

Intuitively, the factor by which we increase the radius should be \textit{inversely} related to the number of points found so far.
When the LFD at the radius scale from the previous iteration is high, this suggests that the data are densely populated in that region.
Thus, a small increase in the radius would likely encounter many more points, so a smaller radial increase would suffice to find $k$ neighbors.
Conversely, when the LFD at the radius scale from the previous iteration is low, this suggests that the data are sparsely populated in that region.
In such a region, a small increase in the radius would likely encounter vacant space, so a larger radial increase is needed.
Thus, the factor of radius increase should also be \textit{inversely} related to the LFD.
However, we should not increase the radius too drastically with any one iteration because we assume that the LFD is only \textit{locally} uniform.
A large increase in the radius would likely break out of the local region and potentially encounter too many new clusters, which would make the subsequent step computationally expensive.

Once $\sum_{C \in Q} |C| \geq k$, we are guaranteed to have found at least $k$ neighbors, and so we perform \textit{leaf-search} to find the $k$ nearest neighbors among the points in the clusters found after the last tree-search.

\subsubsection{Complexity of Repeated \texorpdfstring{$\rho$}{p}-NN}
\label{sec:methods:knn-search:repeated-rnn-complexity}

\begin{theorem} Let $X$ be a dataset and $q$ a query sampled from the same distribution (i.e., arising from the same generative process) as $X$. Then time complexity of performing Repeated $\rho$-NN search on $X$ with query $q$ is \begin{gather}
        \mathcal{O}
        \Bigg(
            \underbrace{
                \log~\overbrace{\mathcal{N}_{\hat{r}}(X)}^{\textrm{metric entropy}}
            }_{\textrm{tree-search}}
            \ + \
            \underbrace{
                \overbrace{k}^{\textrm{output size}} \cdot
                \overbrace{\bigg( 1 + 2 \cdot \Big( \frac{\hat{|C|}}{k} \Big) ^ {d^{-1}} \bigg)^d}^{\textrm{scaling factor}}
            }_{\textrm{leaf-search}}
        \Bigg)
        \label{eq:methods:repeated-rnn-complexity}
    \end{gather}
    where $\mathcal{N}_{\hat{r}}(X)$ is the metric entropy of the dataset, $d$ is the LFD of the dataset, and $k$ is the number of nearest neighbors.
    \label{thm:methods:rnn-complexity}
\end{theorem}

\begin{proof} We consider the \textit{tree-search} and \textit{leaf-search} stages of search separately.
Tree-search refers to the process of identifying clusters that overlap with the query ball, or in other words, clusters that might contain one of the $k$ nearest neighbors by doing repeated iterations of the  $\rho$-NN search Algorithm~\ref{alg:methods:rnn-search:tree-search}.
In ~\cite{ishaq2019clustered}, we showed that the complexity of $\rho$-NN search is
\begin{gather}
    \mathcal{O}
    \Bigg(
        \underbrace{
            \log~\overbrace{\mathcal{N}_{\hat{r}}(X)}^{\textrm{metric entropy}}
        }_{\textrm{tree-search}}
        \ + \
        \underbrace{
            \overbrace{ \big| B(q, \rho) \big|}^{\textrm{output size}}
            \overbrace{ \left( \frac{\rho + 2 \cdot \hat{r}}{ \rho} \right) ^ d}^{\textrm{scaling factor}}
        }_{\textrm{leaf-search}}
    \Bigg)
    \label{eq:methods:rnn-search-complexity}
\end{gather}
where $\hat{r}$ is the \textit{mean} radius of leaf clusters, $\mathcal{N}_{\hat{r}}(X)$ is the metric entropy at that radius, $B(q, \rho)$ is a ball of radius $\rho$ around the query $q$, and $d$ is the LFD around the query at the length scale of $\rho$ and $\rho + 2 \cdot \hat{r}$.

To extend Equation~\ref{eq:methods:rnn-search-complexity} to Repeated $\rho$-NN, we must first estimate the number of iterations of tree-search (Algorithm~\ref{alg:methods:rnn-search:tree-search}) needed to find a radius that guarantees at least $k$ neighbors.
Since $q$ is sampled from the same distribution as $X$, the LFD near $q$ should not differ significantly from the (harmonic) mean of the LFDs of clusters near $q$ at the scale of the distance from the query to the $k^{th}$ nearest neighbor.
Given that LFD near $q$ does not differ significantly from that of nearby clusters, Equation~\ref{eq:methods:repeated-rnn-factor} suggests that in the expected case, we need only two iterations of tree-search to find $k$ neighbors:
one iteration to find at least one cluster, and one more to find enough clusters to guarantee $k$ neighbors.
Since this is a constant factor, complexity of tree-search for Repeated $\rho$-NN is the same as that of $\rho$-NN search, i.e.,\,$\mathcal{O}\big(\log\mathcal{N}_{\hat{r}}(X)\big)$.

We proceed to determine the complexity of leaf-search. Let $Q$ be the set of clusters returned by tree-search. We must estimate $\sum_{C \in Q} |C|$, the total cardinality of the clusters returned by tree-search; since we must examine every point in each such cluster, time complexity of leaf-search is linear in this quantity.
Let $\rho_k$ be the distance from the query to the $k^{th}$ nearest neighbor.
Then, we see that $Q$ is expected to be the set of clusters that overlap with a ball of radius $\rho_k$ around the query.
We can estimate this region as a ball of radius $\rho_k + 2\hat{r}$, where $\hat{r}$ is the mean radius of the clusters in $Q$.

The work in~\cite{yu2015entropy} showed that $\sum_{C \in S} |C| \leq \gamma  \left| B(q, \rho_k) \right| \left(\frac{\rho_k + 2 \cdot \hat{r}}{\rho_k} \right)^d$, where $\gamma$ is a constant.
By definition of $\rho_k$, we have that $|B(q, \rho_k)| = k$.
Thus, $\sum_{C \in S} |C| \leq \gamma k \left( 1 + 2 \cdot \frac{\hat{r}}{\rho_k} \right)^d$.
It remains to determine an estimate for $\rho_k$.
We let $\hat{d}$ be the (harmonic) mean LFD of the clusters in $Q$.
While ordinarily we compute LFD by comparing cardinalities of two balls with two different radii centered at \textit{the same} point, in order to estimate $\rho_k$, we instead compare the cardinality of a ball \textit{around the query} of radius $\rho_k$ to the mean cardinality, $\hat{|C|}$, of clusters in $Q$ at a radius equal to the mean of their radii, $\hat{r}$.
Since $q$ is from the same distribution as $X$, the LFD at $q$ should not be significantly different from that at the center of a cluster in $Q$.
By Equation~\ref{eq:methods:lfd-original}, $\hat{d} = \frac{\log{}\frac{\hat{|C|}}{k}}{\log{}\frac{\hat{r}}{\rho_k}}$.
We can rearrange this equation to get $\frac{\hat{r}}{\rho_k} = \left( \frac{\hat{|C|}}{k} \right)^{\hat{d}^{-1}}$.
Using this to simplify the term for leaf-search in Equation~\ref{eq:methods:rnn-search-complexity}, we get:
\begin{equation*}
    k \left( 1 + 2 \cdot \left( \frac{\hat{|C|}}{k} \right) ^ {\hat{d}^{-1}} \right)^d
\end{equation*}

Again using the assumption that since $q$ is from the same distribution as $X$, the LFD at $q$ should not be significantly different from that at a cluster in $Q$, we have that $\hat{d} \approx d$.
By combining the bounds for tree-search and leaf-search, we see that Repeated $\rho$-NN has time complexity as claimed in Equation~\ref{eq:methods:repeated-rnn-complexity}.
\end{proof}

We remark that the scaling factor in Equation~\ref{eq:methods:repeated-rnn-complexity} should be close to 1 unless LFD is highly variable in the region around the query (i.e.,\,if $\hat{d}$ differs significantly from $d$).

\subsubsection{Breadth-First Sieve}
\label{sec:methods:knn-search:bredth-first-sieve}

This algorithm (described in Algorithm~\ref{alg:methods:bredth-first-sieve}) performs a breadth-first traversal of the tree $\mathcal{T}$, pruning clusters by using a modified version of the QuickSelect algorithm~\cite{hoare1961algorithm} at each level.

We begin by letting $Q$ be a set of 3-tuples $(p, \delta^{+}_{p}, m)$, where $p$ is either a cluster or a point, $\delta^{+}_{p}$ is the $\delta^{+}$ of $p$ as illustrated in Figure~\ref{fig:methods:deltas}, and $m$ is the multiplicity of $p$ in $Q$.
During the breadth-first traversal, for every cluster $C$ we encounter, we add $(C, \delta^{+}_{C}, |C| - 1)$ and $(c, \delta_{c}, 1)$ to $Q$, where $c$ is the center of $C$.
Recall that by the definitions of $\delta$ and $\delta^{+}$ given in Section~\ref{sec:methods:knn-search}, since $c$ is a point, $\delta_{C} = \delta_{c} = \delta^{+}_{c} = \delta^{-}_{c}$.

\begin{minipage}{0.5125\textwidth}
    \begin{algorithm}[H]\small
        \caption{Breadth-First Sieve($\mathcal{R}$, $q$, $k$)} 
        \label{alg:methods:bredth-first-sieve} 
        \begin{algorithmic} 
            \Require $\mathcal{R}$, the root cluster
            \Require $q$, a query
            \Require $k$, the number of neighbors to find
            \State $c \Leftarrow$ \textit{center} of $\mathcal{R}$
            \State $Q \Leftarrow$ \{ ($\mathcal{R}$, $\delta^{+}_{\mathcal{R}}$, $|\mathcal{R}| - 1$), ($c$, $\delta_{\mathcal{R}}$, 1) \}
            \While{$\sum_{(\_, \_, m) \in Q} m \neq k$}
                \State $\tau \Leftarrow$ QuickSelect($Q$, $k$)
                \State $Q^{'} \Leftarrow \emptyset$
                \For{$(C, \_, \_) \in Q$}
                    \If{$\delta^{-}_{C} \leq \tau$}
                        \If{$C$ is a point}
                            \State $Q^{'} \Leftarrow Q^{'} \cup \{ (C, \delta_{C}, 1) \}$
                        \ElsIf{$C$ is a leaf}
                            \State $Q^{'} \Leftarrow Q^{'} \cup \{ (p, \delta_{p}, 1)$ for $p \in C \}$
                        \Else
                            \State $[L, R] \Leftarrow$ children of $C$
                            \State $l, r \Leftarrow$ centers of $L, R$
                            \State $Q^{'} \Leftarrow Q^{'} \cup \{ (L, \delta^{+}_{L}, |L| - 1), (l, \delta_{L}, 1) \}$
                            \State $Q^{'} \Leftarrow Q^{'} \cup \{ (R, \delta^{+}_{R}, |R| - 1), (l, \delta_{R}, 1) \}$
                        \EndIf
                    \EndIf
                \EndFor
                \State $Q \Leftarrow Q^{'}$
            \EndWhile
            \State QuickSelect($Q$, $k$)
            \State \textbf{return} The first $k$ points in $Q$
        \end{algorithmic}
    \end{algorithm}
\end{minipage}
\hfill
\begin{minipage}{0.425\textwidth}
    \begin{algorithm}[H]\small
        \caption{Depth-First Sieve($\mathcal{R}$, $q$, $k$)}
        \label{alg:methods:depth-first-sieve}
        \begin{algorithmic}
            \Require $\mathcal{R}$, the root cluster
            \Require $q$, a query
            \Require $k$, the number of neighbors to find
            \State{$Q \Leftarrow [\mathcal{R}]$, a min-heap by $\delta^{-}$}
            \State{$H \Leftarrow [\ ]$, a max-heap by $\delta$ of size $k$}
            \While{$|H| < k$ \textbf{or} $H.peek.\delta \geq Q.peek.\delta^{-}$}
                \While{$Q.peek$ is not a leaf}
                    \State{$C \Leftarrow Q.pop$, the closest cluster}
                    \State{$[L, R] \Leftarrow$ children of $C$}
                    \State{$Q.push(L)$}
                    \State{$Q.push(R)$}
                \EndWhile
                \State{$leaf \Leftarrow Q.pop$}
                \For{$p \in leaf$}
                    \State{$H.push(p)$}
                \EndFor
            \EndWhile
            \State \textbf{return} $H$
        \end{algorithmic}
    \end{algorithm}
\end{minipage}

\vskip 0.1in

We then use the QuickSelect algorithm, modified to account for multiplicities and to reorder $Q$ in-place, to find the element in $Q$ with the $k^{th}$ smallest $\delta^{+}$; in other words, we find $\tau$, the smallest $\delta^{+}$ in $Q$ such that $\left| B(q, \tau) \right| \geq k$.
Since this step may require a binary search for the correct pivot element to find $\tau$ and reordering with a new pivot takes linear time in the size of the input list, this version of QuickSelect has $\mathcal{O}\left(|Q| \log |Q|\right)$ time complexity.

Next, we go over all items in $Q$, skipping over any for which $\delta^{-} > \tau$ because such elements cannot contain (or be) one of the $k$ nearest neighbors.
If the item corresponds to a point, we keep it.
If the item corresponds to a leaf cluster, we add all its points to $Q$ with a multiplicity of 1 each.
If the item corresponds to a non-leaf cluster, we to $Q$ the pairs of 3-tuples corresponding to its child clusters.
We continue this process until the sum of multiplicities in $Q$ is exactly $k$.
We then use the QuickSelect algorithm one last time to reorder $Q$ and return the $k$ nearest neighbors.

\subsubsection{Depth-First Sieve}
\label{sec:methods:knn-search:depth-first-sieve}

This algorithm (described in Algorithm~\ref{alg:methods:depth-first-sieve}) is quite similar to a depth-first traversal of the tree $\mathcal{T}$, using two priority queues to track clusters and hits, and to prioritize which branch of $\mathcal{T}$ to explore next.

Let $Q$ be a min-queue of clusters prioritized by $\delta^{-}$ and $H$ be a max-queue (with capacity $k$ and the ability to maintain its maximum size) of points prioritized by $\delta$.
$Q$ starts containing only the root cluster $\mathcal{R}$ while $H$ starts empty.
So long as $H$ is not full or the top element in $H$ has $\delta$ greater than or equal to the top element in $Q$, we take the following steps:

\begin{itemize}
    \item While the top-priority element is not a leaf, remove it from $Q$ and add its children to $Q$.
    \item Remove the top-priority element (a leaf) from $Q$ and add all its points to $H$.
\end{itemize}

This process terminates when $H$ is full and the top element in $H$ has $\delta$ less than the top element in $Q$, i.e.,\,the theoretically closest point left to be considered in $Q$ is farther from the query than the $k^{th}$ nearest neighbor in $H$.
This leaves $H$ containing exactly the $k$ nearest neighbors to the query.

Note that this algorithm is not truly a depth-first traversal of $\mathcal{T}$ in the classical sense, because we use $Q$ to prioritize which branch of $\mathcal{T}$ we descend into.
Indeed, we expect this algorithm to often switch which branch of $\mathcal{T}$ is being explored at greater depth.

\subsubsection{Complexity of Sieve Methods}
\label{sec:methods:knn-search:complexity-of-sieve-methods}

Due to their similarity, we combine the complexity analyses of both Sieve methods.
For these methods we again use the terminology of tree-search and leaf-search.
Tree-search navigates the cluster tree and adds clusters to $Q$.
Leaf-search exhaustively searches \textit{some} of the clusters in $Q$ to find the $k$ nearest neighbors.

\begin{theorem}
    Let $X$ be a dataset and $q$ a query sampled from the same distribution (i.e., arising from the same generative process) as $X$. Let $T \coloneqq \mathcal{O} \big( \lceil d \rceil \cdot \log \mathcal{N}_{\hat{r}}(X) \big)$ and   $L \coloneqq \mathcal{O} \left( k \cdot \bigg( 1 + 2 \cdot \Big( \frac{\hat{|C|}}{k} \Big) ^ {d^{-1}} \bigg)^d \right)$, where $\mathcal{N}_{\hat{r}}(X)$ is the metric entropy of $X$, $d$ is the LFD of $X$, $\hat{|C|}$ is the mean cardinality of clusters overlapping the query ball, and $k$ is the number of nearest neighbors. Then, for dataset $X$ and query $q$, the time complexity of performing Breadth-First Sieve search is \begin{equation}
        \mathcal{O} \Big( (T + L ) \log (T + L ) \Big)
        \label{eq:methods:breadth-first-sieve-complexity}
    \end{equation} and the the time complexity of performing Depth-First Sieve search is \begin{equation}
        \mathcal{O} \Big( T \log T + L \log k \Big).
        \label{eq:methods:depth-first-sieve-complexity}
    \end{equation}
    \label{thm:methods:sieve-complexity}
\end{theorem}

\begin{proof}
Since $q$ is sampled from the same distribution as $X$, the LFD near $q$ should not differ significantly from the LFDs of clusters near $q$ at the scale of the distance from the query to the $k^{th}$ nearest neighbor.

Consider leaf clusters with cardinalities near $k$. Let $d$ be the LFD in this region, and $Q$ denote the set of candidate points and clusters.
Then the number of leaf-clusters in $Q$ is bounded above by $2d$, where the bound is achieved if we have a cluster overlapping the query ball at each end of each of $\lceil d \rceil$ mutually-orthogonal axes.
In the worst-case scenario for tree-search, these leaf clusters would all come from different branches of the tree, and so tree-search looks at $2 \cdot \lceil d \rceil \cdot \log \mathcal{N}_{\hat{r}}(X)$ clusters.
Thus, the asymptotic complexity is $T \coloneqq \mathcal{O} \big( \lceil d \rceil \cdot \log \mathcal{N}_{\hat{r}}(X) \big)$.
For leaf-search, the output size and scaling factor are the same as in Repeated $\rho$-NN, and so the asymptotic complexity is $L \coloneqq \mathcal{O} \left( k \cdot \bigg( 1 + 2 \cdot \Big( \frac{\hat{|C|}}{k} \Big) ^ {d^{-1}} \bigg)^d \right)$.

The asymptotic complexity of Breadth-First Sieve is dominated by the QuickSelect algorithm to calculate $\tau$.
Since this method is log-linear in the length of $Q$, and $Q$ contains the clusters from tree-search and the points from leaf-search, we see that the asymptotic complexity is as claimed in Equation~\ref{eq:methods:breadth-first-sieve-complexity}.

For Depth-First Sieve, since we use two priority queues, the asymptotic complexity is as claimed in Equation~\ref{eq:methods:depth-first-sieve-complexity}.
\end{proof}

\subsection{Auto-Tuning}
\label{sec:methods:auto-tuning}

We perform some simple auto-tuning to select the optimal $k$-NN algorithm to use with a given dataset.
We start by taking the center of every cluster at a low depth (e.g.,\,10) in $\mathcal{T}$ as a query.
This gives us a small, representative sample of the dataset.
Using these clusters' centers as queries, and a user-specified value of $k$, we record the time taken for $k$-NN search on the sample using each of the three algorithms described in Section~\ref{sec:methods:knn-search}.
We select the fastest algorithm over all the queries as the optimal algorithm for that dataset and value of $k$.
Note that even though we select the optimal algorithm based on use with some user-specified value of $k$, we still allow search with any value of $k$.

\subsection{Synthetic Data}
\label{sec:methods:synthetic-data}

Based on our asymptotic complexity analyses, we expect CAKES to perform well on datasets with low LFD, and for its performance to scale sub-linearly with the cardinality of the dataset.
To test this hypothesis, we use some datasets from the ANN-benchmarks suite~\cite{aumuller2020ann} and synthetically augment them to generate similar datasets with exponentially larger cardinalities.
We do the same with a large random dataset of uniformly distributed points in a hypercube.
We then compare the performance of CAKES to that of other algorithms on the original datasets and the synthetically augmented datasets.

To elaborate on the augmentation process, we start with an original dataset from the ANN-benchmarks suite.
Let $X$ be the dataset, $d$ be its dimensionality, $\epsilon$ be a user-specified noise level, and $m$ be a user-specified integer multiplier.
For each datum $\mathbf{x} \in X$, we create $m - 1$ new data points within a distance $\epsilon$ of $\mathbf{x}$.
We construct a random vector $\mathbf{r}$ of $d$ dimensions in the hyper-sphere of radius $\epsilon$ centered at the origin.
We then add $\mathbf{r}$ to $\mathbf{x}$ to get a new point $\mathbf{x}'$.
Since $||\mathbf{r}|| \leq \epsilon$, we have that $||\mathbf{x} - \mathbf{x}'|| \leq \epsilon$ (i.e.,\,$\mathbf{x}'$ is within a distance $\epsilon$ of $\mathbf{x}$).
This produces a new dataset $X'$ with $|X'| = m \cdot |X|$.
This augmentation process does not add to the overall topological structure of the dataset, but it does increase its cardinality by a factor of $m$.
This allows us to isolate the effect of cardinality on search performance from that of other factors such as dimensionality, choice of metric, or the topological structure of the dataset.

%% file: sections/3-datasets.tex
\section{Datasets And Benchmarks}
\label{sec:datasets-and-benchmarks}

\subsection{ANN-Benchmark Datasets}
\label{sec:datasets-and-benchmarks:ann-benchmark-datasets}

We benchmark on a variety of datasets from the ANN-benchmarks suite~\cite{aumuller2020ann}.
Table~\ref{tab:datasets:summary} summarizes these datasets.
All benchmarks were conducted on an Intel Xeon E5-2690 v4 CPU @ 2.60GHz with 512GB RAM.
The OS kernel was Manjaro Linux 5.15.164-1-MANJARO.
The Rust compiler was Rust 1.83.0, and the Python interpreter version was 3.9.18.

\begin{table}
    \caption{Datasets used in benchmarks.}
    \label{tab:datasets:summary}
    \begin{center}
        \begin{sc}
            \begin{tabular}{|l|l|l|l|}
                \hline
                \textbf{Dataset} & \textbf{Distance Function}  &\textbf{Cardinality}  & \textbf{Dimensionality}  \\
                \hline
                Fashion-Mnist    & Euclidean                   & 60,000             & 784                    \\
                \hline
                Glove-25         & Cosine                      & 1,183,514          & 25                     \\
                \hline
                Sift             & Euclidean                   & 1,000,000          & 128                    \\
                \hline
                Random           & Euclidean                   & 1,000,000          & 128                    \\
                \hline
                SILVA            & Levenshtein                 & 2,224,640          & 3,712         \\
                \hline
                RadioML          & Dynamic Time Warping        & 97,920             & 1,024                  \\
                \hline
            \end{tabular}
        \end{sc}
    \end{center}
    \vskip -0.1in
\end{table}

\subsection{Random Datasets and Synthetic Augmentation}
\label{sec:datasets-and-benchmarks:random-datasets}

In addition to benchmarks on datasets from the ANN-Benchmarks suite, we also benchmarked on synthetic augmentations of these real datasets, using the process described in \ref{sec:methods:synthetic-data}.
In particular, we use a noise tolerance $\epsilon = 0.01$ and explore the scaling behavior as the cardinality multiplier (referred to as ``Mult.'' in Tables~\ref{tab:results:qps-and-recall-fmn},~\ref{tab:results:qps-and-recall-glove},~\ref{tab:results:qps-and-recall-sift} and~\ref{tab:results:qps-and-recall-random}) increases.
We also benchmarked on purely randomly-generated datasets of various cardinalities.
For this, we used a base cardinality of 1,000,000 and a dimensionality of 128 to match the Sift dataset; hereafter, we refer to the random dataset with cardinality 1,000,000 and dimensionality 128 as ``Random.''
This benchmark allows us to isolate the effect of a manifold structure (which we expect to be absent in a purely random dataset) on the performance of the CAKES' algorithms.

In order to calculate recall on the augmented datasets, we perform linear search in Rust and save the results to disk.
We verified that linear search on the original dataset produced neighbors with perfect recall compared to those provided by the ANN-Benchmarks suite.
We then use the results of linear search on the augmented datasets as ground truth for calculating recall of CAKES's algorithms in Rust and read them in Python for calculating recall of HNSW, ANNOY, and FAISS-IVF.

\subsection{SILVA 18S}
\label{sec:datasets-and-benchmarks:silva-18s}

To demonstrate CAKES with a more exotic distance function, we also benchmarked on the SILVA 18S ribosomal RNA dataset~\cite{10.1093/nar/gks1219}.
This dataset contains ribosomal RNA sequences of 2,224,640 genomes, with the longest sequence having 3,712 letters.
We held out a set of 1,000 random sequences from the dataset to use as queries for benchmarking.
We use Levenshtein~\cite{levenshtein1966binary} distance on the unaligned sequences to build the tree and to perform $k$-NN search.
We note that the sequences in this dataset were provided in a multiple sequence alignment with a width of 50,000 characters.
We could have used Hamming distance on the aligned sequences, but we chose to use Levenshtein distance on the unaligned sequences to help demonstrate the flexibility of CAKES in handling exotic distance functions.
Under Hamming distance, we could consider this dataset to have an embedding dimension of 50,000, but for Levenshtein distance, we state the dimensionality as the length of the longest sequence, i.e.\, 3,712.

\subsection{Radio ML}
\label{sec:datasets-and-benchmarks:radio-ml}

As another example of an exotic distance function, we benchmarked CAKES on the Radio-ML dataset~\cite{oshea2018radioml} under Dynamic Time Warping~\cite{muller2007dynamic} as the distance function.
This dataset contains samples of synthetically generated signal captures of different modulation modes over a range of signal-to-noise ratio (SNR) levels.
Specifically, it comprises 24 modulation modes at 26 different SNRs ranging from -20 dB to 30 dB, with 4,096 samples at each combination of modulation mode and SNR level.
Thus, it contains $24 \cdot 26 \cdot 4096 = 2,555,504$ samples in total.
Each sample is a 1,024-dimensional complex-valued vector, representing a signal capture (a time-series of complex-valued numbers).
We used a subset of this dataset, containing $97,304$ samples at 10dB SNR and used another $1,000$ samples at the same SNR as a hold-out set of queries.

\subsection{Other Algorithms}
\label{sec:datasets-and-benchmarks:other-algorithms}

We benchmarked the three CAKES algorithms against a na\"ive linear search implementation in Rust.
We also benchmarked against three state-of-the-art similarity search algorithms: HNSW, ANNOY, and FAISS-IVF in Python.
We verified that our implementation of linear search produces the same neighbors as provided by the ANN-Benchmarks suite for Fashion-MNIST, Glove-25 and Sift datasets.
We then used this linear search implementation to find and store the ground-truth for the augmented versions of the datasets.
We used this ground-truth to calculate recall for CAKES's algorithms in Rust and for HNSW, ANNOY, and FAISS-IVF in Python.
We plot the results of these benchmarks in Figure~\ref{fig:results:scaling-plots}.

%% file: sections/4-results.tex
\section{Results}
\label{sec:results}

The algorithms presented in this paper rely heavily on the local fractal dimension (LFD) being much smaller than than the embedding dimension of the dataset.
As such, we begin by examining our assumptions about the LFD of real-world datasets (see Section~\ref{sec:results:lfd-of-datasets} and Figure~\ref{fig:results:lfd-plots}).

We continue by comparing the performance of the CAKES algorithms against HNSW, ANNOY and FAISS-IVF on the Fashion-MNIST, Glove-25, Sift and Random datasets.
We also compare how this performance scales to the augmented versions of these datasets.
We also report the performance of CAKES on the Silva and RadioML datasets (see Section~\ref{sec:results:scaling-behavior-and-recall}, Figure~\ref{fig:results:scaling-plots}, and Tables~\ref{tab:results:qps-and-recall-fmn},~\ref{tab:results:qps-and-recall-glove},~\ref{tab:results:qps-and-recall-sift} and~\ref{tab:results:qps-and-recall-random}).

Finally, we test our intuitions about unbalanced clustering being better for search
than balanced clustering (see Section~\ref{sec:results:clustering-strategies-and-number-of-distance-computations} and Figure~\ref{fig:results:distance-counts}).

\subsection{Local Fractal Dimension of Datasets}
\label{sec:results:lfd-of-datasets}

\begin{figure}
    \captionsetup[subfigure]{aboveskip=-15pt,belowskip=-3pt}
    \begin{subfigure}[b]{0.5\textwidth}
        \includegraphics[width=0.99\textwidth]{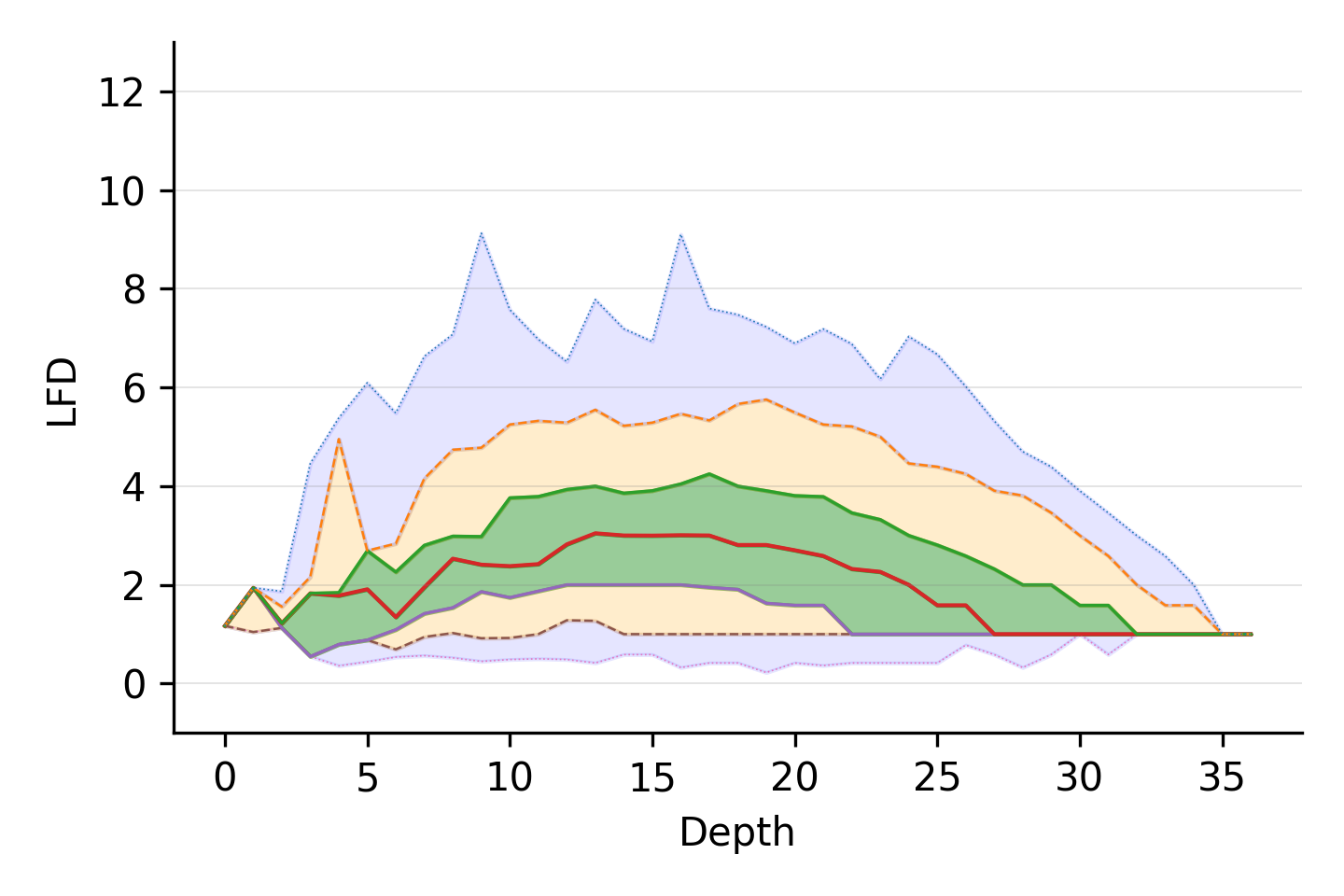}\\
        \subcaption{Fashion-MNIST}
        \label{fig:results:fashion-mnist-lfd}
    \end{subfigure}%
    \begin{subfigure}[b]{0.5\textwidth}
        \includegraphics[width=0.99\textwidth]{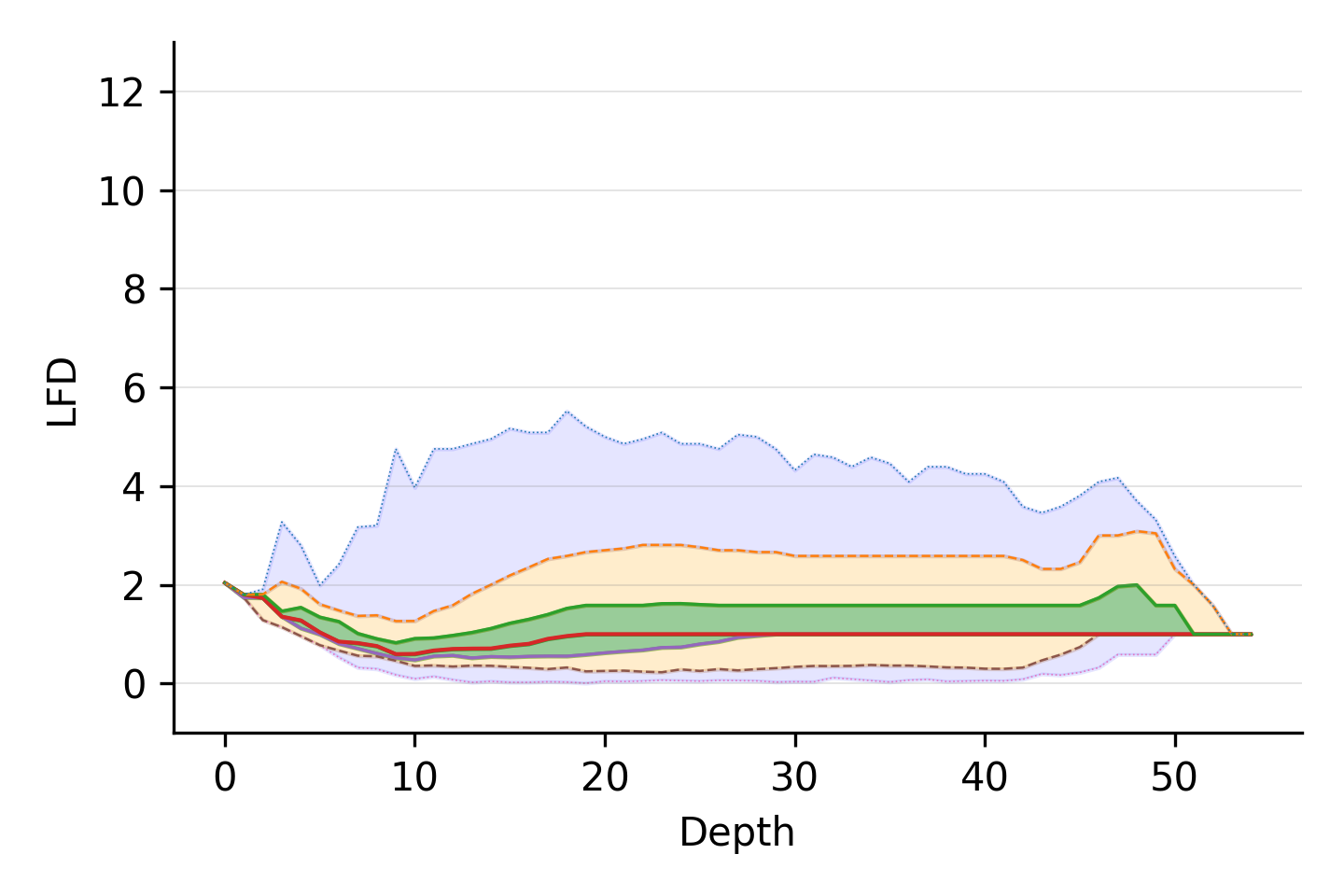}\\
        \subcaption{Glove-25}
        \label{fig:results:glove-25-lfd}
    \end{subfigure}
    \\
    \begin{subfigure}[b]{0.5\textwidth}
        \includegraphics[width=0.99\textwidth]{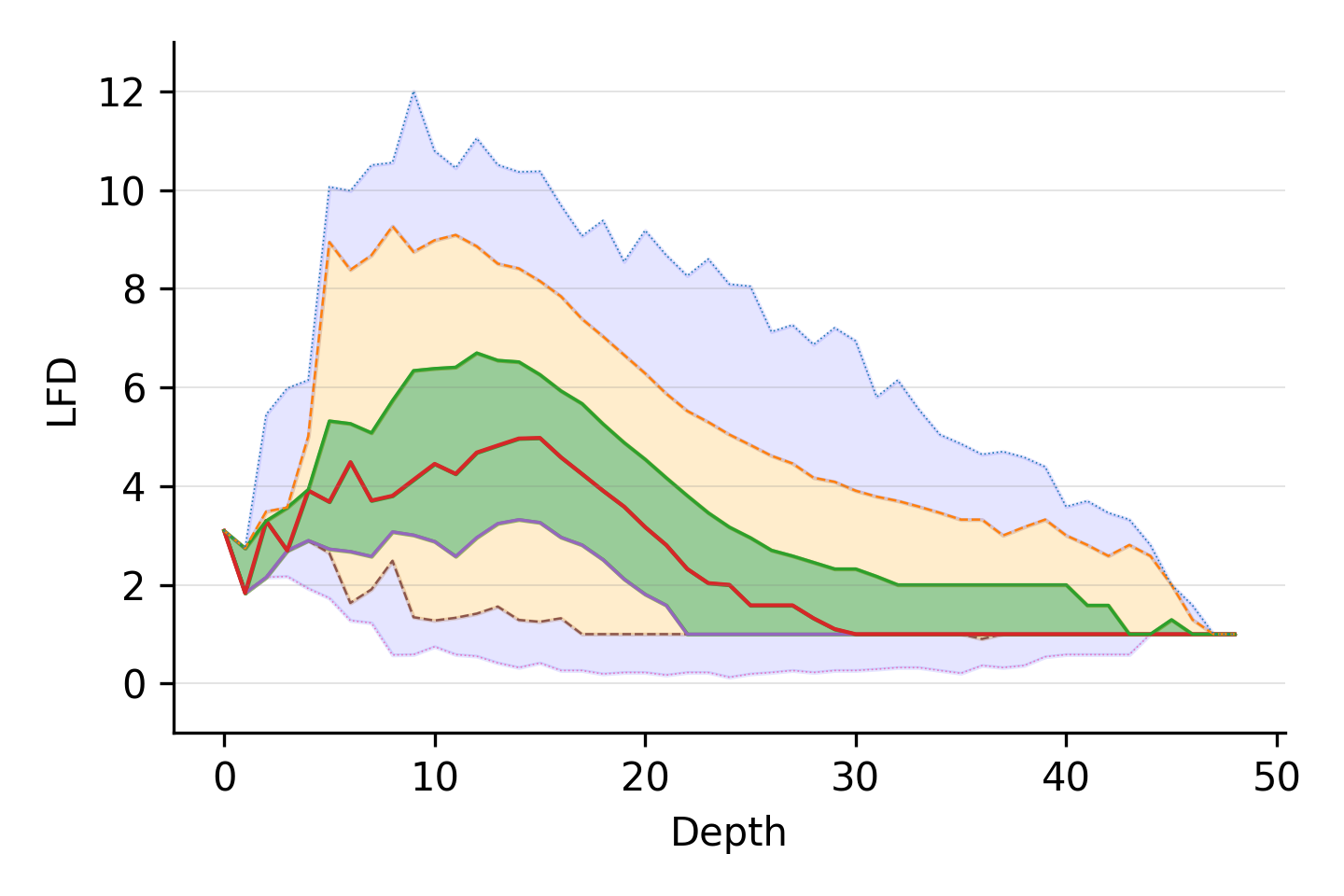}\\
        \subcaption{Sift}
        \label{fig:results:sift-lfd}
    \end{subfigure}%
    \begin{subfigure}[b]{0.5\textwidth}
        \includegraphics[width=0.99\textwidth]{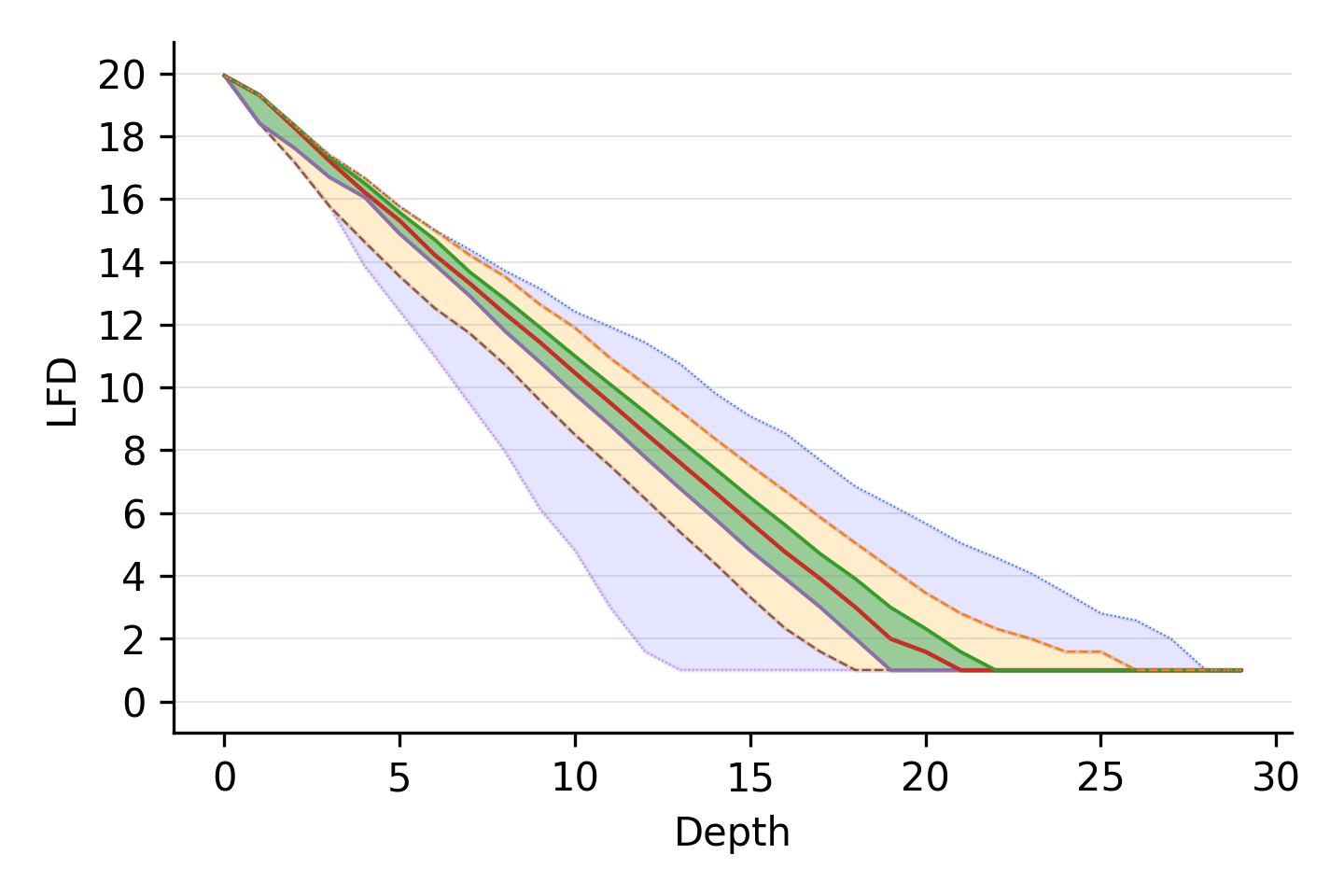}\\
        \subcaption{A random dataset}
        \label{fig:results:random-lfd}
    \end{subfigure}
    \\
    \begin{subfigure}[b]{0.5\textwidth}
        \includegraphics[width=0.99\textwidth]{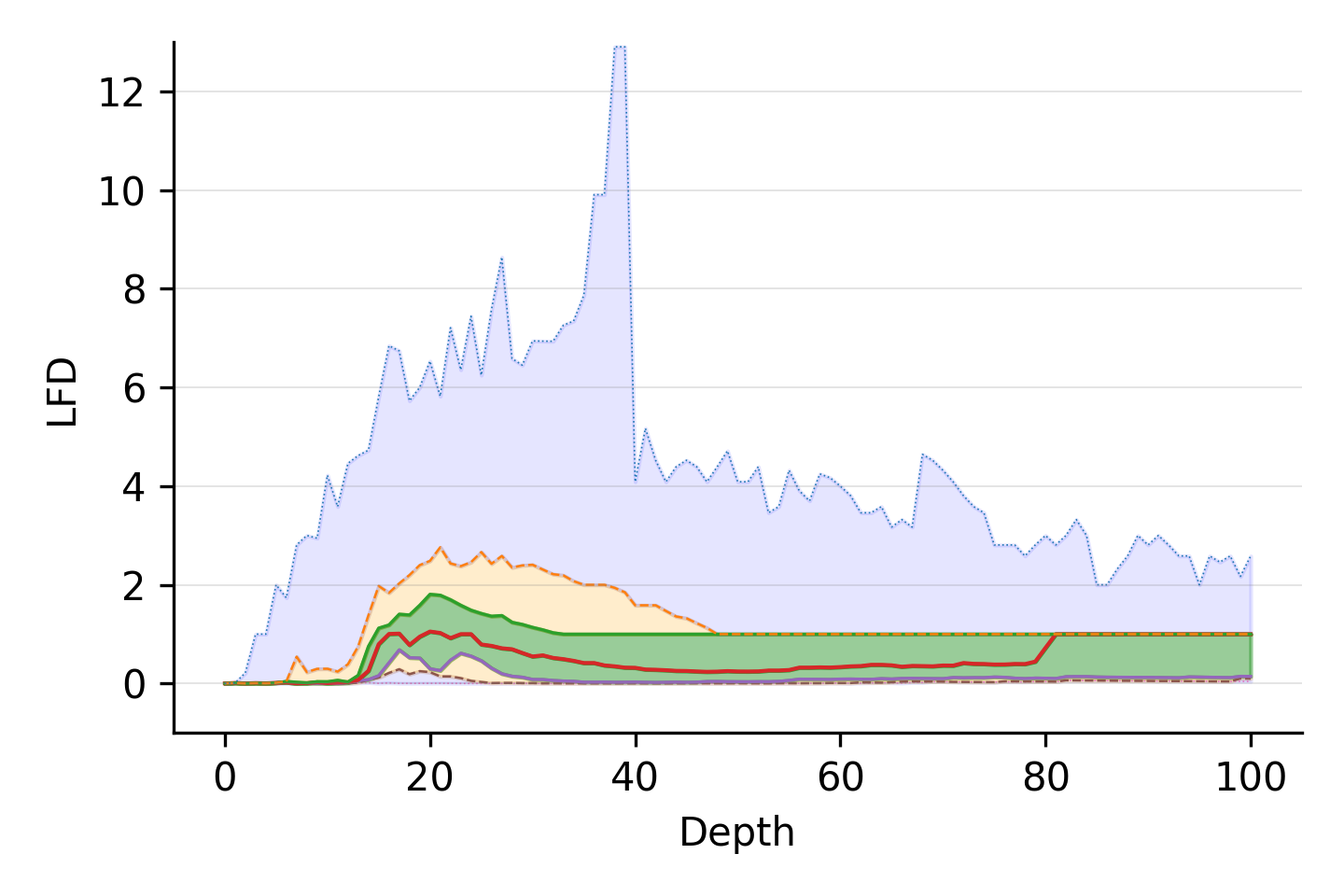}\\
        \subcaption{Silva 18S}
        \label{fig:results:silva-lfd}
    \end{subfigure}%
    \begin{subfigure}[b]{0.5\textwidth}
        \includegraphics[width=0.99\textwidth]{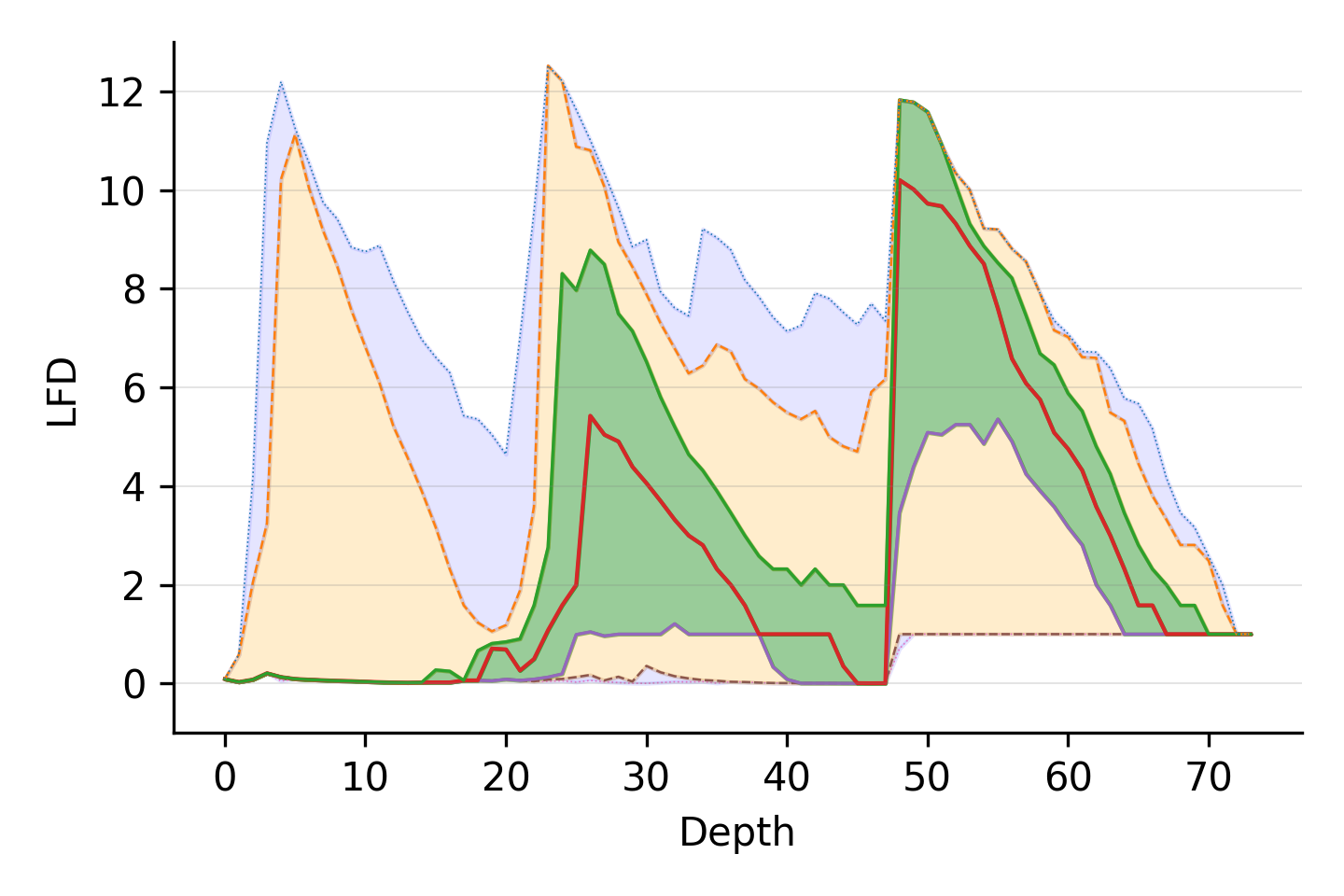}\\
        \subcaption{RadioML}
        \label{fig:results:radioml-lfd}
    \end{subfigure}%
    \\
    \vskip -0.1in
    \begin{subfigure}[b]{0.94\textwidth}
        \centering
        \includegraphics[width=0.7\textwidth]{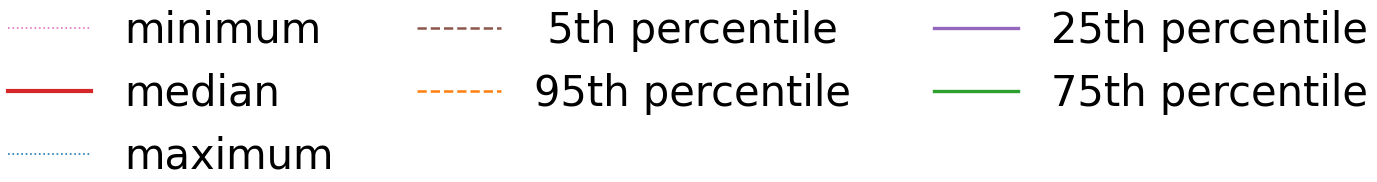}
        \label{fig:results:lfd-legend}
    \end{subfigure}%
    \vskip -0.1in
    \caption{Local fractal dimension vs. cluster depth across six datasets. The `random' dataset is randomly generated according to the procedure in Section~\ref{sec:datasets-and-benchmarks:random-datasets}; note that the y-axis is different for this dataset. In each plot, the horizontal axis denotes depth in the cluster tree, and the vertical axis denotes the LFD of clusters at that depth. We show lines for the 5$^{th}$, 25$^{th}$, 50th, 75$^{th}$ and 95$^{th}$ percentiles of LFD, as well as the minimum and maximum LFD at each depth. So that plots best reflect the distribution of LFDs across the entire \textit{dataset}, we count each cluster as many times as its cardinality. For example, if, for some dataset, the 95$^{th}$ percentile of LFD at depth 40 is 3, this means that 95\% of the points in clusters at depth 40 belong to a cluster whose LFD is at most 3.}
    \label{fig:results:lfd-plots}
    \vskip -0.4in
\end{figure}

Since the time complexity of CAKES algorithms scales with the LFD of the dataset, we examine the LFD of each dataset we used for benchmarks.
Figure~\ref{fig:results:lfd-plots} illustrates the trends in LFD for Fashion-MNIST, Glove-25, Sift, Random, Silva 18S, and Radio-ML.
In this section, when we discuss trends in LFD, unless otherwise noted, we are referring to the 95$^{th}$ percentile of LFD.

The Fashion-MNIST dataset has an embedding dimension of 784 and uses the Euclidean distance metric.
In Figure~\ref{fig:results:fashion-mnist-lfd} we observe that until approximately depth 5, Fashion-MNIST's LFD is low (i.e., less than 4).
It then starts increasing, reaching a peak of about 6 near depth 20, before decreasing to 1 at the maximum depth.

The Glove-25 dataset has an embedding dimension of 25 and uses the cosine distance function which, notably, is not a metric.
Relative to Fashion-MNIST, Glove-25 has low LFD, as shown in Figure~\ref{fig:results:glove-25-lfd}.
All percentile lines for Glove-25 are flatter and lower, indicating that the LFD is lower across the entire dataset, and that the LFD does not vary as much by depth.
In particular, Glove-25's LFD is less than 3 for all depths.

The Sift dataset has an embedding dimension of 128 and uses the Euclidean distance metric.
Figure~\ref{fig:results:sift-lfd} shows the LFD by depth for Sift, which has higher LFD relative to Fashion-MNIST and Glove-25.
It increases sharply to a peak of 9 around a depth of 10.
It then decreases smoothly until reaching the deepest leaves in the tree.

We generated the Random dataset to have the same cardinality and dimensionality as Sift.
We used a uniform distribution in a 128-dimensional unit-hypercube to generate the points and the Euclidean metric to measure distances among them.
Figure~\ref{fig:results:random-lfd} shows that the character of this dataset is significantly different from the others.
The LFD starts at 20 at depth 0 and all percentile lines decrease linearly with depth until reaching the leaves of the tree.
The spread in LFD starts very small for the first few clusters and increases as depth increases.
The LFD of approximately 20 for the root cluster $\mathcal{R}$ is what we expect for this random dataset.
To elaborate, the distribution of points in such a dataset should reflect the curse of dimensionality, i.e.,\,the fact that in high dimensional spaces, the minimum and maximum pairwise distances between any two points are approximately equal.
As a result, $\mathcal{R}$'s radius $r$, which reflects the maximum distance between the center $c$ and any other point, should not differ significantly from the distance between the center and its closest point.
A consequence of this is that, with high probability, for every point in $\mathcal{R}$, its distance from $c$ is greater than $\tfrac{r}{2}$;
in other words, $B(c, \tfrac{r}{2})$ contains only $c$ while $B(c, r)$ contains the entire dataset.
Given our definition of LFD in Equation~\ref{eq:methods:lfd-half}, this means that the LFD of $\mathcal{R}$ is approximately $\log_2(\frac{|X|}{1}) = \log_2(1,000,000) \approx 20$, which is what we observe in Figure~\ref{fig:results:random-lfd}.
Theoretically, the LFD of this dataset should be 128, i.e.\, it should be the same as the embedding dimension.
This reflects the difference between how we empirically measure the LFD and the value we would expect.
With sample sizes larger than 1,000,000, we would expect the LFD to approach 128 until we have sampled $2^{128}$ points, at which point the LFD would be 128 and would stay at 128 for even larger sample sizes.
Unfortunately, such a large sample is practically impossible to generate.

The Silva-18S dataset consists of genomic sequences whose unaligned lengths are at-most 3,712.
As such the embedding dimension is 3,712, though as discussed previously, the embedding dimension would be 50,000 in a multiple sequence alignment.
We use the Levenshtein edit distance (a metric) to measure distances between sequences.
This dataset, as shown in Figure~\ref{fig:results:silva-lfd}, exhibits consistently low LFD.
In particular, LFD is less than 3 for all depths, hovering near 1 for clusters at depth 40 and deeper.

The Radio-ML dataset consists of measurements of radio-frequency signals using 1,024 dimensional complex-valued vectors.
We use the Dynamic Time Warping distance metric on this dataset.
This dataset is synthetic~\cite{oshea2018radioml} but uses a far more elaborate generation process than our Random dataset.
The LFD values show three distinct peaks around an LFD of 12 at or near depths of 8, 25 and 50.
Each peak is followed by a linear decrease until encountering a sharp spike for the next peak.
Within each of the tree portions, this dataset has a character very similar to that of the Random dataset.
This suggests that the dataset obeys the manifold hypothesis at some scales, but that it is not ``scale free,'' as the LFD varies significantly by depth.
This is likely the result of a piecewise uniform sampling strategy used to generate the different modulation modes present in the dataset.

\subsection{Indexing and Tuning Time}
\label{sec:results:indexing-and-tuning-time}

For each of the ANN-benchmark datasets and the Random dataset, we report the time taken for each algorithm to build the index and to tune the hyper-parameters for these indices to achieve the highest possible recall. For the sake of brevity, these results are reported in the supplement.

\subsection{Scaling Behavior and Recall}
\label{sec:results:scaling-behavior-and-recall}

We benchmark the CAKES algorithms, na\"{i}ve linear search, HNSW, ANNOY, and FAISS-IVF on the Fashion-MNIST, Glove-25, Sift, and Random datasets.
We augment these datasets with synthetic points to examine how performance scales with cardinality.
We also benchmark the CAKES algorithms on the Silva and RadioML datasets, and we subsample these datasets instead of augmenting them to examine how performance scales with cardinality.
For each dataset, we use the corresponding distance function (see Table~\ref{tab:datasets:summary}).
For HNSW, ANNOY, and FAISS-IVF, we allow for a hyper-parameter search to tune their index for maximum recall.
For CAKES, we build the tree and use our auto-tuning approach (see Section~\ref{sec:methods:auto-tuning}) to select the fastest algorithm for each dataset and cardinality.
We show results for $k=10$ in the main text, but results for $k=100$ can be found in the Supplement.
Figure~\ref{fig:results:scaling-plots} shows the scaling behavior of these algorithms on these datasets while Tables~\ref{tab:results:qps-and-recall-fmn},~\ref{tab:results:qps-and-recall-glove},~\ref{tab:results:qps-and-recall-sift} and~\ref{tab:results:qps-and-recall-random} report the throughput and recall of CAKES's algorithms at each augmented cardinality for Fashion-MNIST, Glove-25, Sift, and Random respectively.

On the Fashion-MNIST, Glove-25, and Sift datasets (in Figures~\ref{fig:results:fashion-mnist-scaling},~\ref{fig:results:glove-25-scaling},~and~\ref{fig:results:sift-scaling} respectively), we observe that as cardinality increases, the Depth-First Sieve algorithm is consistently the fastest CAKES algorithm with a throughput that is constant in the cardinality of the dataset.
The Breadth-First Sieve algorithm is the usually the second fastest, also with a nearly constant throughput across all cardinalities.
The Repeated $\rho$-NN algorithm, however, falls off in throughput as cardinality increases.
All three CAKES algorithms exhibit perfect recall on the Fashion-MNIST and Sift datasets, and near-perfect recall on the Glove-25 dataset (which uses cosine distance).
In contrast, HNSW and ANNOY are faster than CAKES's algorithms for all cardinalities and have near constant throughput as cardinality increases, but their recall degrades quickly as cardinality increases.
FAISS-IVF exhibits linearly decreasing throughput as cardinality increases, which is expected from the algorithm given that we tune the hyper-parameters to maximize recall.

On the Random dataset, HNSW and ANNOY are still the fastest algorithms but exhibit recall values near 0.
CAKES's algorithms show linearly decreasing throughput as cardinality increases and are also slower than na\"{i}ve linear search.

Tables~\ref{tab:results:qps-and-recall-fmn},~\ref{tab:results:qps-and-recall-glove},~\ref{tab:results:qps-and-recall-sift} and~\ref{tab:results:qps-and-recall-random} show the throughput and recall of CAKES's algorithms at each augmented cardinality for Fashion-MNIST, Glove-25, Sift, and Random respectively.
Though the plots in Figure~\ref{fig:results:scaling-plots} present results for each of CAKES's three algorithms separately, the results in the CAKES column in these tables represent the fastest CAKES algorithm at that dataset and cardinality only.
We used our auto-tuning approach (see Section~\ref{sec:methods:auto-tuning}) for each new tree (i.e.,\,at each cardinality), and this approach was always able to select the fastest algorithm for each dataset at each cardinality.
Since we also allow for tuning hyper-parameters for the other algorithms, and we allow for different sets of hyper-parameters at each cardinality, it is a fair comparison for these tables to only list the performance of the fastest CAKES algorithm.
When reporting recall, we use $1.000*$ to denote that the recall is imperfect, but rounds to $1.000$.

For Silva and RadioML, we benchmarked only CAKES's algorithms because HNSW, ANNOY and FAISS support neither the required distance functions nor, in the case of RadioML, complex-valued data.
Due to the massive sizes of these datasets and challenges in generating plausible augmentations, we took random sub-samples ranging up to the entirety of the dataset---rather than augmented versions of the dataset---to examine how performance scales with cardinality.
For the Silva dataset (see Figure~\ref{fig:results:silva-scaling}), we observe that throughput initially decreases linearly as cardinality increases, but levels off at higher cardinalities.
Depth-First Sieve is, once again, the fastest CAKES algorithm.
For Radio-ML (see Figure~\ref{fig:results:radioml-scaling}), however, we see that throughput declines nearly linearly with cardinality, and that the three CAKES algorithms are, again, slower that na\"{i}ve linear search.

\begin{figure}
    \captionsetup[subfigure]{aboveskip=-5pt,belowskip=-0.5pt}
    \begin{subfigure}[b]{0.5\textwidth}
        \includegraphics[width=0.95\textwidth]{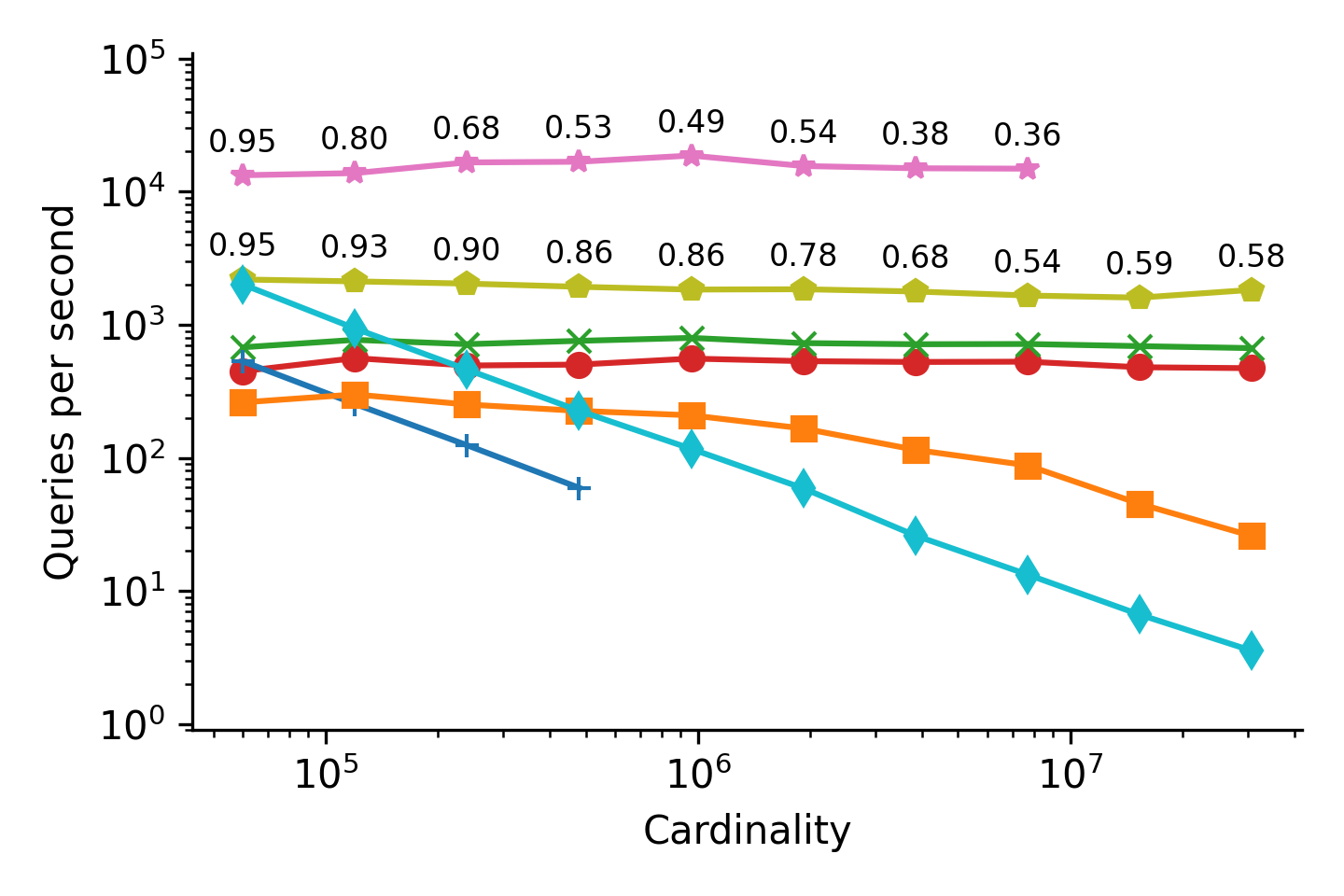}
        \subcaption{Fashion-MNIST for $k=10$.}
        \label{fig:results:fashion-mnist-scaling}
    \end{subfigure}%
    \begin{subfigure}[b]{0.5\textwidth}
        \includegraphics[width=0.95\textwidth]{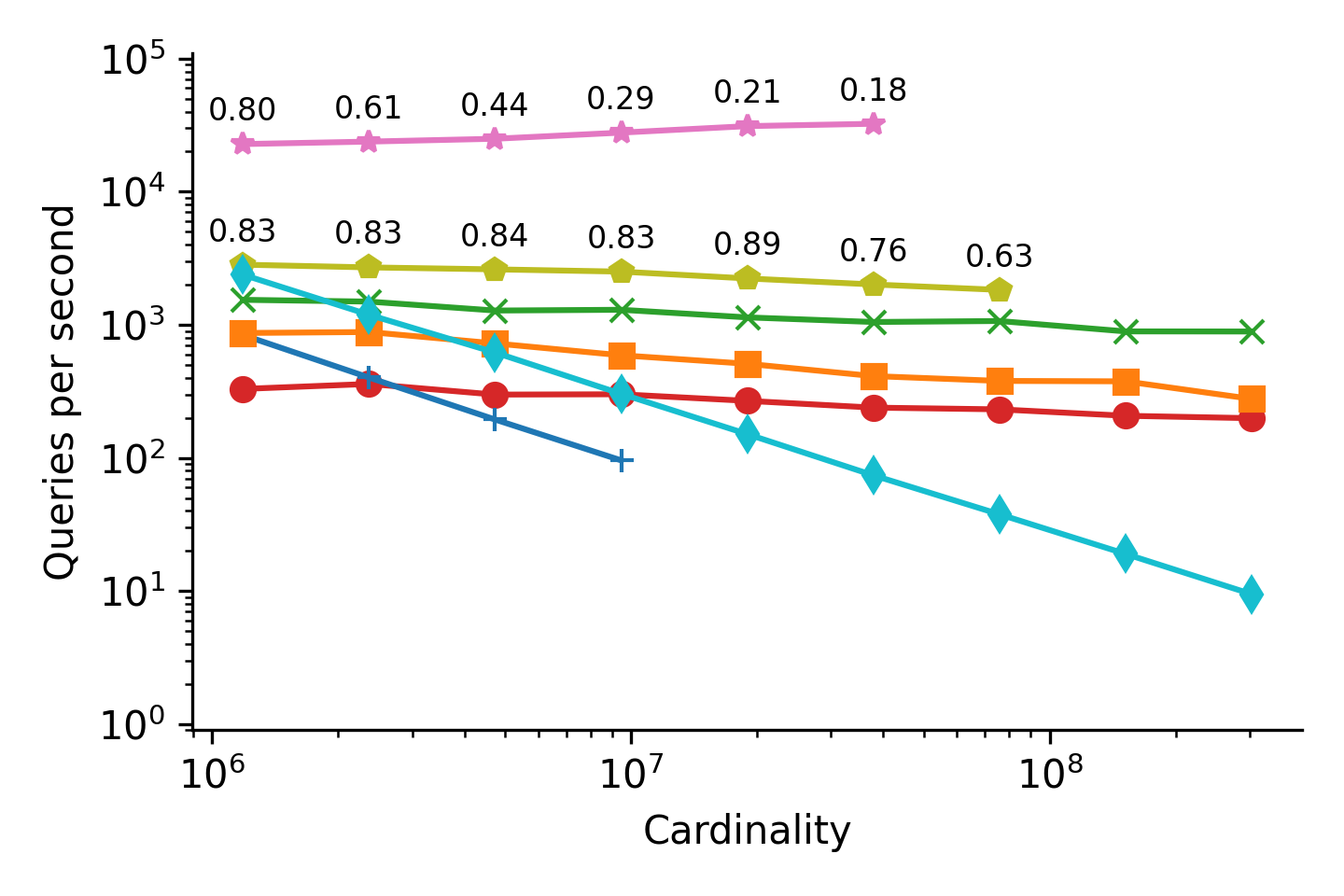}
        \subcaption{Glove-25 for $k=10$.}
        \label{fig:results:glove-25-scaling}
    \end{subfigure}%
    \\
    \begin{subfigure}[b]{0.5\textwidth}
        \includegraphics[width=0.95\textwidth]{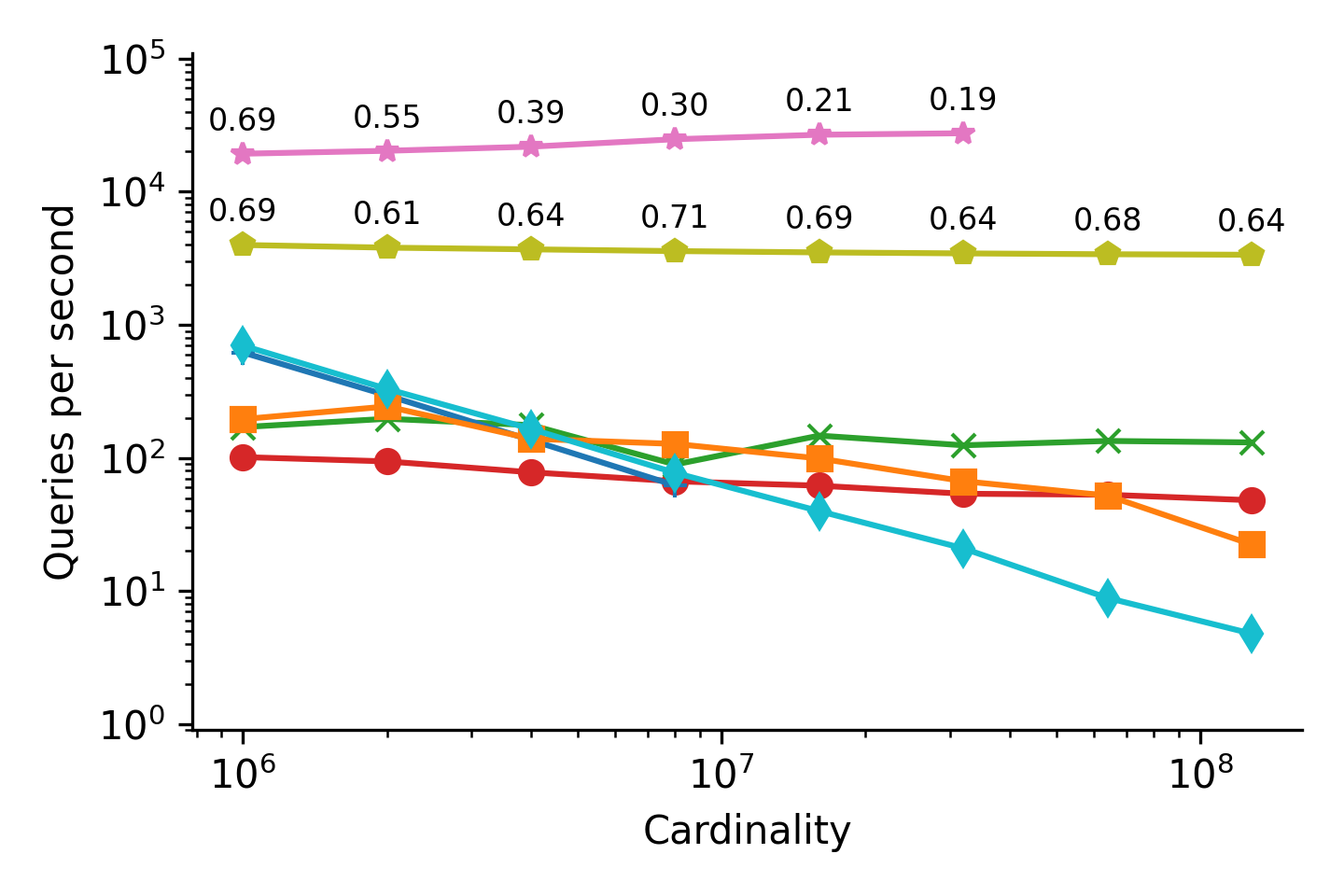}
        \subcaption{Sift for $k=10$.}
        \label{fig:results:sift-scaling}
    \end{subfigure}%
    \begin{subfigure}[b]{0.5\textwidth}
        \includegraphics[width=0.95\textwidth]{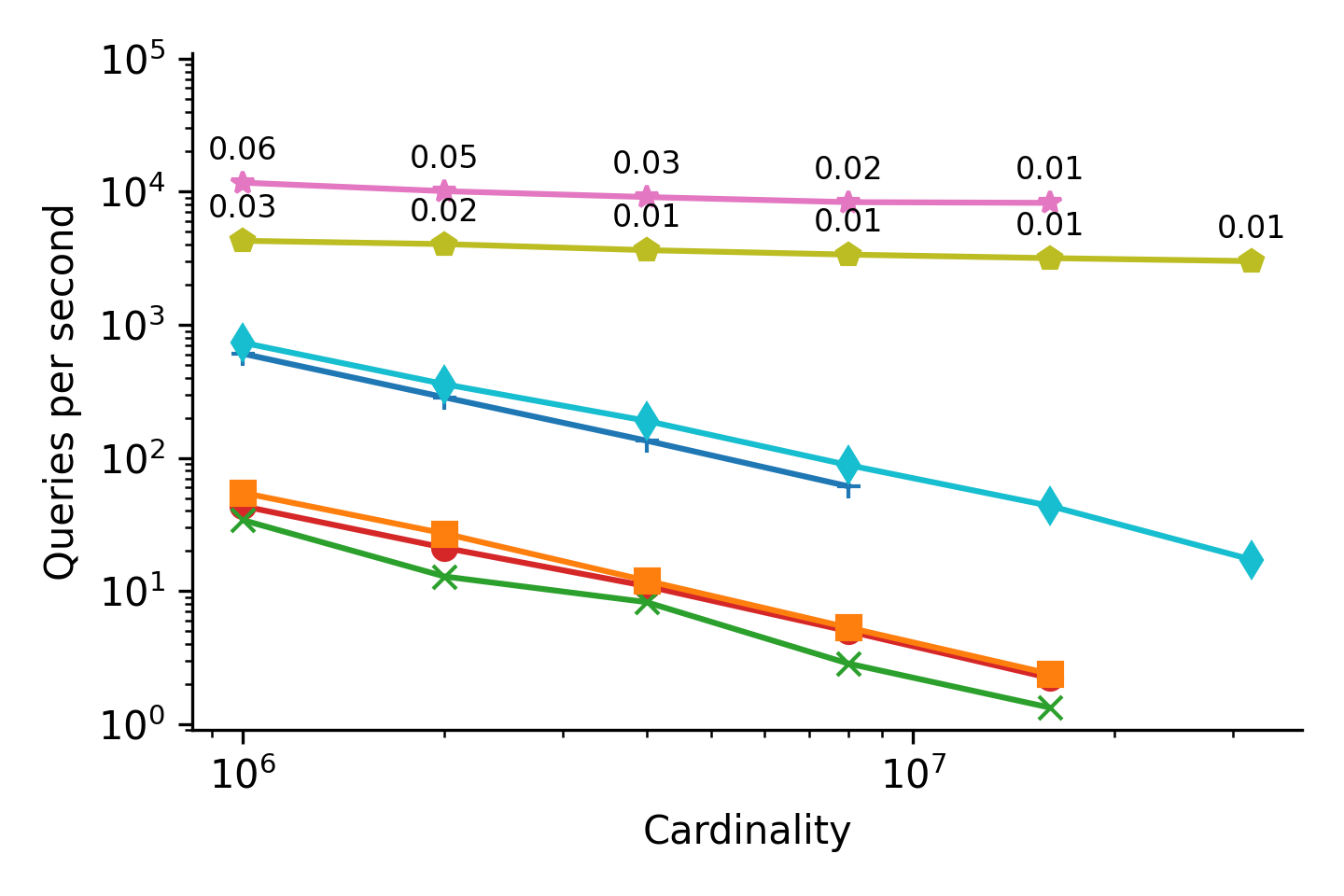}
        \subcaption{A random dataset for $k=10$.}
        \label{fig:results:random-scaling}
    \end{subfigure}%
    \\
    \begin{subfigure}[b]{0.5\textwidth}
        \includegraphics[width=0.95\textwidth]{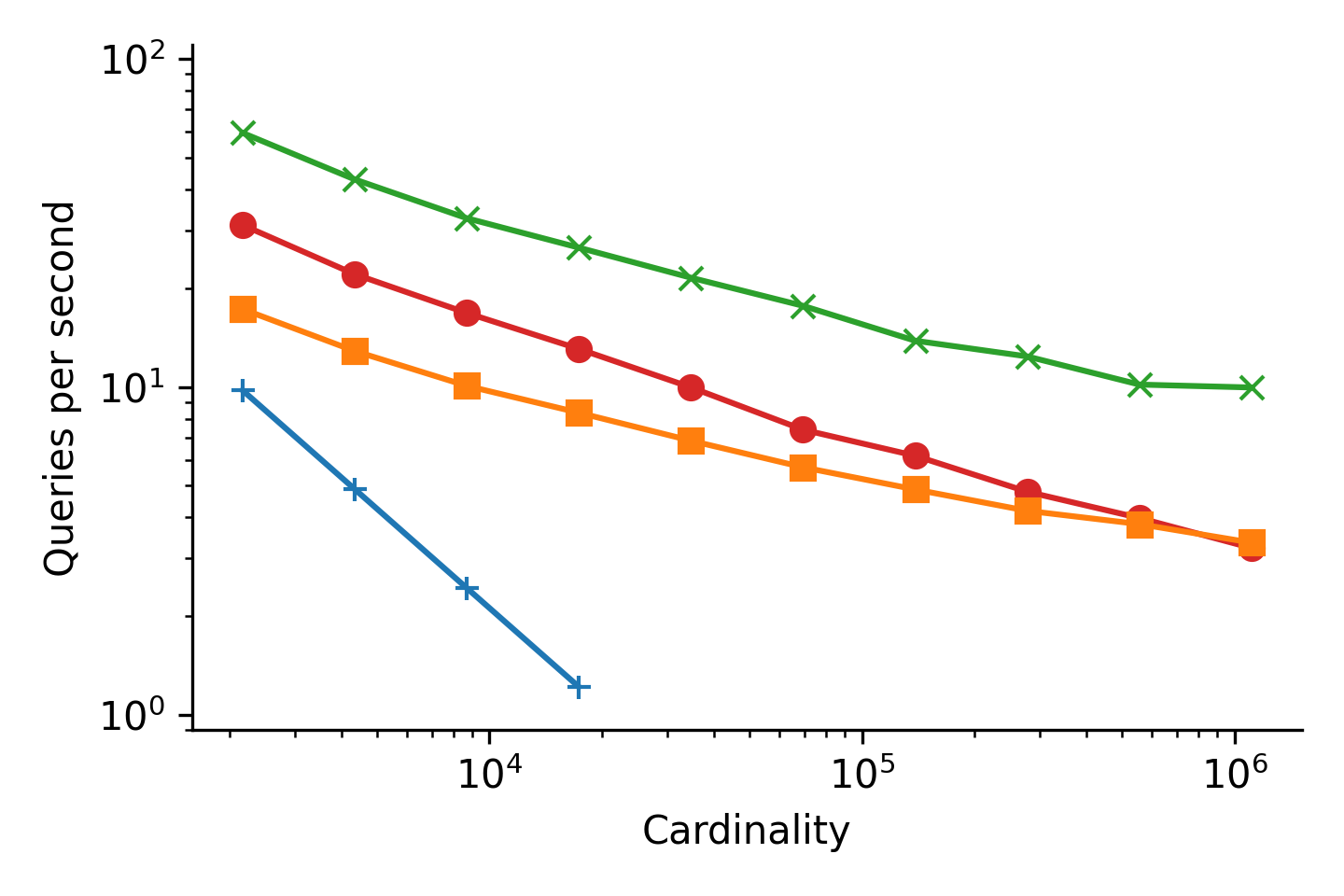}
        \subcaption{Silva for $k=10$.}
        \label{fig:results:silva-scaling}
    \end{subfigure}%
    \begin{subfigure}[b]{0.5\textwidth}
        \includegraphics[width=0.95\textwidth]{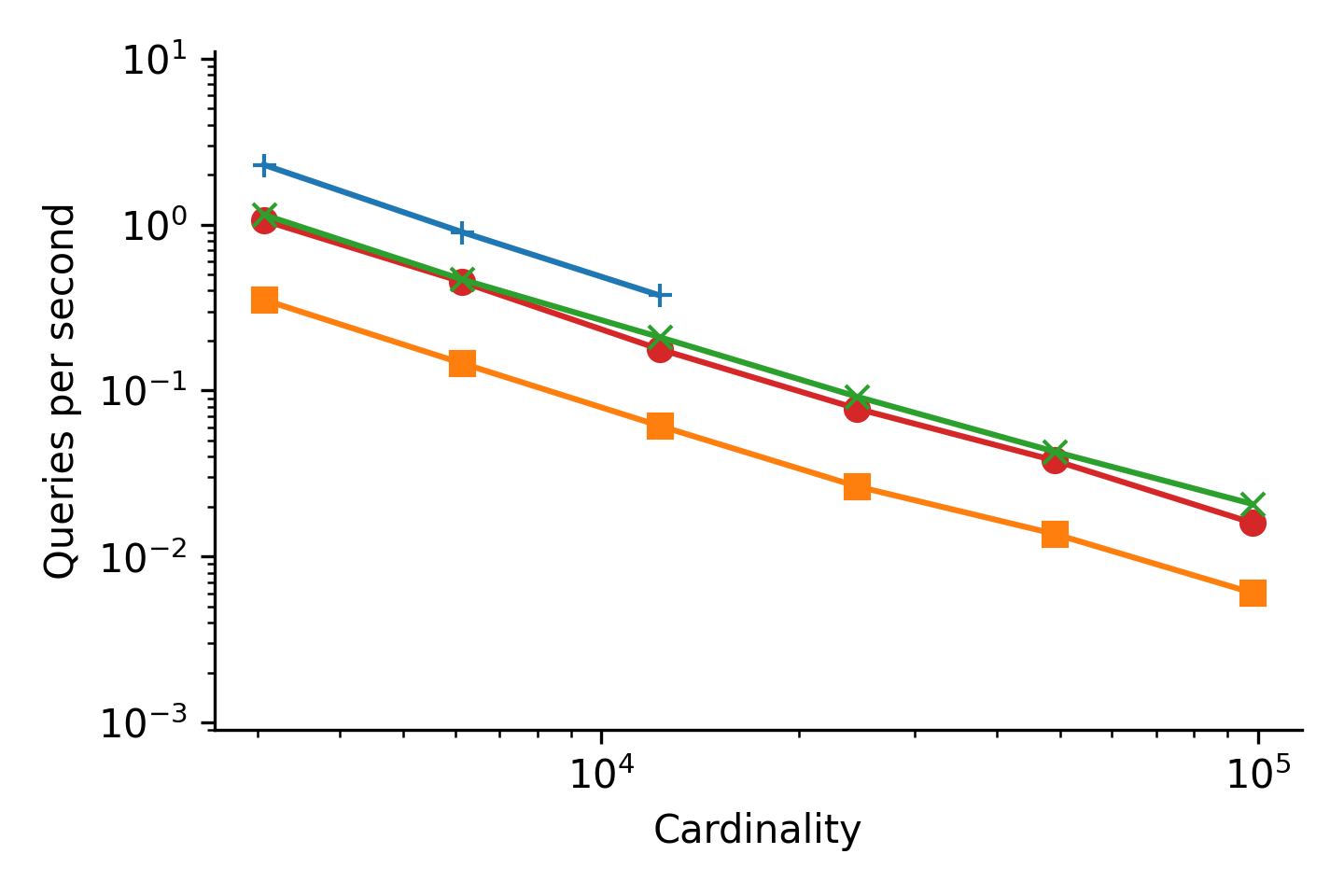}
        \subcaption{RadioML for $k=10$ at SnR = 10dB.}
        \label{fig:results:radioml-scaling}
    \end{subfigure}%
    \\
    \begin{subfigure}[b]{0.94\textwidth}
        \centering
        \includegraphics[width=0.7\textwidth]{plots/legend.png}
        \label{fig:results:scaling-legend}
    \end{subfigure}%
    \caption{Throughput across six datasets, including a randomly-generated dataset.
    In each plot, the horizontal axis represents increasing cardinality of the dataset, while the vertical axis represents the throughput in queries per second (higher is better).
    For Fashion-MNIST, Glove-25, and Sift, as cardinality increases, the CAKES algorithms become faster than linear search but the cardinality at which this occurs differs by dataset.
    For Fashion-MNIST and Glove-25, Depth-First Sieve is consistently fastest, while for Sift, Repeated $\rho$-NN is the fastest for smaller cardinalities and Depth-First Sieve is the fastest for larger cardinalities.
    With Silva, we observe that for all algorithms, throughput initially seems to linearly decrease as cardinality increases, but that it starts to level off at higher cardinalities.
    Depth-First Sieve is consistently the fastest algorithm on the Silva dataset.
    For Radio-ML and Random, we see that all three CAKES algorithms are slower than na\"{i}ve linear search, and that their throughput decreases linearly with cardinality.
    HNSW and ANNOY are the fastest algorithms on all four datasets we benchmarked them on, but their recall degrades quickly as cardinality increases on all datasets.
    In particular, on the Random dataset, HNSW and ANNOY have near-zero recall.}
    \label{fig:results:scaling-plots}
\end{figure}

\begin{table}
    \caption{Throughput (QPS) and Recall on the Fashion-MNIST dataset.
    A recall value of $1.000*$ denotes imperfect recall that rounds to 1.000. While recall for CAKES's does \emph{not} degrade with cardinality, recall for HNSW and ANNOY degrades with cardinality.
    In contrast, CAKES has perfect recall on Fashion-MNIST, since it used with Euclidean distance, which is a metric.}
    \label{tab:results:qps-and-recall-fmn}
    \begin{tabular}{|l|p{1.55cm}|p{1.1cm}|p{1.55cm}|p{1.1cm}|p{1.55cm}|p{1.1cm}|p{1.55cm}|p{1.1cm}|}
        \hline
        \multirow{2}{*}{\textbf{Mult.}} & \multicolumn{2}{c|}{\textbf{HNSW}} & \multicolumn{2}{c|}{\textbf{ANNOY}} & \multicolumn{2}{c|}{\textbf{FAISS-IVF}}  & \multicolumn{2}{c|}{\textbf{CAKES}} \\\cline{2-9}
        & QPS & Recall & QPS & Recall & QPS & Recall & QPS & Recall \\
        \cline{1-9}
        \hline
        1   & \num{1.33e4} & 0.954  & \num{2.19e3} & 0.950  & \num{2.01e3} & 1.000* & \num{3.46e3} & 1.000  \\\cline{1-9}
        2   & \num{1.38e4} & 0.803  & \num{2.12e3} & 0.927  & \num{9.39e2} & 1.000* & \num{3.68e3} & 1.000  \\\cline{1-9}
        4   & \num{1.66e4} & 0.681  & \num{2.04e3} & 0.898  & \num{4.61e2} & 0.997  & \num{3.44e3} & 1.000  \\\cline{1-9}
        8   & \num{1.68e4} & 0.525  & \num{1.93e3} & 0.857  & \num{2.26e2} & 0.995  & \num{3.30e3} & 1.000  \\\cline{1-9}
        16  & \num{1.87e4} & 0.494  & \num{1.84e3} & 0.862  & \num{1.17e2} & 0.991  & \num{3.34e3} & 1.000  \\\cline{1-9}
        32  & \num{1.56e4} & 0.542  & \num{1.85e3} & 0.775  & \num{5.91e1} & 0.985  & \num{2.96e3} & 1.000  \\\cline{1-9}
        64  & \num{1.50e4} & 0.378  & \num{1.78e3} & 0.677  & \num{2.61e1} & 0.968  & \num{3.25e3} & 1.000  \\\cline{1-9}
        128 & \num{1.49e4} & 0.357  & \num{1.66e3} & 0.538  & \num{1.33e1} & 0.964  & \num{2.96e3} & 1.000  \\\cline{1-9}
        256 & --           & --     & \num{1.60e3} & 0.592  & \num{6.65e0} & 0.962  & \num{2.79e3} & 1.000  \\\cline{1-9}
        512 & --           & --     & \num{1.83e3} & 0.581  & \num{3.56e0} & 0.949 & \num{2.84e3} & 1.000 \\
        \hline
    \end{tabular}
    \vskip -0.2in
\end{table}

\begin{table}
    \caption{Throughput (QPS) and Recall on the Glove-25 dataset.
    A recall value of $1.000*$ denotes imperfect recall that rounds to 1.000. While recall for CAKES's does \emph{not} degrade with cardinality, recall for HNSW and ANNOY degrades with cardinality.
    In contrast, CAKES has near-perfect recall on Glove-25, since it used with cosine distance, which is not a metric.}
    \label{tab:results:qps-and-recall-glove}
    \begin{tabular}{|l|p{1.55cm}|p{1.1cm}|p{1.55cm}|p{1.1cm}|p{1.55cm}|p{1.1cm}|p{1.55cm}|p{1.1cm}|}
        \hline
        \multirow{2}{*}{\textbf{Mult.}} & \multicolumn{2}{c|}{\textbf{HNSW}} & \multicolumn{2}{c|}{\textbf{ANNOY}} & \multicolumn{2}{c|}{\textbf{FAISS-IVF}}  & \multicolumn{2}{c|}{\textbf{CAKES}} \\\cline{2-9}
        & QPS & Recall & QPS & Recall & QPS & Recall & QPS & Recall \\
        \cline{1-9}
        \hline
        1   & \num{2.28e4} & 0.801 & \num{2.83e3} & 0.835 & \num{2.38e3} & 1.000* & \num{1.54e3} & 1.000* \\\cline{1-9}
        2   & \num{2.38e4} & 0.607 & \num{2.70e3} & 0.832 & \num{1.19e3} & 1.000* & \num{1.49e3} & 1.000* \\\cline{1-9}
        4   & \num{2.50e4} & 0.443 & \num{2.61e3} & 0.839 & \num{6.19e2} & 1.000* & \num{1.28e3} & 1.000* \\\cline{1-9}
        8   & \num{2.78e4} & 0.294 & \num{2.51e3} & 0.834 & \num{3.03e2} & 1.000* & \num{1.30e3} & 1.000* \\\cline{1-9}
        16  & \num{3.11e4} & 0.213 & \num{2.23e3} & 0.885 & \num{1.51e2} & 1.000* & \num{1.14e3} & 1.000* \\\cline{1-9}
        32  & \num{3.24e4} & 0.178 & \num{2.01e3} & 0.764 & \num{7.40e1} & 0.999  & \num{1.05e3} & 1.000* \\\cline{1-9}
        64  & --           & --    & \num{1.99e3} & 0.631 & \num{3.77e1} & 0.997  & \num{1.07e3} & 1.000* \\\cline{1-9}
        128 & --           & --    & --           & --    & \num{1.90e1} & 0.998  & \num{8.92e2} & 1.000* \\\cline{1-9}
        256 & --           & --    & --           & --    & \num{9.47e0} & 0.998  & \num{8.91e2} & 1.000* \\
        \hline
    \end{tabular}
    \vskip -0.2in
\end{table}

\begin{table}
    \caption{Throughput (QPS) and Recall on the Sift dataset.
    A recall value of $1.000*$ denotes imperfect recall that rounds to 1.000. While recall for CAKES's does \emph{not} degrade with cardinality, recall for HNSW and ANNOY degrades with cardinality.
    In contrast, CAKES has perfect recall on Sift, since it is used with Euclidean distance, which is a metric.}
    \label{tab:results:qps-and-recall-sift}
    \begin{tabular}{|l|p{1.55cm}|p{1.1cm}|p{1.55cm}|p{1.1cm}|p{1.55cm}|p{1.1cm}|p{1.55cm}|p{1.1cm}|}
        \hline
        \multirow{2}{*}{\textbf{Mult.}} & \multicolumn{2}{c|}{\textbf{HNSW}} & \multicolumn{2}{c|}{\textbf{ANNOY}} & \multicolumn{2}{c|}{\textbf{FAISS-IVF}}  & \multicolumn{2}{c|}{\textbf{CAKES}} \\\cline{2-9}
        & QPS & Recall & QPS & Recall & QPS & Recall & QPS & Recall \\
        \cline{1-9}
        \hline
        1   & \num{1.93e4} & 0.782 & \num{3.98e3} & 0.686 & \num{6.98e2} & 1.000* & \num{6.20e2} & 1.000 \\\cline{1-9}
        2   & \num{2.03e4} & 0.552 & \num{3.80e3} & 0.614 & \num{3.30e2} & 1.000* & \num{2.95e2} & 1.000 \\\cline{1-9}
        4   & \num{2.18e4} & 0.394 & \num{3.69e3} & 0.637 & \num{1.65e2} & 1.000* & \num{1.76e2} & 1.000 \\\cline{1-9}
        8   & \num{2.48e4} & 0.298 & \num{3.58e3} & 0.710 & \num{7.72e1} & 1.000* & \num{1.27e2} & 1.000 \\\cline{1-9}
        16  & \num{2.68e4} & 0.210 & \num{3.50e3} & 0.690 & \num{3.98e1} & 1.000* & \num{1.47e2} & 1.000 \\\cline{1-9}
        32  & \num{2.75e4} & 0.193 & \num{3.44e3} & 0.639 & \num{2.09e1} & 0.999  & \num{1.24e2} & 1.000 \\\cline{1-9}
        64  & --           & --    & \num{3.39e3} & 0.678 & \num{8.87e0} & 0.997  & \num{1.34e2} & 1.000 \\\cline{1-9}
        128 & --           & --    & \num{3.36e3} & 0.643 & \num{4.78e0} & 0.993  & \num{1.31e2} & 1.000 \\
        \hline
    \end{tabular}
    \vskip -0.2in
\end{table}

\begin{table}
    \caption{Throughput (QPS) and Recall on the Random dataset.
    A recall value of $1.000*$ denotes imperfect recall that rounds to 1.000. In contrast with the results on the ANN Benchmark datasets reported above, with the Random dataset, we observe that CAKES's algorithms perform quite slowly.
    As with the real datasets, HNSW and ANNOY are the fastest algorithms, and CAKES exhibits perfect recall at all cardinalities.
    HNSW and ANNOY exhibit \textit{much} lower recall on this random dataset than on any of the ANN benchmark datasets.
    }
    \label{tab:results:qps-and-recall-random}
    \begin{tabular}{|l|p{1.55cm}|p{1.1cm}|p{1.55cm}|p{1.1cm}|p{1.55cm}|p{1.1cm}|p{1.55cm}|p{1.1cm}|}
        \hline
        \multirow{2}{*}{\textbf{Mult.}} & \multicolumn{2}{c|}{\textbf{HNSW}} & \multicolumn{2}{c|}{\textbf{ANNOY}} & \multicolumn{2}{c|}{\textbf{FAISS-IVF}}  & \multicolumn{2}{c|}{\textbf{CAKES}} \\\cline{2-9}
        & QPS & Recall & QPS & Recall & QPS & Recall & QPS & Recall \\
        \cline{1-9}
        \hline
        1  & \num{1.17e4} & 0.060 & \num{4.28e3} & 0.028 & \num{7.342} & 1.000* & \num{6.06e2} & 1.000 \\\cline{1-9}
        2  & \num{1.01e4} & 0.048 & \num{4.04e3} & 0.021 & \num{3.582} & 1.000* & \num{2.75e2} & 1.000 \\\cline{1-9}
        4  & \num{9.12e3} & 0.031 & \num{3.64e3} & 0.014 & \num{1.902} & 1.000* & \num{1.35e2} & 1.000 \\\cline{1-9}
        8  & \num{8.35e3} & 0.022 & \num{3.37e3} & 0.013 & \num{8.841} & 1.000* & \num{6.13e1} & 1.000 \\\cline{1-9}
        16 & \num{8.25e3} & 0.008 & \num{3.17e3} & 0.006 & \num{4.361} & 1.000* & \num{2.82e1} & 1.000 \\\cline{1-9}
        32 & --           & --    & \num{3.01e3} & 0.007 & \num{1.721} & 1.000* & \num{1.31e1} & 1.000 \\
        \hline
    \end{tabular}
    \vskip -0.2in
\end{table}

\subsection{Clustering Strategies and Number of Distance Computations}
\label{sec:results:clustering-strategies-and-number-of-distance-computations}

In addition to the scaling experiments, we also explore how four different clustering strategies affect the performance of search.
These strategies are the Cartesian product of balanced vs. unbalanced clustering, and the presence vs. absence of depth-first reordering as described in Section~\ref{sec:methods:clustering:depth-first-reordering}.
To help with this analysis, we also added some instrumentation to our implementation of CAKES to count the number of distance computations performed during search.
We note that this instrumentation significantly slows down the wall-clock time of search, so it would not be used in a real-world application.

\begin{figure}
    \begin{subfigure}[b]{0.5\textwidth}
        \includegraphics[width=0.99\textwidth]{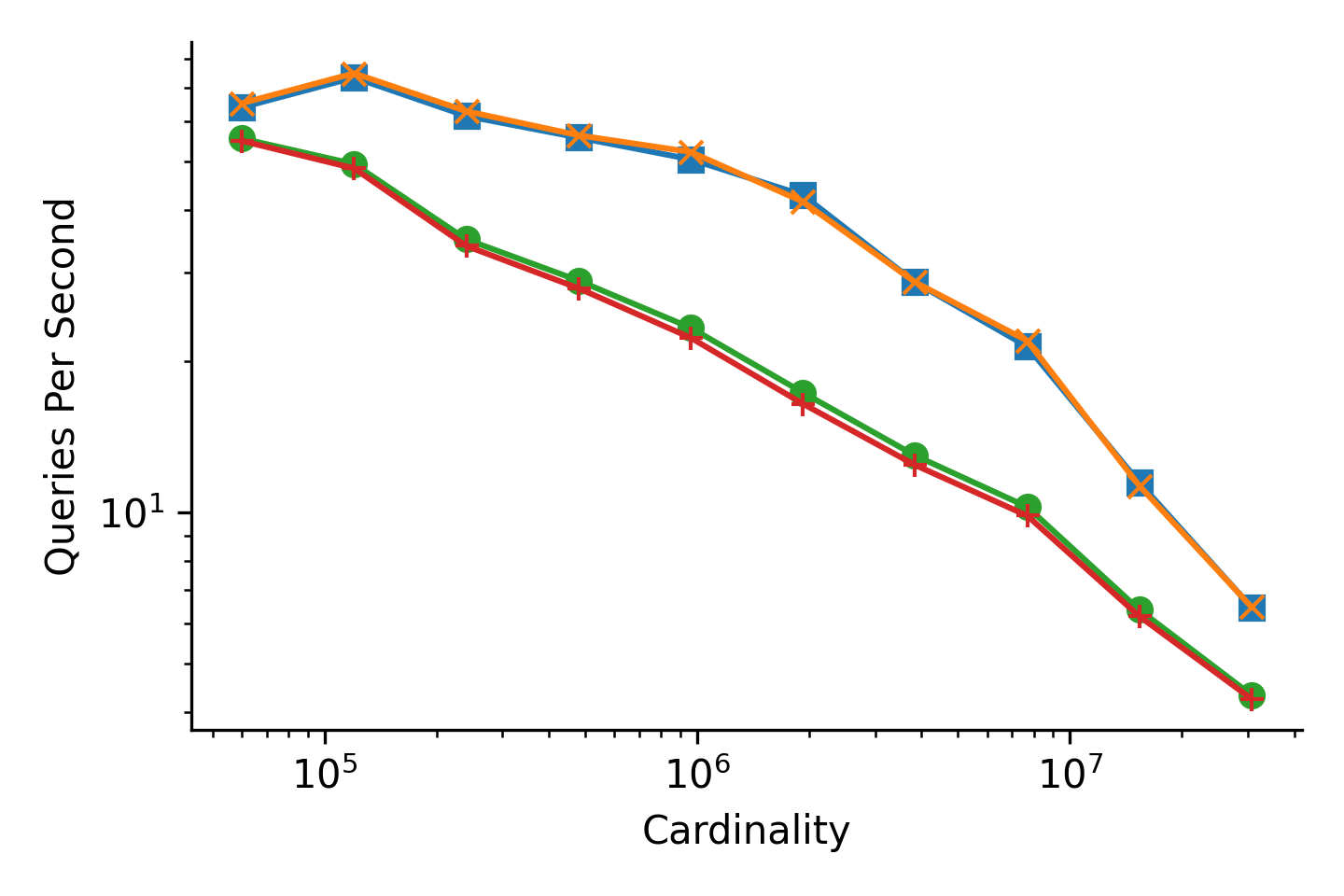}
        \subcaption{Repeated $\rho$-NN}
        \label{fig:results:fashion-mnist-counts-throughput}
    \end{subfigure}%
    \begin{subfigure}[b]{0.5\textwidth}
        \includegraphics[width=0.99\textwidth]{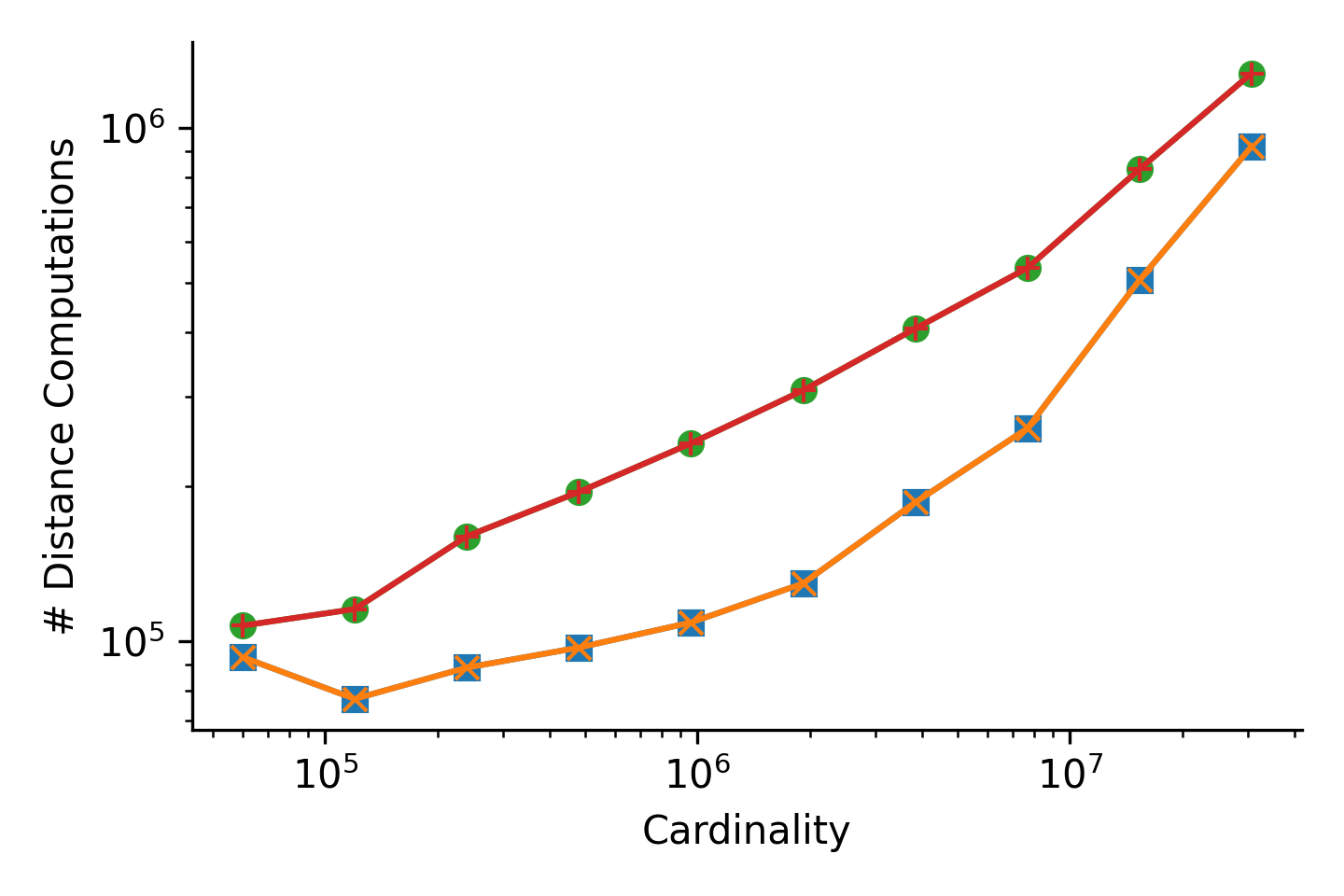}
        \subcaption{Repeated $\rho$-NN}
        \label{fig:results:glove-25-counts-counts}
    \end{subfigure}%
    \\
    \begin{subfigure}[b]{0.5\textwidth}
        \includegraphics[width=0.99\textwidth]{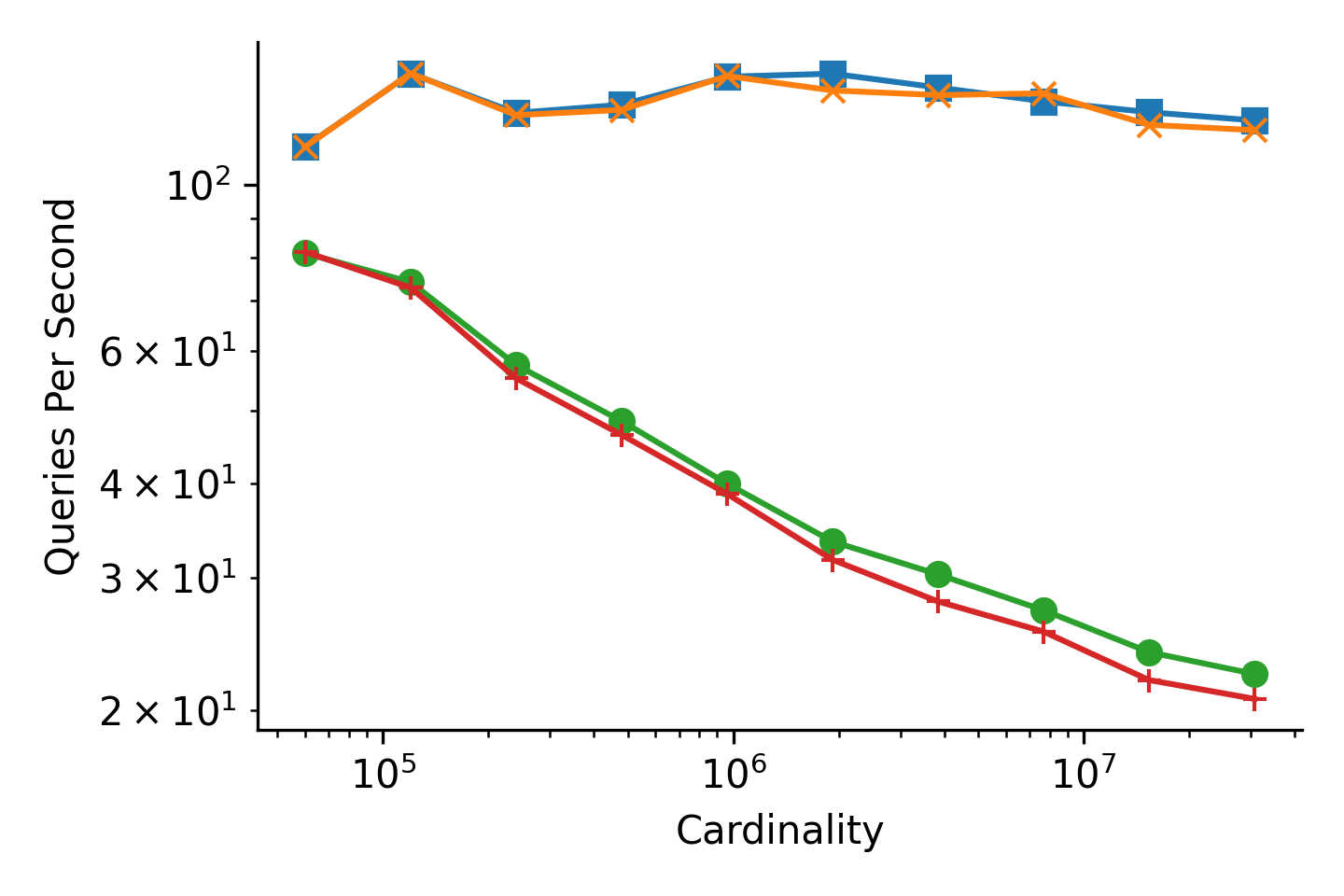}
        \subcaption{Breadth First Sieve}
        \label{fig:results:sift-counts-throughput}
    \end{subfigure}%
    \begin{subfigure}[b]{0.5\textwidth}
        \includegraphics[width=0.99\textwidth]{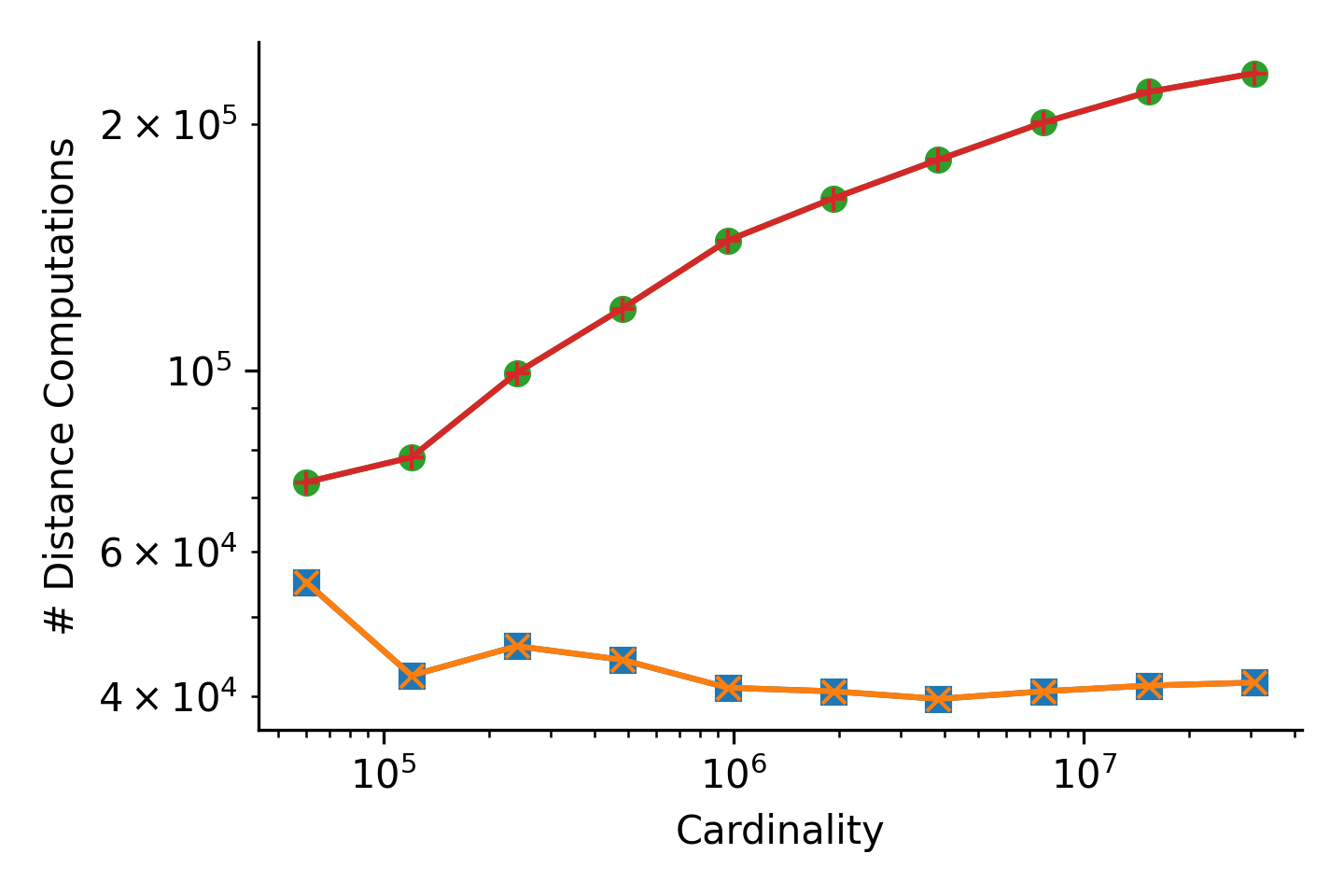}
        \subcaption{Breadth First Sieve}
        \label{fig:results:random-counts-counts}
    \end{subfigure}%
    \\
    \begin{subfigure}[b]{0.5\textwidth}
        \includegraphics[width=0.99\textwidth]{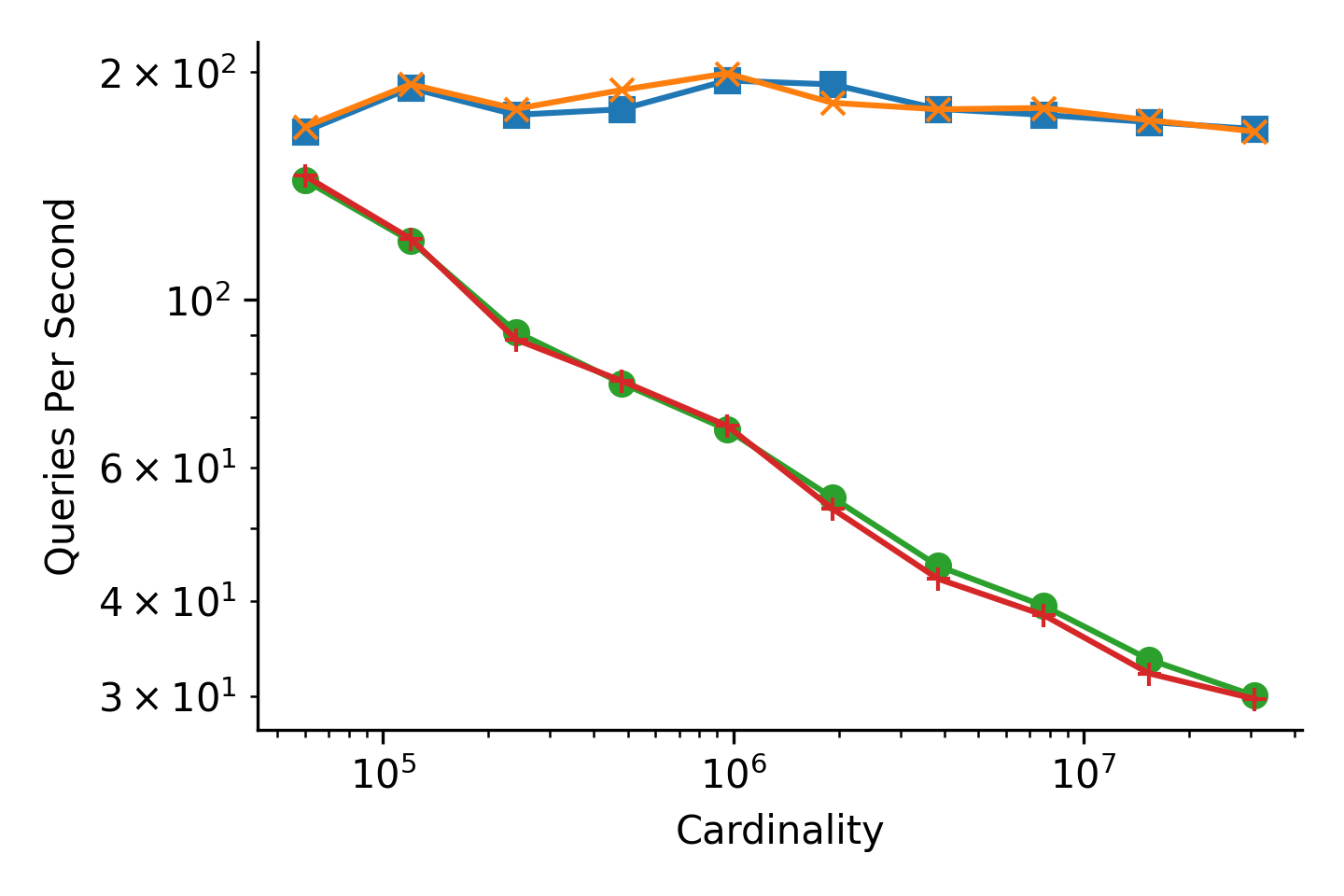}
        \subcaption{Depth First Sieve}
        \label{fig:results:silva-counts-throughput}
    \end{subfigure}%
    \begin{subfigure}[b]{0.5\textwidth}
        \includegraphics[width=0.99\textwidth]{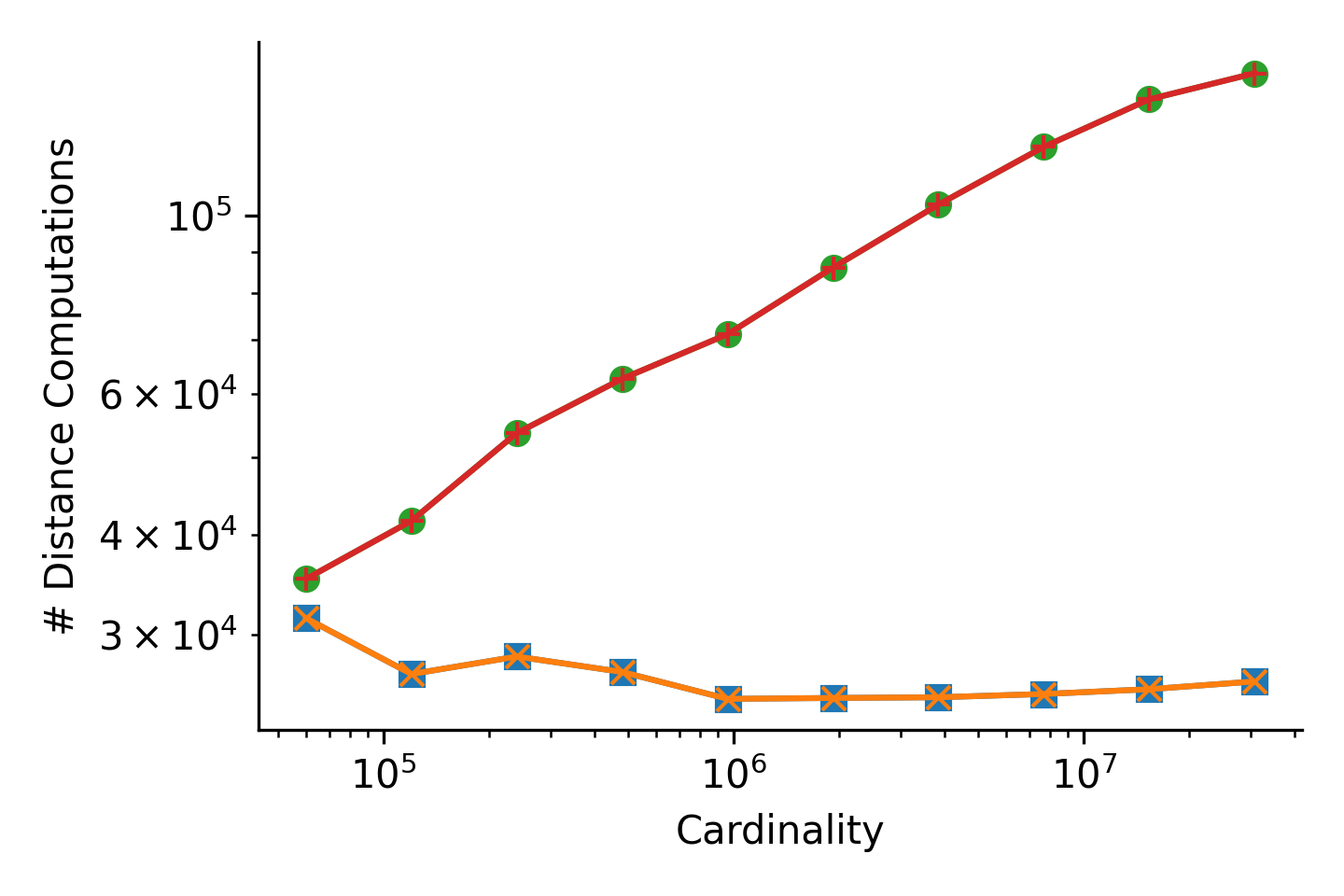}
        \subcaption{Depth First Sieve}
        \label{fig:results:radioml-counts-counts}
    \end{subfigure}%
    \\
    \begin{subfigure}[b]{0.94\textwidth}
        \centering
        \includegraphics[width=0.6\textwidth]{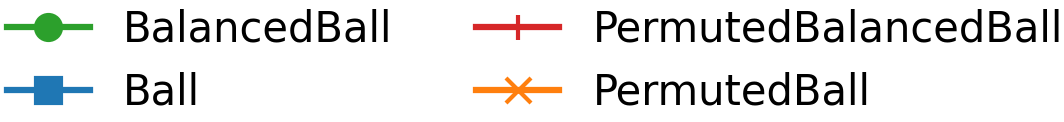}
        \label{fig:results:counts-legend}
    \end{subfigure}%
    \caption{Number of distance computations across four clustering strategies and three search algorithms on the Fashion-MNIST dataset.
    Adding the instrumentation to count the number of distance computations had the side-effect of significantly slowing down the search algorithms compared to those reported in Figure~\ref{fig:results:scaling-plots}.
    The left column shows the throughput in queries per second, while the right column shows the mean number of distance computations per query.
    The x-axis represents increasing cardinality of the dataset.}
    \label{fig:results:distance-counts}
\end{figure}

These four comparisons (referred to as Ball, BalancedBall, PermutedBall, and, PermutedBalancedBall, where depth-first-reordering is referred to as Permuted for brevity) for all three search algorithms on the Fashion-MNIST dataset are shown in Figure~\ref{fig:results:distance-counts}.
Notably, performance is not only strictly worse for the balanced approaches, but the asymptotic behavior is significantly worse;
the algorithms (such as Depth-First Sieve) that seemed to exhibit constant-time behavior in Figure~\ref{fig:results:scaling-plots}, which shows results using PermutedBall, no longer do so under a balanced clustering.
We note that on this dataset, depth-first reordering (the Permuted variants) seems to have little effect on throughput; depth-first reordering is primarily intended to improve space complexity.

%% file: sections/5-discussion.tex
\section{Discussion and Future Work}
\label{sec:discussion-and-future-work}

We have presented CAKES, a set of three algorithms for fast $k$-NN search that are generic over a variety of distance functions.
CAKES's algorithms are exact when the distance function in use is a metric (see definition in Section~\ref{sec:methods}).
Even under cosine distance, which is not a metric, CAKES's algorithms exhibit nearly perfect recall.
CAKES's algorithms are designed to be most effective when the data uphold the manifold hypothesis, or in other words, when the data are constrained to a low-dimensional manifold even when embedded in a high-dimensional space.
CAKES uses an unbalanced clustering algorithm to build a binary tree that better captures the manifold structure of the data and provides for faster search performance.
As a consequence, these algorithms do not perform well on data with random distributions, because of the absence of a manifold structure.
In Figure~\ref{fig:results:lfd-plots}, we show the extent to which each dataset we tested exhibits a manifold structure, as quantified by the percentiles of LFD.

\subsection{Performance}

As expected, our results from Figure~\ref{fig:results:scaling-plots} indicate that CAKES's algorithms scale sub-linearly with cardinality on real-world datasets with low LFD, and scale linearly on synthetic datasets with high LFD.
CAKES's performance on Sift against that on the Random dataset illustrates this phenomenon well.
As seen in Figures~\ref{fig:results:random-lfd}~and~\ref{fig:results:sift-lfd}, the LFD of the Random dataset is much higher than that of Sift, even though both datasets have the same cardinality and dimensionality.
This is unsurprising, given that the Random dataset is uniformly distributed, while Sift is not.
On the Random dataset, CAKES's speed decreases linearly as the multiplier increases, while with Sift, Depth-First Sieve and Breadth-First Sieve both exhibit nearly constant throughput.
While Repeated $\rho$-NN does not exhibit near-constant throughput with Sift, throughput decreases much more slowly than it does with the Random dataset.
Since these two datasets have the same cardinality and embedding dimension, the aforementioned discrepancies highlight how manifold structure affects algorithm performance.

We emphasize that although CAKES's algorithms are slower on the Random dataset, their recalls remain perfect.
Figure~\ref{fig:results:scaling-plots} also shows that HNSW and ANNOY have near-constant throughput as the multiplier increases, but their recall continues to decrease as the multiplier increases.
This, coupled with the increasing time and space cost of building the indices for HNSW and ANNOY, makes these algorithms unsuitable for real-world applications in which the datasets are growing exponentially.
CAKES's tree is orders of magnitude faster and less memory-intensive to build, and CAKES's algorithms, while slower than HNSW and ANNOY, still show near-constant scaling as well as perfect recall.

When written with the same notation as used in Section~\ref{sec:methods:knn-search:complexity-of-sieve-methods}, we see that the time complexity of Repeated $\rho$-NN is $\mathcal{O}(\tfrac{T}{\lceil d \rceil} + L)$, where $T$ is the time complexity of tree-search and $L$ is the time complexity of leaf-search.
Even though Repeated $\rho$-NN has the least time cost (compared to $\mathcal{O}\left((T + L)\log{(T+L)}\right)$ for Breadth-First Sieve and $\mathcal{O}(T\log{T} + L\log{k})$ for Depth-First Sieve) of the three CAKES algorithms, it is not always the fastest algorithm empirically.
We believe that some of this discrepancy can be explained by the fact that Repeated $\rho$-NN can significantly ``overshoots'' the correct radius for $k$ hits ($\rho_k$) during tree-search.
This causes leaf-search to exhaustively search many more clusters than necessary, rendering the true scaling factor higher than that in Equation~\ref{eq:methods:repeated-rnn-complexity}.
This overshot can occur when the LFDs of clusters near the query are not concentrated around their expectation.
For example, if most clusters near the query have very low LFD except for one anomaly with very high LFD,
the harmonic mean LFD $\mu$ can still be low, so the factor of radial increase in (Equation)~\ref{eq:methods:repeated-rnn-factor} may be much larger than necessary for guaranteeing $k$ hits.
This suggests that rather than using the reciprocal of the harmonic mean LFD in~\ref{eq:methods:repeated-rnn-factor}, we may achieve better results with a mean that is more sensitive to high outliers, such as the geometric mean.
We leave it as an avenue for future work to characterize when Repeated $\rho$-NN significantly overestimates the correct radius $\rho_k$ and to improve upon the factor of radial increase in Equation~\ref{eq:methods:repeated-rnn-factor} so that this occurs less frequently and with less severity.

While the fastest CAKES algorithm can vary by dataset and cardinality, we note that on all three ANN-benchmark datasets (Fashion-MNIST, Glove-25, and Sift), Depth-First Sieve and Breadth-First Sieve show nearly constant throughput as the cardinality increases.
This observation warrants further investigation, but it is especially promising given that both algorithms consistently outperform linear search for high cardinalities.
Finally, we emphasize that the variation in performance of our algorithms across different datasets and cardinalities supports our use of an auto-tuning function (as discussed in Section~\ref{sec:methods:auto-tuning}) to select the best algorithm for a given dataset.
Future work is needed to better characterize those datasets for which each algorithm performs best, but our results suggest that a more sophisticated auto-tuning function could be developed to select the fastest algorithm based on the properties of the dataset.

We also stress that CAKES's algorithms are designed to work well for \textit{big} data, and our results support this claim;
we observe that while linear search can outperform CAKES on datasets with small cardinality, CAKES always overtakes linear search at a large enough cardinality for each of the ANN-benchmark datasets we tested.
Moreover, we observe that CAKES's algorithms outperform linear search at a lower cardinality when the dataset has a lower LFD.
For example, Sift has much higher LFD than Fashion-Mnist and Glove-25, and we observe that CAKES overtakes linear search at a cardinality of $10^7$ for Sift, as opposed to about $10^5$ and $10^6$ for Fashion-Mnist and Glove-25, respectively.
For the Random dataset, which has the highest LFD, CAKES's algorithms \textit{never} outperform linear search.
These observations support our claim that CAKES performs well on datasets that arise from constrained generating phenomena, even as the cardinalities of these datasets grow exponentially.

\subsection{Applicability}

We note that despite the genericity of CAKES, effectively applying it to real-world datasets may require domain-specific knowledge.
As an example, consider the domain of protein sequence search; protein sequences are of highly variable length (30--1000 amino acids are typical, with some outliers) and typically comprise a 20-letter alphabet.
There have been numerous protein sequence search algorithms~\cite{kim2021entrance, daniels2013compressive, yu2015entropy, steinegger2018clustering}, and the one thing they have in common besides the application domain is that their development required deep domain-specific knowledge.
While simple edit distance may be a poor choice for clustering when sequence lengths exhibit such variance, other choices of distance metrics and preprocessing could be used by CAKES, such as representing each sequence as a set of $k$-mers (short substrings of length $k$) and choosing Jaccard distance on these sets as the metric; this approach was chosen by~\cite{kim2021entrance}.
If the distance function uses a global sequence alignment algorithm, such as Needleman-Wunsch~\cite{needleman1970general}, the dataset may also be preprocessed into a number of mutually exclusive subsets based on the lengths of the sequences, and a separate CAKES tree may be built for each subset.
At search time, the query sequence would be assigned to the appropriate tree based on its length, and the search would be performed on that tree, and on a small number of trees on adjacent lengths' bins.

\subsection{Future Work}

The questions raised by this study suggest several additional avenues for future work.
A comparison across more datasets is in order, as is further analysis with other distance functions that existing methods, such as FAISS, HNSW and ANNOY, do not support.
These include Wasserstein distance~\cite{vallender1974calculation} for probability distributions (particularly for high dimensional distributions) and Tanimoto distance~\cite{bajusz2015tanimoto} for comparing molecular structures by their maximal common subgraphs.
Incorporating new distance functions in CAKES requires only that a Rust implementation of the distance function be provided.

Additionally, we intend to improve the data augmentation process described in Section~\ref{sec:methods:synthetic-data} to better preserve the topological structure of the underlying dataset.
In particular, we expect that favoring augmentation along the first few principal components of the local manifold may allow for more realistic data augmentation.

We also plan to investigate hierarchical data compression by representing differences at each level of the binary tree, particularly for string or genomic data, where all differences are discrete.
Such an encoding would allow us to store the data in a compressed format, and to perform search on the compressed data without decompressing it in its entirety.
This would allow us to perform fast search on datasets that are otherwise too large to fit in memory.
An exploration of compression and compressed search is another avenue for future work.

We also would like to explore the use of CAKES in a streaming environment.
This would require the ability to perform ``online'' updates to the tree as points are added to or deleted from the dataset.
Such online updates would take advantage of the fast search algorithms provided by CAKES.  
CAKES can also be used to extend anomaly detection in CHAODA~\cite{ishaq2021clustered}.
We could add to CHAODA's ensemble of graph-based anomaly detection methods by using the distribution of distances among the $k$ nearest neighbors of cluster centers.

\subsection{Availability}

CLAM and CAKES are implemented in the Rust programming language, and the source code is available under an MIT license at https://github.com/URI-ABD/clam.

%% file: paper.bbl
\newcommand{\noop}[1]{}
\begin{thebibliography}{10}

\bibitem{faissivf}
{\em Faiss-ivf}.
\newblock \url{ https://github.com/facebookresearch/faiss/wiki/Faiss-indexes },
  2016.

\bibitem{aumuller2020ann}
{\sc M.~Aum{\"u}ller, E.~Bernhardsson, and A.~Faithfull}, {\em Ann-benchmarks:
  A benchmarking tool for approximate nearest neighbor algorithms}, Information
  Systems, 87 (2020), p.~101374.

\bibitem{bajusz2015tanimoto}
{\sc D.~Bajusz, A.~R{\'a}cz, and K.~H{\'e}berger}, {\em Why is tanimoto index
  an appropriate choice for fingerprint-based similarity calculations?},
  Journal of cheminformatics, 7 (2015), pp.~1--13.

\bibitem{annoy}
{\sc E.~Bernhardsson}, {\em Annoy}.
\newblock \url{https://github.com/spotify/annoy}, 2015.

\bibitem{boguna2009navigability}
{\sc M.~Boguna, D.~Krioukov, and K.~C. Claffy}, {\em Navigability of complex
  networks}, Nature Physics, 5 (2009), pp.~74--80.

\bibitem{2020arXiv200514165B}
{\sc T.~B. {Brown}, B.~{Mann}, N.~{Ryder}, M.~{Subbiah}, J.~{Kaplan},
  P.~{Dhariwal}, A.~{Neelakantan}, P.~{Shyam}, G.~{Sastry}, A.~{Askell},
  S.~{Agarwal}, A.~{Herbert-Voss}, G.~{Krueger}, T.~{Henighan}, R.~{Child},
  A.~{Ramesh}, D.~M. {Ziegler}, J.~{Wu}, C.~{Winter}, C.~{Hesse}, M.~{Chen},
  E.~{Sigler}, M.~{Litwin}, S.~{Gray}, B.~{Chess}, J.~{Clark}, C.~{Berner},
  S.~{McCandlish}, A.~{Radford}, I.~{Sutskever}, and D.~{Amodei}}, {\em
  {Language Models are Few-Shot Learners}}, arXiv e-prints,  (2020),
  \url{https://doi.org/10.48550/arXiv.2005.14165}.

\bibitem{cover1967nearest}
{\sc T.~M. Cover, P.~Hart, et~al.}, {\em Nearest neighbor pattern
  classification}, IEEE Transactions on Information Theory, 13 (1967),
  pp.~21--27.

\bibitem{daniels2013compressive}
{\sc N.~M. Daniels, A.~Gallant, J.~Peng, L.~J. Cowen, M.~Baym, and B.~Berger},
  {\em Compressive genomics for protein databases}, Bioinformatics, 29 (2013),
  pp.~i283--i290.

\bibitem{dosovitskiy2020image}
{\sc A.~Dosovitskiy, L.~Beyer, A.~Kolesnikov, D.~Weissenborn, X.~Zhai,
  T.~Unterthiner, M.~Dehghani, M.~Minderer, G.~Heigold, S.~Gelly, et~al.}, {\em
  An image is worth 16x16 words: Transformers for image recognition at scale},
  arXiv preprint arXiv:2010.11929,  (2020).

\bibitem{fefferman2016testing}
{\sc C.~Fefferman, S.~Mitter, and H.~Narayanan}, {\em Testing the manifold
  hypothesis}, Journal of the American Mathematical Society, 29 (2016),
  pp.~983--1049.

\bibitem{fix1952discriminatory}
{\sc E.~Fix and J.~L. Hodges~Jr}, {\em Discriminatory analysis-nonparametric
  discrimination: Small sample performance}, tech. report, California Univ
  Berkeley, 1952.

\bibitem{gao2023high}
{\sc J.~Gao and C.~Long}, {\em High-dimensional approximate nearest neighbor
  search: with reliable and efficient distance comparison operations},
  Proceedings of the ACM on Management of Data, 1 (2023), pp.~1--27.

\bibitem{gold2018dynamic}
{\sc O.~Gold and M.~Sharir}, {\em Dynamic time warping and geometric edit
  distance: Breaking the quadratic barrier}, ACM Transactions on Algorithms
  (TALG), 14 (2018), pp.~1--17.

\bibitem{hoare1961algorithm}
{\sc C.~A. Hoare}, {\em Algorithm 65: find}, Communications of the ACM, 4
  (1961), pp.~321--322.

\bibitem{hu2016distance}
{\sc L.-Y. Hu, M.-W. Huang, S.-W. Ke, and C.-F. Tsai}, {\em The distance
  function effect on k-nearest neighbor classification for medical datasets},
  SpringerPlus, 5 (2016), pp.~1--9.

\bibitem{ishaq2021clustered}
{\sc N.~Ishaq, T.~J. Howard, and N.~M. Daniels}, {\em Clustered hierarchical
  anomaly and outlier detection algorithms}, in 2021 IEEE International
  Conference on Big Data (Big Data), IEEE, 2021, pp.~5163--5174.

\bibitem{ishaq2019clustered}
{\sc N.~Ishaq, G.~Student, and N.~M. Daniels}, {\em Clustered hierarchical
  entropy-scaling search of astronomical and biological data}, in 2019 IEEE
  International Conference on Big Data (Big Data), IEEE, 2019, pp.~780--789.

\bibitem{johnson2019billion}
{\sc J.~Johnson, M.~Douze, and H.~J{\'e}gou}, {\em Billion-scale similarity
  search with {GPUs}}, IEEE Transactions on Big Data, 7 (2019), pp.~535--547.

\bibitem{kahn2011future}
{\sc S.~D. Kahn}, {\em On the future of genomic data}, Science, 331 (2011),
  pp.~728--729.

\bibitem{kent1990signature}
{\sc A.~Kent, R.~Sacks-Davis, and K.~Ramamohanarao}, {\em A signature file
  scheme based on multiple organizations for indexing very large text
  databases}, Journal of the american society for information science, 41
  (1990), pp.~508--534.

\bibitem{kim2021entrance}
{\sc Y.~Kim, Z.~Guo, J.~A. Robertson, B.~Reidys, Z.~Zhang, and L.~S. Heath},
  {\em Entrance: Exploration of entropy scaling ball cover search in protein
  sequences}, bioRxiv,  (2021),
  \url{https://doi.org/10.1101/2021.05.31.446458},
  \url{https://www.biorxiv.org/content/early/2021/05/31/2021.05.31.446458},
  \url{https://arxiv.org/abs/https://www.biorxiv.org/content/early/2021/05/31/2021.05.31.446458.full.pdf}.

\bibitem{kleinberg2000navigation}
{\sc J.~M. Kleinberg}, {\em Navigation in a small world}, Nature, 406 (2000),
  pp.~845--845.

\bibitem{levenshtein1966binary}
{\sc V.~I. Levenshtein et~al.}, {\em Binary codes capable of correcting
  deletions, insertions, and reversals}, in Soviet physics doklady, vol.~10,
  Soviet Union, 1966, pp.~707--710.

\bibitem{malkov2016hnsw}
{\sc Y.~Malkov and D.~A. Yashunin}, {\em Efficient and robust approximate
  nearest neighbor search using hierarchical navigable small world graphs},
  IEEE Transactions on Pattern Analysis and Machine Intelligence, 42 (2016),
  pp.~824--836, \url{https://api.semanticscholar.org/CorpusID:8915893}.

\bibitem{muller2007dynamic}
{\sc M.~M{\"u}ller}, {\em Dynamic time warping}, Information retrieval for
  music and motion,  (2007), pp.~69--84.

\bibitem{needleman1970general}
{\sc S.~B. Needleman and C.~D. Wunsch}, {\em A general method applicable to the
  search for similarities in the amino acid sequence of two proteins}, Journal
  of molecular biology, 48 (1970), pp.~443--453.

\bibitem{OpenAI2023GPT4TR}
{\sc OpenAI}, {\em Gpt-4 technical report}, ArXiv, abs/2303.08774 (2023).

\bibitem{oshea2018radioml}
{\sc T.~J. O’Shea, T.~Roy, and T.~C. Clancy}, {\em Over-the-air deep learning
  based radio signal classification}, IEEE Journal of Selected Topics in Signal
  Processing, 12 (2018), pp.~168--179,
  \url{https://doi.org/10.1109/JSTSP.2018.2797022}.

\bibitem{10.1093/nar/gks1219}
{\sc C.~Quast, E.~Pruesse, P.~Yilmaz, J.~Gerken, T.~Schweer, P.~Yarza,
  J.~Peplies, and F.~O. Glöckner}, {\em {The SILVA ribosomal RNA gene database
  project: improved data processing and web-based tools}}, Nucleic Acids
  Research, 41 (2012), pp.~D590--D596,
  \url{https://doi.org/10.1093/nar/gks1219}.

\bibitem{radford2021learning}
{\sc A.~Radford, J.~W. Kim, C.~Hallacy, A.~Ramesh, G.~Goh, S.~Agarwal,
  G.~Sastry, A.~Askell, P.~Mishkin, J.~Clark, et~al.}, {\em Learning
  transferable visual models from natural language supervision}, in
  International Conference on Machine Learning, PMLR, 2021, pp.~8748--8763.

\bibitem{sacks1987multikey}
{\sc R.~Sacks-Davis, A.~Kent, and K.~Ramamohanarao}, {\em Multikey access
  methods based on superimposed coding techniques}, ACM Transactions on
  Database Systems (TODS), 12 (1987), pp.~655--696.

\bibitem{steinegger2018clustering}
{\sc M.~Steinegger and J.~S{\"o}ding}, {\em Clustering huge protein sequence
  sets in linear time}, Nature communications, 9 (2018), p.~2542.

\bibitem{suyanto2022knnclassifier}
{\sc S.~Suyanto, P.~E. Yunanto, T.~Wahyuningrum, and S.~Khomsah}, {\em A
  multi-voter multi-commission nearest neighbor classifier}, Journal of King
  Saud University - Computer and Information Sciences, 34 (2022),
  pp.~6292--6302, \url{https://doi.org/10.1016/j.jksuci.2022.01.018}.

\bibitem{Touvron2023Llama2O}
{\sc H.~Touvron, L.~Martin, K.~R. Stone, P.~Albert, A.~Almahairi, Y.~Babaei,
  N.~Bashlykov, S.~Batra, P.~Bhargava, S.~Bhosale, D.~M. Bikel, L.~Blecher,
  C.~C. Ferrer, M.~Chen, G.~Cucurull, D.~Esiobu, J.~Fernandes, J.~Fu, W.~Fu,
  B.~Fuller, C.~Gao, V.~Goswami, N.~Goyal, A.~S. Hartshorn, S.~Hosseini,
  R.~Hou, H.~Inan, M.~Kardas, V.~Kerkez, M.~Khabsa, I.~M. Kloumann, A.~V.
  Korenev, P.~S. Koura, M.-A. Lachaux, T.~Lavril, J.~Lee, D.~Liskovich, Y.~Lu,
  Y.~Mao, X.~Martinet, T.~Mihaylov, P.~Mishra, I.~Molybog, Y.~Nie, A.~Poulton,
  J.~Reizenstein, R.~Rungta, K.~Saladi, A.~Schelten, R.~Silva, E.~M. Smith,
  R.~Subramanian, X.~Tan, B.~Tang, R.~Taylor, A.~Williams, J.~X. Kuan, P.~Xu,
  Z.~Yan, I.~Zarov, Y.~Zhang, A.~Fan, M.~Kambadur, S.~Narang, A.~Rodriguez,
  R.~Stojnic, S.~Edunov, and T.~Scialom}, {\em Llama 2: Open foundation and
  fine-tuned chat models}, ArXiv, abs/2307.09288 (2023),
  \url{https://api.semanticscholar.org/CorpusID:259950998}.

\bibitem{ukey2023survey}
{\sc N.~Ukey, Z.~Yang, B.~Li, G.~Zhang, Y.~Hu, and W.~Zhang}, {\em Survey on
  exact knn queries over high-dimensional data space}, Sensors, 23 (2023),
  p.~629.

\bibitem{vallender1974calculation}
{\sc S.~Vallender}, {\em Calculation of the wasserstein distance between
  probability distributions on the line}, Theory of Probability \& Its
  Applications, 18 (1974), pp.~784--786.

\bibitem{yu2015entropy}
{\sc Y.~W. Yu, N.~Daniels, D.~C. Danko, and B.~Berger}, {\em Entropy-scaling
  search of massive biological data}, Cell Systems, 1 (2015), pp.~130--140.

\bibitem{zhang2022imbalanced}
{\sc J.~Zhang, T.~Wang, W.~W. Ng, and W.~Pedrycz}, {\em Ensembling
  perturbation-based oversamplers for imbalanced datasets}, Neurocomput., 479
  (2022), p.~1–11, \url{https://doi.org/10.1016/j.neucom.2022.01.049}.

\end{thebibliography}
